\documentclass{article}
\usepackage{arxiv}

\usepackage[numbers]{natbib}
\usepackage{amsmath}
\usepackage{amsfonts}
\usepackage{dsfont}
\usepackage{rotating}
\usepackage{adjustbox}  
\usepackage{graphicx}
\usepackage{url}
\usepackage{caption}
\usepackage[ruled, vlined]{algorithm2e}
\usepackage{bbold} 
\usepackage{mathtools}
\usepackage{color}
\usepackage[latin9]{inputenc}

\DeclareMathOperator*{\argmin}{argmin}

\numberwithin{equation}{section}

\title{On a simultaneous parameter inference and missing data imputation for nonstationary autoregressive models}%

\author{%
  Dimitri Igadlov \\
  Universit\'a della Svizzera italiana, Switzerland\\
  Lugano, Switzerland \\  \smallskip
  \texttt{dimitri.igdalov@usi.ch} \\
  \And
  Olga Kaiser \\
  Universit\'a della Svizzera italiana, Switzerland (current NNAISENSE) \\
  Lugano, Switzerland \\  \smallskip
  \texttt{olga.kaiser@nnaisense.com}
  \And
  Illia Horenko   \\
  Universit\'a della Svizzera italiana, Switzerland\\
  Lugano, Switzerland \\  \smallskip
  \texttt{illia.horenko@usi.ch} \\
}

\begin{document}
\maketitle

\begin{abstract}
This work addresses the problem of missing data in time-series analysis focusing on (a) estimation of model parameters in the presence of missing data and (b) reconstruction of missing data. Standard approaches used to solve these problems like the maximum likelihood estimation or the Bayesian inference rely on a priori assumptions like the Gaussian or stationary behavior of missing data and might lead to biased results where these assumptions are unfulfilled. In order to go beyond, we extend the Finite Element Methodology (FEM) for Vector Auto-Regressive models with eXogenous factors and bounded variation of the model parameters (FEM-VARX) towards handling the missing data problem. The presented approach estimates the model parameters and reconstructs the missing data in the considered time series and in the involved exogenous factors, simultaneously. The resulting computational framework was compared to the state-of-art methodologies on a set of test-cases and is available as open-source software. %
\end{abstract}


%
%
\section{Introduction}  
\label{sec:intro}
Many scientific and industrial data sets exhibit missing measurements. Examples can be found, among others, in social science, medical studies and in climatological/meteorological measurements. 
Until the 1970s the problem of missing data was primarily addressed either by deleting the missing values, or by filling in the gaps by applying, for instance, the "hot deck" imputation, which replaces a missing variable by a similar observed one from the same sample~\citep{graham2009missing,van2012flexible,garcia2015data}. %
Today, the widely used concept for statistical analysis in presence of missing data is based on the "missing data mechanism" and was introduced in 1976 by Rubin~\citep{RUBIN01121976}. This mechanism describes the relationship between the observed variables and their probability of being missing. According to Rubin~\citep{RUBIN01121976}, each value in the observed data set has a probability to be missing. The underlying process that assigns these probabilities is called the "missing data mechanism"~\citep{RUBIN01121976, van2012flexible}. Resulting from different mechanisms the missing data can then be classified into on of the three categories: the missing data is said to be (i) Missing Completely At Random (MCAR) if the probability of being missing is the same for all the variables in the considered data set~\citep{enders2010applied, van2012flexible}; (ii) Missing At Random (MAR), if the probability of being missing is the same only within some subsets characterized by observed data; (iii) Missing Not At Random (MNAR), if the missing data is either MCAR nor MAR. %
Rubin's classification is important for understanding how different missing data mechanisms affect the performance of different methodologies for handling the problem of missing data. For instance, in the case when the missing data is MCAR or MAR, maximum likelihood estimation and multiple imputation techniques provide unbiased estimators~\citep{RUBIN01121976,enders2010applied}. However, in real applications there is little or no information available about the missing data mechanism~\citep{kaambwa2012methods,rubin2004multiple,enders2010applied}. For more discussion on how to determine the correct missing data mechanism please refer to~\citep{kaambwa2012methods,cohen2013applied}. In the following, the state-of-the-art techniques for handling the missing data problem are reviewed.\\ %
For a given sample $Y = \left(Y_1,\dots,Y_n\right)$, with $Y_i\in\mathbb R^d,\,i=1,\dots,n$, one of the goals in statistical analysis is to find a descriptive model for $Y$, i.e., the probability density function $f(Y;\theta)$ characterized by the model parameter $\theta$. In order to obtain the optimal model parameter $\theta$ for $Y$, Maximum Likelihood (ML) approach is often applied. ML is based on the idea that the original model parameter maximizes the likelihood, or rather the log-likelihood $l\left(\theta;Y\right)$, of the joint occurrence of $Y_1,\dots,Y_n$. %
An alternative way for inference of the model parameter is provided by the Bayesian statistics. Thereby, a priori knowledge about the parameter model, the prior distribution $\pi\left(\theta\right)$, is incorporated for the inference by applying the Bayes Theorem. The resulting posterior $f(\theta|y)$ can be used for instance for the estimation of the expected value and the variance of the model parameter $\theta$. For further details on ML and Bayesian statistics please refer, for instance, to~\citep{davison2003statistical}.\\%
In cases when $Y$ contains missing data, the log-likelihood can not be directly evaluated. One possible approach is to delete all the dimensions with missing data and to obtain the optimal model parameters on the reduced sample. A more efficient way is provided by the Full Information Maximization Likelihood (FIML) estimator proposed by Finkbeiner in 1975~\citep{enders2001performance,enders2001relative}. FIML computes the likelihood for each $i=1,\dots,n$ separately, by reducing $Y_i$ to a vector $y_i\in\mathbb R^{v_i}$ which contains only the observed dimensions of $Y_i$ ($v_i$ is the number of observed dimensions for fixed $i$). For instance, assuming multivariate Gaussian distribution of $Y$, the log-likelihood for the reduced vector is given by %
\begin{align}
	\label{eq:dimwise_ll}
	l_i = -\frac{1}{2}v_i\log(2\Pi) -\frac{1}{2}\log(|\Sigma_i|) -\frac{1}{2}\left(y_i-\mu_i\right)^{\dagger}\Sigma_i^{-1}\left(y_i-\mu_i\right),
\end{align} 
where $\Sigma_i$ and $\mu_i$ are the reduced covariance and the mean, respectively. Please note, that for some cases reduced covariance matrix $\Sigma_i$ might be not positive definite, and the corresponding $l_i$ can not be evaluated. The overall log-likelihood is the sum of $l_i$. FIML is a parametric approach that provides an efficient estimator of the Gaussian model parameters exploiting full available information. Deploying parametric representation for $\mu$, FIML can be applied to a wide range of linear models, e.g., the widely used multiple linear regression~\citep{enders2001performance,enders2001relative, graham2009missing}. However, in many real applications the normality assumption might be too restrictive. \\%
A more general approach is to preserve the concepts of both, the ML estimation procedure and the Bayesian inference. The idea here is to split the data into observed and missing part, i.e., $Y = \left(Y_{obs},Y_{mis}\right)$. Then, the distribution of $f(Y|\theta)$ equals to the joint distribution $f\left(Y_{obs},Y_{mis}|\theta\right)$~\citep{little2002statistical}. With this, the marginal distribution for observed data only can be evaluated as%
\begin{align}
	\label{eq:Marginal_Yobs}
	f\left(Y_{obs}|\theta\right)=\int f\left(Y_{obs},Y_{mis}|\theta\right)dY_{mis}.
\end{align}
Rubin showed in 1976 that if the missing data is MAR, the missing data mechanism can be ignored. Then, the correct likelihood for $\theta$, denoted by $L_{ign}(\theta|Y_{obs})$, fulfills%
\begin{align}
	\label{eq:Likelihood_ign}
	L_{ign}(\theta|Y_{obs})\propto f\left(Y_{obs}|\theta\right),
\end{align}
the subscript "ign" points to the fact that the likelihood was obtained by ignoring the missing data mechanism~\citep{little2002statistical}. The corresponding ML estimates are obtained by maximizing (\ref{eq:Likelihood_ign}) with respect to $\theta$. Analogously, in the case when the missing data is MAR, the "ignorable Bayesian inference" for the model parameter $\theta$ is based on the posterior with%
\begin{align}
	\label{eq:posterior_ign}
	p(\theta|Y_{obs})\propto \pi(\theta)L_{ign}(\theta|Y_{obs}),
\end{align}
which involves observed data only. A detailed derivation of (\ref{eq:Likelihood_ign}-\ref{eq:posterior_ign}) with examples can be found in~\citep{little2002statistical}.\\ %
The direct application of this extensions for inference of $\theta$ in presence of missing data is limited by the fact that often there is no closed parametric form for $f\left(Y_{obs}|\theta\right)$ in (\ref{eq:Marginal_Yobs}) and numerical computational methods are required. One of such methods is the iterative Expectation Maximization (EM) algorithm, first introduced by Sundberg~\citep{sundberg1974maximum,sundberg1976iterative} and generalized by Dempster et al.~\citep{dempster1977maximum}. The EM algorithm maximizes $L_{ign}(\theta|Y_{obs})$ in (\ref{eq:Likelihood_ign}) by iteratively maximizing the likelihood for complete data~\citep{dempster1977maximum,meng1993maximum}. Thereby, in the E-step of the algorithm the expected log-likelihood for complete data is estimated conditioned on the distribution of $Y_{mis}$ and for given $Y_{obs}$ and a current estimate of $\theta$. In the M-step the parameter $\theta$ is updated by maximizing the expected log-likelihood obtained from the E-step~\citep{little2002statistical}.\\ %
Despite of the conceptual simplicity of the EM algorithm, the evaluation depends on the chosen distributions for observed and missing data. So, in the M-step the expected log-likelihood might be not assessable analytically. To overcome this, Monte Carlo based approaches can be deployed to provide an approximation to the expected log-likelihood~\citep{wei1990monte}. The simplicity of EM is lost if there is no closed form for the maximizer in M-step. Then, numerical maximization techniques are required, for instance, the Generalized EM (GEM) algorithm and the Expectation/Conditional Maximization (ECM) algorithm~\citep{dempster1977maximum,meng1993maximum}. Accelerated versions are obtained by incorporating gradient-based techniques. More details and further alternative variants of the EM algorithm can be found in~\citep{little2002statistical,mclachlan2007algorithm}. A disadvantage of all the maximum likelihood based methods in a presence of missing data is that they tend to underestimate the standard errors of the model parameters~\citep{garcia2015data}.\\%
In order to address this problem, Rubin introduced the idea of Multiple Imputation (MI) in 1987. Assuming a data model, MI based approaches first create $m$ different complete data sets, then analyze these using standard techniques and lastly combine the $m$ different results to obtain the "averaged" model parameters~\citep{rubin2004multiple,schafer1997analysis}. The choice of an appropriate data model for MI affects directly the final results and leads to statistical or machine learning based MI~\citep{rubin2004multiple,van2012flexible,garcia2015data}.\\ %
Statistical (or model based) MI iteratively fills in the missing values and estimates the model parameters in line with the ML approach. In contrast to ML (which refers to single imputation), MI imputes the missing values multiple times by involving, for instance, linear regression with Gaussian distributed residuals. Thereby, prior estimates for the mean value and the covariance matrix are used to formulate the regression equation~\citep{enders2010applied, van2012flexible}. The Bayesian formulation of this approach is more general: the missing values are drawn from a conditional distribution dependent on the previous estimates of the model parameters and the observed data. For this purpose Markov Chain Monte Carlo (MCMC) based methods can be used~\citep{enders2010applied,garcia2015data}. \\%
The idea of imputation methods based on machine learning is to construct a predictive model for the missing data based on observed data only. Different approaches exist, for instance, imputation can be carried out by the (weighted) K-Nearest Neighbor (KNN) approach, where each missing value is reconstructed by the average of its (weighted) K nearest neighbors. An extension of KNN imputation are clustering techniques like K-Means/Fuzzy K-Means. Trough clustering the data are ordered according to their similarity, such that the nearest neighbors are now defined as data from the same cluster~\citep{garcia2015data}. Regression and autoregression based techniques represent a further class of imputations, for instance,  regressive models  the Least Squares (LS) for linear regression and Support Vector Machine (SVM) or Neuronal Networks (NN) based techniques for non-linear regression~\citep{little1992regression, gupta1996estimating, rubin2004multiple, van2012flexible, garcia2015data, nelwamondo2007missing, bashir2018handling}. A detailed classification of machine learning based methods for handling the missing data problem is given in~\citep{garcia2015data}.\\ %
%
%
A further class of models that are able to handle missing data are the so called State Space Models (SSM). A linear state space model assumes that the underlying dynamics of a system is characterized by a linear process $x(t)$ which can not be observed directly. Instead, its modified/disturbed behavior, denoted by $y(t)$, is observed. The relation between $x(t)$ and $y(t)$ is described by a state space model. For instance, a linear Gaussian formulation of a state space model is given by  %
\begin{align}
	x(t) = Ax\left(t-1\right) + Bu(t)+ \eta_t,\,\text{ with }\eta_t\sim\mathcal N\left(0,\Sigma_{\eta}(t)\right),\label{eq:state_eqI}\\%
	y(t) = Cx(t) + \epsilon_t,\,\text{ with }\epsilon_t\sim\mathcal N\left(0,\Sigma_t\right)\,t=1,\dots,N_T\label{eq:space_eqII},%
\end{align}
where $x(t)\in\mathbb R^{dimx}$ is denoted as the state vector and $y(t)\in\mathbb R^{dimy}$ as the output. Matrix $A\in\mathbb R^{dimx\times dimx}$ describes the interaction between the elements of $x(t)$, $u(t)\in\mathbb R^{dimu}$ is the vector of external factors and $B\in\mathbb R^{dimx\times dimu}$ describes how $u(t)$ controls the elements of $x(t)$. Matrix $C\in\mathbb R^{dimy\times dimx}$ is the so called output matrix. The primary goal of state space models is to investigate the dynamics of the state vector based on the output observations only, i.e., to obtain estimates of $x(t)$ based on $y(t)$. State space modeling addresses problems like prediction, data assimilation,  signal extraction, and estimation of parameters. Hereby, a variety of different models have a state space model formulation, e.g., autoregressive moving average models, general regression, and dynamic factor models~\citep{durbin2012time}. %
In order to solve the linear state space models, i.e., to obtain an estimate of $x(t)$ for given $y(t)$, the Kalman Filter algorithm is used~\citep{kalman1960new,durbin2012time}. The "filtering" refers to an update of the knowledge about the investigating system $x(t)$ each time a new observation of $y(t)$ is made. Kalman Filter operates in two steps: prediction and update. In the prediction step estimates of the state vector $\hat x(t)$ and the output vector $\hat y(t)$ are made conditioned on the information obtained so far. In the update step $\hat x(t)$ is corrected based on the difference between $\hat y(t)$ and the true observed vector $y(t)$. \\%
The Kalman filter can easily handle missing values in $y(t)$. Therefore, the matrices in (\ref{eq:state_eqI}-\ref{eq:space_eqII}) are constructed in a way that the missing dimension is  omitted for the prediction and the update steps by assigning the corresponding matrix entries to be equal to zero~\citep{isaksson1993identification,rubin2004multiple,durbin2012time,holmes2013derivation}. After the update step, the missing values in $y(t)$ can be reconstructed by evaluating equation (\ref{eq:space_eqII}). This approach considers only observations before the missing value occurs. An alternative is to involve all known data via the "Fixed-Interval Smoothing" approach: the Kalman filter is applied forwards in time and a fixed-point smoother backwards in time~\citep{isaksson1993identification}. Also missing values in covariates $u(t)$ can be handled, for instance, by modeling $u(t)$ by another linear Gaussian state space model~\citep{naranjo2013extending}. %
If the underlying dynamics of considered data is nonlinear and nonstationary, then the nonstationary, nonlinear and non-Gaussian formulation of the state space model~\citep{durbin2012time} handles the problem of missing data, i.e., estimation of the model parameters and reconstruction of the missing values. Concluding, state space models in combination with Kalman filter provide a flexible parametric approach for simultaneous model identification and reconstruction of the missing data in time series analysis. However, standard packages available in programming environments like Matlab and R are focusing on the more easy-to-handle stationary, linear/nonlinear, Gaussian state space representation~\citep{peng2011state,holmes2012marss}.\\%
In some special cases the main goal is the reconstruction of the missing data only. Then, the iterative and nonparametric Singular Spectral Analysis (SSA) can be deployed~\citep{kondrashov2006spatio}. SSA operates in an alternating order: (a) temporal embedding of the time series under consideration in a higher dimensional space, (b) followed by the reconstruction of the missing data exploiting the dominant temporal stationary Principal Components (PCs) of the embedded series. The PCs can describe the principal/major characteristics like an oscillating behavior of the underlying dynamics. The optimal memory depth and the total number of required PCs are estimated via cross-validation. %
A further approach based on spectral decomposition assumes that if a multidimensional time series changes its behavior (e.g., exhibiting peaks and valleys) in all dimensions simultaneously, then they are correlated~\citep{khayati2012rebom}. In this case, Singular Value Decomposition (SVD) can provide the most correlated dimensions which can then be used to reconstruct the missing data. Thus, if all dimensions of the considered time series are uncorrelated, the approach can not be applied. Additionally, standard methods based on spectral decomposition rely on the validity of the a priori stationarity and Gaussianity assumptions. \\%
Summarizing, the two main goals of the missing data problems are the reconstruction/imputation of the missing data (for the subsequent analysis of these data) and/or the inference of the underlying dynamics in the presence of missing data. In both cases, it is preferable to proceed without too strong a priori assumptions, in order to avoid possible distortion and bias that they could induce. From this perspective, the standard techniques described above can be classified into parametric and nonparametric approaches. The parametric ones, where the set of model parameters is fixed, describe the involved parameters by some predefined functions, referring for instance to linear regression, or make a priori assumptions about the underlying distributions, e.g., the Gaussian distribution~\citep{Hastie2009Elements, bishop2006pattern}. Thus, the results will depend on the particular choice of the appropriate function/distribution. %
Nonparametric approaches aim to describe of the underlying dynamics, rather then to fit an a priori fixed distribution. Following, the dimension of the involved model parameters is dependent on the complexity of the data~\citep{Hastie2009Elements, bishop2006pattern}. An examples here-fore, in the context of the missing data problem, is the widely used smoothing based regression~\citep{little1992regression, rubin2004multiple}. Being part of the kernel smoothing methods, smoothing based regression is a local approach and can not resolve jump behavior of the considered process~\citep{wahba1990spline, Hastie2009Elements, bishop2006pattern}, that are often observed in meteorological applications~\citep{majda2006distinct,franzke2008hidden,horenko2008metastable}. An other example is the nonparametric spectral decomposition based approach, with the assumption that the underlying dynamics exhibits stationary behavior, i.e., the principal characteristics of the considered process are not changing over the observed period~\citep{kondrashov2006spatio}. %
Here, we will go beyond these assumptions and limitations while addressing the missing data problem for time series data which can be described by nonstationary Vector Auto-Regressive models with eXogenous factors (VARX). %
For this purpose, we extend the family of nonstationary FEM-BV-VARX models (Finite Element based vector autoregressive time series analysis with bounded variation of the model parameters)~\citep{horenko2010identification,horenko2011nonstationarity} towards handling the missing values in the involved data (i.e., in the time series as well as in the involved exogenous factors). The presented approach will be denoted by FEMM-VARX. The FEMM-VARX approach simultaneously estimates the FEM-VARX model parameters and reconstructs the missing values by applying alternating optimization. FEMM-VARX is a semi-parametric approach and goes beyond standard a priori assumption like stationarity and/or locality for the underlying parameter process. The performance of the approach is compared to the state-of-the-art methodologies for handling the missing data problem on a generic set of test cases. \\%
The work is organized as follows: In Section~\ref{sec:new_theory} the FEM-VARX and its extension towards FEMM-VARX are presented. Section~\ref{sec:fem_framework} describes the computational framework behind the FEMM-VARX approach. The performance of the FEMM-VARX framework is presented in the Section~\ref{sec:numerical_examples} by comparing the reconstruction of the missing values with the state-of-the-art methods like the multiple imputation approach and nonparametric methods based on Singular Spectral Analysis (SSA). The results of this numerical comparison are summarized in Section~\ref{sec:conclusion}. %
%
%
\section{THEORY} 
\label{sec:new_theory}
Consider a continuous, multivariate time series $X_t\in\mathbb R^{dimx}$ and a series of exogenous factors (external factors/covariates) $U_t\in\mathbb R^{dimu}$ for $t=1,\dots,T$, where $dimx$ and $dimu$ denote the dimension of $X_t$ and of $U_t$. In order to describe the dynamics of $X_t$ dependent on $U_t$, we will use the FEM-BV-VARX approach~\citep{horenko2010identification}. That is, the dynamics of $X_t$ is approximated by a linear combination of $K\geq 1$ local stationary VARX models with time-dependent weights $\gamma_{k,t}$, for $k=1,\dots,K$:%
\begin{align}
	\label{eq:VARX}
	X_t \approx \sum\limits_k^K\gamma_{k,t}\left(c_{k} + \sum\limits^Q_{q=1}A_{k,q}X_{t-q} + \sum\limits^P_{p=0}B_{k,p}U_{t-p} + \epsilon_{k}\right), %
\end{align}
where $c_k\in\mathbb R^{dimx}$ is the constant/offset vector, $A_{k,q}\in\mathbb R^{dimx \times dimx}$, $q=1,\dots,Q$ are the interaction matrices, $B_{k,p}\in\mathbb R^{dimx \times dimu}$, $p=0,\dots,P$ are the control matrices and $\epsilon_{k}\sim \mathcal N(0,\Sigma_{k})$  is the normal distributed noise with the covariance matrix $\Sigma_k\in\mathbb R^{dimx \times dimx}$, for $k=1,\dots,K$. The time-dependent weights $\gamma_{k,t}$ satisfy the convexity constraints:%
\begin{align}
	&\sum\limits_{k=1}^K\gamma_{k,t} = 1,\quad \gamma_{k,t} \geq 0,\quad \forall k, t.\label{eq:VARX_inverse_contr_convex}%
\end{align}
In the following, the FEM-BV-VARX model parameters are summarized by:%
\begin{align}
	&\Gamma_t = \left(\gamma_{1,t},\dots,\gamma_{K,t}\right), \\
	& \Theta = \left(\theta_1,\dots,\theta_K\right),%
\end{align}
where $\theta_k$, for $k=1,\dots,K$, are the local stationary VARX model parameters with: %
\begin{align}
	\theta_k = \left(c_k, A_{k,1},\dots,A_{k,Q}, B_{k,0},\dots,B_{k,P}\right).%
\end{align}
The vector $\Gamma_t$ is denoted as the hidden switching process. FEM-BV-VARX controls its persistency by involving%
\begin{align}
	\|\gamma_{k,t}\|_{BV} = \sum_{t=t_{st}}^{T-1}|\gamma_{k,t+1}-\gamma_{k,t}|\leq C \quad\forall k \label{eq:VARX_inverse_contr_BV},%
\end{align}
where $C$ is a predefined constant. The accuracy of the FEM-BV-VARX approximation is measured by the following overall model distance function:%
\begin{align}
	&\mathcal L \left(\Theta,\Gamma_t\right) = \sum\limits_{t = t_{st}}^T\sum\limits_{k=1_{\vphantom{s_i}}}^K\gamma_{k,t}g\left(\theta_k;X_t, U_t\right).\label{eq:VARX_inverse_mv}%
\end{align}
with $t_{st} = mem+1$ and $mem = \max(Q,P)$ and the Euclidean model distance function 
\begin{align}
	g\left(\theta_k;X_t, U_t\right) = || X_t - c_k- \sum\limits_{q=1}^QA_{k,q}X_{t-q} - \sum\limits_{p=0}^PB_{k,p}U_{t-p}||_2^2. %
\end{align}
In this work, we incorporate Lasso shrinkage~\citep{Hastie2009Elements} on the model parameter $\Theta$ by involving%
\begin{align}
	\label{eq:lasso_theta}
	\| \theta_k\|_1 \leq C_{\Theta}, 
\end{align}
for $k=1,\dots,K$, where $C_{\Theta}$ is a fixed constant. First, Lasso shrinkage regularize the corresponding QP, for the special cases when the estimation ill-posed. Second, Lasso shrinkage helps to identify the most significant covariates, as it forces elements that are not significant towards zero. Then, for given $X_t$ and $U_t$ the optimal FEM-BV-VARX model parameters $\left(\Gamma_t,\Theta\right)$ are obtained as%
\begin{align}
	\label{eq:optGT}
	\left(\Theta^{opt},\Gamma_t^{opt}\right) = \argmin\limits_{\Theta,\Gamma_t}\mathcal L \left(\Theta,\Gamma_t\right)%
	\text{ wrt. (\ref{eq:VARX_inverse_contr_convex})-(\ref{eq:VARX_inverse_contr_BV})-(\ref{eq:lasso_theta})}.	%
\end{align}
There exists no analytical global solution of the constrained minimization problem in (\ref{eq:optGT}). A local optimal solution for fixed $K,\,C$ and $C_{\Theta}$ can be obtained through a frequently used Alternating Optimization (AO) approach~\citep{bezdek2002some,jain2013low} with respect to $\Theta$ and $\Gamma$. The global optimum is then obtained by restarting the AO multiple times with random initializations. Further, in order to find the optimal constants $K,\,C$ and $C_{\Theta}$, the global optimization is restarted for different combinations of possible values of $K,\,C$ and $C_{\Theta}$. The overall optimal combination can then be obtained either through cross-validation or by applying information criteria like the Akaike Information Criteria (AIC) or the Bayesian Information Criteria (BIC).\\%
In this manuscript the FEM-BV-VARX approach will be extended towards handling the problem of missing data. For this purpose, the missing data in $X_t$ and in $U_t$ are involved as additional variables, $X_{miss}$ and $U_{miss}$, into the FEM-BV-VARX approach. Such that, in a presence of missing data the objective functional $\mathcal L \left(\Theta,\Gamma_t\right)$ will be extended towards $\mathcal L \left(\Theta,\Gamma_t, X_{miss}, U_{miss}\right)$. Within the FEM-BV-VARX framework this extension will result in two additional steps in the AO: for fixed $\left(\Theta,\Gamma_t,U_{miss}\right)$ the optimal $X_{miss}$ will be obtained by minimizing the objective functional $\mathcal L \left(\Theta,\Gamma_t, X_{miss}, U_{miss}\right)$ with respect to $X_{miss}$ - and analogously for $U_{miss}$. In both cases the corresponding minimization can be reduced to a QP. This is discussed in detail in Section~\ref{sub:missing_in_xt} and Section~\ref{sub:missing_in_ut}.%
%
%
\subsection{Missing Data in $X_t$} 
\label{sub:missing_in_xt}
\begin{figure}[ht!]
	\centering
	\includegraphics[scale=0.4]{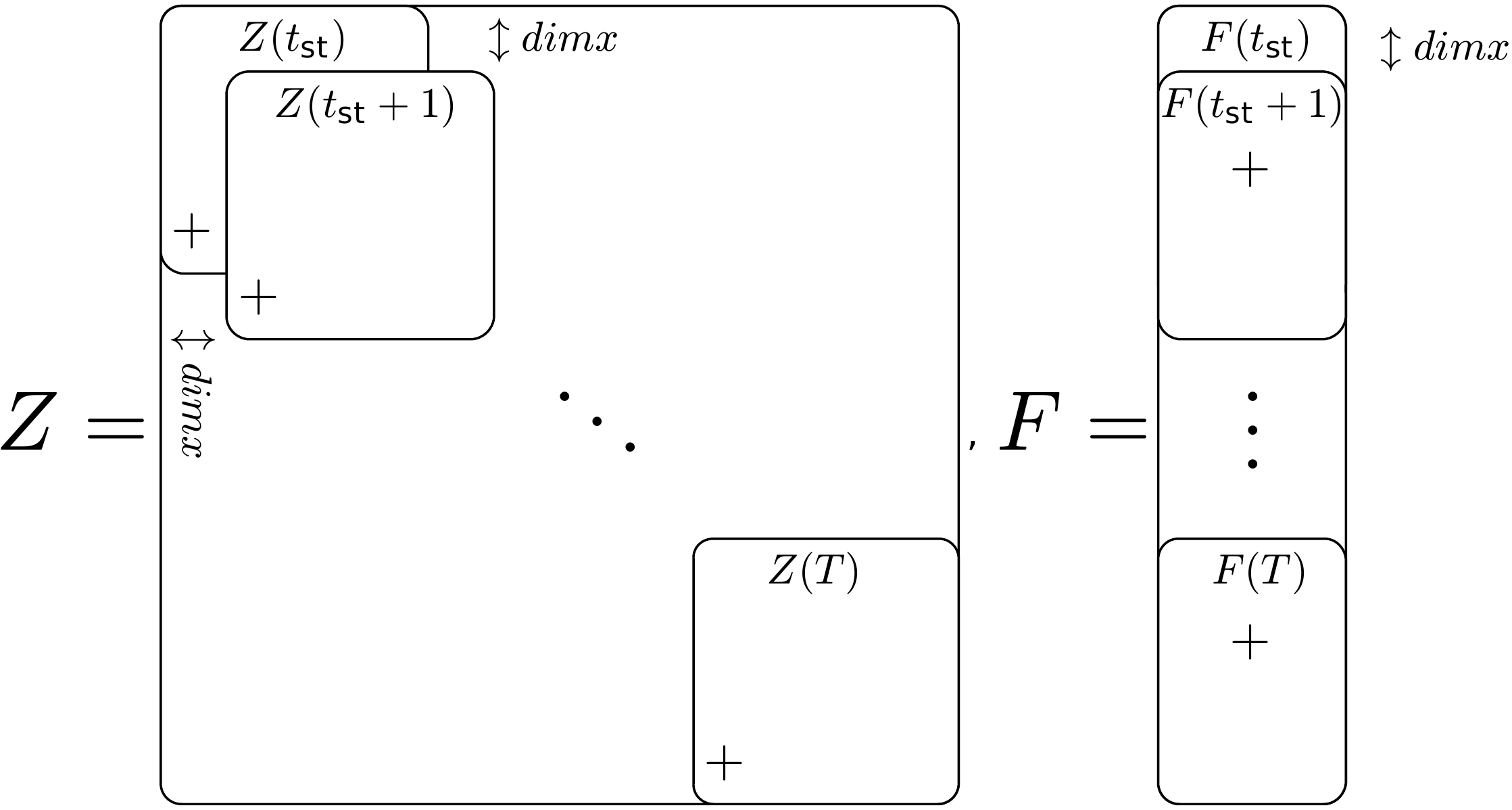}
	\caption{The construction of matrix $Z$ and vector $F$. The subsequent elements $Z_t$ and $F_t$ are added up on top of each other with a shift of $dimx$.} %
	\label{fig:ZF}
\end{figure}

In this section, we rewrite the objective functional $\mathcal L \left(\Theta,\Gamma_t, X_{miss}, U_{miss}\right)$ into a QP with respect to $X_{miss}$. First, we use temporal embedding to define a new variable:%
\begin{align}
	\label{eq:newX}
	\tilde X_{t} = \left(X_{t-Q}^{\dagger},\dots, X_{t-1}^{\dagger}, X_{t}^{\dagger}\right)^{\dagger}\in\mathbb R^{dimx(Q+1)\times 1}%
\end{align}
Then, it can be easily verified that for a fixed time step $t$, for $t=t_{st},\dots,T$, following equation holds: %
\begin{align}
	\label{eq:VARX_QP_t}
	\sum\limits_{k=1}^K\gamma_{k,t}g\left(\theta_k;X_t, U_t\right) = \tilde X_{t}^{\dagger} Z_t^{\vphantom{\dagger}} \tilde X_{t}^{\vphantom{\dagger}} + F_t^{\dagger}\tilde X_{t}^{\vphantom{\dagger}} + c_t^{\vphantom{\dagger}}, %
\end{align}
where the matrix $Z_t\in\mathbb R^{dimx(Q+1)\times dimx(Q+1)}$ and the vector $F_t\in\mathbb R^{dimx(Q+1)\times 1}$ are given by%
\begin{align}
	&Z_t =
		\sum\limits_{k=1}^K\gamma_{k,t}\begin{bmatrix} -A^{\dagger}_{k,Q}\\ \dots\\ -A^{\dagger}_{k,1}\\\mathbb{1}\end{bmatrix} %
		\begin{bmatrix} -A_{k,Q},\dots, -A_{k,1},\mathbb{1}\end{bmatrix},\label{eq:matCt_vecft}\\%
	&F_t =  -2 \sum\limits_{k=1}^K\gamma_{k,t}\begin{bmatrix} -A^{\dagger}_{k,Q}\\ \dots\\ -A^{\dagger}_{k,1}\\\mathbb{1}\end{bmatrix}\left(\mu_k + \begin{bmatrix} B_{k,0},\dots, B_{k,P}\end{bmatrix} \begin{bmatrix} U_{t}\\ \vdots\\ U_{t-P}\end{bmatrix}\right),\label{eq:matCt_vecft2}%
\end{align}
with the identity matrix $\mathbb{1}\in\mathbb R^{dimx\times dimx}$. A detailed derivation can be found in Appendix~\ref{app:qp_xt_all}. As the constant $c_t$ in (\ref{eq:VARX_QP_t}) can be neglected during minimization, it is omitted in the following. With this, the overall model distance functional defined in equation (\ref{eq:VARX_inverse_mv}) can be rewritten as%
\begin{align}
	\label{eq:tmp_qp_x}
	\mathcal L \left(\Theta,\Gamma_t\right) =\sum\limits_{t = t_{st}}^T \tilde X_{t}^{\dagger} Z_t^{\vphantom{\dagger}} \tilde X_{t}^{\vphantom{\dagger}} +F_t^{\dagger}\tilde X_{t}^{\vphantom{\dagger}}.%
\end{align}
This formulation, in turn, can be displayed in a matrix-vector form with respect to $X = \left(X_{1}^{\dagger},\dots,X_T^{\dagger}\right)^{\dagger}$ as follows: %
\begin{align}
	\label{eq:VARX_qp}
	\mathcal L \left(\Theta,\Gamma_t\right) =X^{\dagger}Z X + F^{\dagger}X.
\end{align}
The construction of the symmetric matrix $Z\in\mathbb R^{dimxT\times dimxT}$ and the vector $F\in\mathbb R^{dimxT\times 1}$ is visualized in Figure~\ref{fig:ZF}.%
Finally, we define the objective functional, which has to be minimized for the reconstruction of $X_{miss}$, by reducing the QP on the right hand side of equation (\ref{eq:VARX_qp}) with respect to $X_{miss}$:%
\begin{align}
	\label{eq:VARX_qp_miss}
	\mathcal L \left(X_{miss}\right) = X_{miss}^{\dagger}Z_{miss}^{\vphantom{\dagger}} X_{miss}^{\vphantom{\dagger}} + F_{miss}^{\dagger}X_{miss}^{\vphantom{\dagger}}.%
\end{align}
The construction of $Z_{miss}$ and $F_{miss}$ is shown in Appendix~\ref{app:qp_xt_all}. An estimate for $X_{miss}$ is obtained either analytically by evaluating%
\begin{align}
	\label{eq:recXmiss}
	X^{opt}_{miss} = Z^{-1}_{miss}F_{miss}^{\vphantom{\dagger}},
\end{align}
in case there are no constraints on $X_{miss}$, or by solving a constrained QP otherwise. The accuracy of $X^{opt}_{miss}$ depends on the condition number of $Z_{miss}$. Thus, in the special cases when the condition of $Z_{miss}$ is bad or the matrix is not invertible, we regularize the QP by Ridge shrinkage. That is, the objective functional (\ref{eq:VARX_qp_miss}) is minimized with respect to: %
\begin{align}
	\label{eq:ridge_xt}
	\| X_{miss}\|_2 \leq C_{X_{miss}}, 
\end{align}
where $C_{X_{miss}}$ is a fixed constant. By incorporating this constraint into the QP via a Lagrange multiplier $\lambda$, we redefine the objective functional with respect to $X_{miss}$ towards%
\begin{align}
	\label{eq:VARX_qp_miss_reg}
	\mathcal L \left(X_{miss}\right) = X_{miss}^{\dagger}\left(Z_{miss}^{\vphantom{\dagger}}+\lambda\mathbb{1}^{\vphantom{\dagger}}_{\vphantom{miss}} \right)X_{miss}^{\vphantom{\dagger}} + F^{\dagger}X_{miss}^{\vphantom{\dagger}},%
\end{align}
the identity matrix $\mathbb{1}$ is of the same dimension as $Z_{miss}$. %
%
%
%
%
\subsection{Missing Data in $U_t$} 
\label{sub:missing_in_ut}
The problem of missing data in $U_t$ is addressed in the exact same manner as the problem of missing data in $X_t$, such that $U_{miss}$ is reconstructed by solving a QP. This section outlines briefly the construction of the corresponding QP. First, we define a new vector %
\begin{align}
	\label{eq:newU}
	U = \left(U_1^{\dagger}, \dots, U_{T}^{\dagger}\right)^{\dagger}\in\mathbb R^{dimuT\times 1}.%
\end{align}
Then, the overall model distance functional defined in equation (\ref{eq:VARX_inverse_mv}) can be rewritten as a QP with respect to $U$:%
\begin{align}
	\label{eq:VARX_qpu}
	\mathcal L \left(\Theta,\Gamma_t\right) = U^{\dagger}D U + G^{\dagger}U.
\end{align}
Hereby, in line with (\ref{eq:VARX_qp}), the matrix $D$ and the vector $G$ is a composition of block-matrices $D_t$ and block-vectors $G_t$, respectively. For a fixed time step $t$, for $t=t_{st},\dots,T$, the blocks are defined by%
\begin{align}
	&D_t=\sum\limits_{k=1}^K\gamma_{k,t}\begin{bmatrix} B^{\dagger}_{k,P}\\ \dots\\ B^{\dagger}_{k,0}\end{bmatrix}\begin{bmatrix} B_{k,P},\dots, B_{k,0}\end{bmatrix},\\%
	&G_t = 2 \begin{bmatrix} 1,X^{\dagger}_t, X^{\dagger}_{t-1}, \dots, X^{\dagger}_{t-Q}\end{bmatrix} \sum\limits_{k=1}^K\gamma_{k,t}%
		 \begin{bmatrix}\mu^{\dagger}_k\\-\mathbb 1\\A^{\dagger}_{k,1} \\ \vdots \\ A^{\dagger}_{k,Q}  \end{bmatrix}%
		 \begin{bmatrix} B_{k,P},\dots, B_{k,0}\end{bmatrix}.%
\end{align}
By reducing the QP in equation (\ref{eq:VARX_qpu}) with respect to $U_{miss}$, we define the objective functional%
\begin{align}
	\label{eq:VARX_qp_Umiss}
	\mathcal L \left(U_{miss}\right) = U_{miss}^{\dagger}D_{miss}^{\vphantom{\dagger}} U_{miss}^{\vphantom{\dagger}} + G_{miss}^{\dagger}U_{miss}^{\vphantom{\dagger}}.%
\end{align}
which is used for $U_{miss}$ reconstruction. The construction of $D_{miss}$ and $G_{miss}$ is shown in Appendix~\ref{app:qp_ut_all}. Following equation (\ref{eq:VARX_qp_Umiss}), an estimate of $U_{miss}$ is obtained either analytically by%
\begin{align}
	\label{eq:recUmiss}
	U_{miss}^{opt} = D_{miss}^{-1} G_{miss}^{\vphantom{\dagger}},
\end{align}
if there are no constraints on $U_{miss}$, or by solving a constrained QP otherwise. The accuracy of $U_{miss}^{opt}$ depends on the condition of $D_{miss}^{\vphantom{opt}}$. In line with Section~\ref{sub:missing_in_xt}, Ridge shrinkage is imposed on the parameter $U_{miss}$ and is incorporated into the corresponding QP via a Lagrange multiplier $\eta$. Then, the objective functional in (\ref{eq:VARX_qp_Umiss}) is redefined: %
\begin{align}
	\label{eq:VARX_qp_Umiss_reg}
	\mathcal L \left( U_{miss}\right) = U_{miss}^{\dagger}\left( D_{miss}^{\vphantom{\dagger}} + \eta\mathbb{1}^{\vphantom{\dagger}}_{\vphantom{miss}} \right)U_{miss}^{\vphantom{\dagger}} + G_{miss}^{\dagger}U_{miss}^{\vphantom{\dagger}}.%
\end{align}
with the identity matrix $\mathbb{1}$ of the same dimension as $D_{miss}$.%
%
%
\subsection{FEMM-VARX} 
\label{sub:femm_varx}
Concluding, FEMM-VARX estimates FEM-VARX model parameters $\left(\Theta,\Gamma_t\right)$ and the missing values $\left(X_{miss},U_{miss}\right)$ by minimizing the resulting objective functional %
\begin{align}
	\label{eq:final_L}
	\mathcal L \left(\Theta,\Gamma_t, U_{miss}, X_{miss}\right)=
	\begin{cases}
		\mathcal L \left(\Theta,\Gamma_t\right) \text{ def. in (\ref{eq:VARX_inverse_mv})}, &X_{miss},\,U_{miss}\text{ fixed,}\\%
		\mathcal L \left(X_{miss}\right) \text{ def. in (\ref{eq:VARX_qp_miss_reg})},& \Theta,\,\Gamma_t,\,U_{miss}\text{ fixed,}\\%
		\mathcal L \left(U_{miss}\right) \text{ def. in (\ref{eq:VARX_qp_Umiss_reg})},&\Theta,\,\Gamma_t,\,X_{miss}\text{ fixed,}%
	\end{cases}
\end{align}
with respect to (\ref{eq:VARX_inverse_contr_convex})-(\ref{eq:VARX_inverse_contr_BV})-(\ref{eq:lasso_theta}). The minimization is approached by AO. The corresponding AO steps and the resulting FEMM-VARX framework are discussed in Section~\ref{sec:fem_framework}.\\%
The presented FEMM-VARX approach reconstructs $X_{miss}$ and $U_{miss}$ without strong a priori probabilistic or deterministic assumption about their underlying dynamics, but through an optimization of the infinite-dimensional distance functional defined in (\ref{eq:final_L}). As FEMM-VARX is a member of the FEM family, the two involved assumptions are: (i) $\Gamma_t$ is considered as a function in some (very broad) function space, i.e, the BV-space and (ii) $\Gamma_t$ is persistent/regular in the corresponding function space~\citep{metzner2012analysis}. In contrast, standard computational techniques, e.g., Gaussian based imputation, reconstruct the missing data by making a priori assumption about the parametric distribution of the data. Nonstationary regression based imputation exploits spline regression, and imposes additional locality and continuity/differentiability assumptions on the parameter process.\\%
Please note, the regularization of the parameters $X_{miss}$/$U_{miss}$ focuses on Ridge shrinkage, rather then Lasso shrinkage. This choice is supported by two reasons. First, Lasso is not considered for $X_{miss}$/$U_{miss}$ because as their dimension increases, the corresponding optimization problem becomes computationally expensive: In order to obtain a differentiable optimization problem, the L1-norm constraint must be rewritten into a linear constraints by introducing slack variables and this almost doubles the parameter space. Second, Lasso forces not significant elements towards zero, while Ridge shrinkage provides a smooth shrinkage of all the elements. The latter seems to be more appropriate in the context of $X_{miss}/U_{miss}$, while Lasso is more appropriate for the parameter $\Theta$ as it approaches the significance of external factors and of the regressive part. And, as the dimension of $\theta_k$ is usually small in comparison to $X_{miss}$/$U_{miss}$, Lasso shrinkage of $\Theta$ does not impose computational drawbacks.%

%
%
\section{FEMM-VARX: Computational Framework} 
\label{sec:fem_framework}
%
%
%
The proposed FEMM-VARX framework is integrated into the existing FEM MATLAB toolbox\footnote{The FEM MATLAB toolbox combines the family of FEM-based methods developed in the working group of Illia Horenko at the Institute of Computational Science (Universitá della Svizerra Italiana) in Lugano. The toolbox was implemented by Dimitri Igdalov and can be obtained on request.}.
 This section discusses the implementation of the FEMM-VARX framework, that is the minimization of the objective functional:%
\begin{align}
	&\mathcal L \left(\Theta,\Gamma_t, U_{miss}, X_{miss}; X_t, U_t\right) = \sum\limits_{t = t_{st}}^T\sum\limits_{k=1}^K\gamma_t^{(k)}g\left(\theta_k, U_{miss}, X_{miss}; X_t, U_t\right),\label{eq:FEMMVARX_inverse}\\%
	&\text{with respect to }\sum\limits_{k=1}^K\gamma^{(k)}_t = 1,\quad \gamma^{(k)}_t \geq 0,\quad \|\gamma_t^{(k)}\|_{BV}\leq C \quad\forall k, t. \label{eq:FEMMVARX_inverse_contr}%
\end{align}
%
%
\begin{small}
\IncMargin{1em}
\begin{algorithm}[hbt]
\DontPrintSemicolon
\SetKwData{Left}{left}
\SetKwData{This}{this}
\SetKwData{Up}{up}
\SetKwFunction{Union}{Union}
\SetKwFunction{FindCompress}{FindCompress}
\SetKwInOut{Input}{input}
\SetKwInOut{Output}{output}
\caption{Restarting and Alternating Optimization}
\Input{$X_t,\,U_t,\,K,\,C$, $Tol$, $maxRestart$, $maxAlternate$ }
\Output{global optimal parameter set $\left(\Gamma^*_t,\Theta^*,X_{miss}^*, U_{miss}^*\right)$}
\BlankLine
\label{algo:FEM-annealing}
\lnl{1}$\mathcal L\left(\Theta^* ,\Gamma^*_t,X_{miss}^*, U_{miss}^*;X_t,U_t\right) = inf$\;
\lnl{lines:annealing}\For{a = 1:maxRestart}{
		\BlankLine
		\lnl{lines:gold}$\Gamma^{opt}_t$ random initialization wrt. (\ref{eq:FEMMVARX_inverse_contr})\;%
		\lnl{lines:XUtold}$X_{miss}^{opt}$ and $U_{miss}^{opt}$ initialization, e.g., via interpolation\;%
		\lnl{lines:told}$\Theta^{opt} = \argmin\limits_{\Theta}\mathcal L\left(\Theta,\Gamma_t^{opt},X_{miss}^{opt}, U_{miss}^{opt};X_t,U_t\right)$\;%
		\BlankLine
		\lnl{lines:subspace}\While{not convergency or maxAlternate}%
		{
				\BlankLine
				\lnl{lines:subspace1}\textbf{Step1}:			
				$ \Gamma_t^{opt} = \argmin\limits_{\Gamma_t}\mathcal L\left(\Theta^{opt},\Gamma_t,X^{opt}_{miss}, U^{opt}_{miss};X_t,U_t \right)$; \\ Constrained minimization wrt. $\Gamma_t$ results in a linear problem.\;%
				\BlankLine
				\lnl{lines:subspace2}\textbf{Step2}:
				$X_{miss}^{opt} =  \argmin\limits_{X_{miss}} \mathcal L\left(\Theta^{opt},\Gamma_t^{opt}, X_{miss}, U_{miss}^{opt};X_t,U_t \right)$;\\ An analytical solution exists, if there are no constraints on $X_{miss}$.%
				\BlankLine
				\lnl{lines:subspace3}\textbf{Step3}:
				$U_{miss}^{opt} =  \argmin\limits_{U_{miss}} \mathcal L\left(\Theta^{opt},\Gamma_t^{opt}, X^{opt}_{miss}, U_{miss};X_t,U_t \right)$; \\An analytical solution exists, if there are no constraints on $U_{miss}$.%
				\BlankLine
				\lnl{lines:subspace4}\textbf{Step4}: 
				$\Theta^{opt} = \argmin\limits_{\Theta} \mathcal L\left(\Theta,\Gamma_t^{opt}, X^{opt}_{miss}, U^{opt}_{miss};X_t,U_t \right)$; \\An analytical solution exists.%
				}
		\BlankLine
		\lnl{ifsubspace}\If{$ \mathcal L\left(\Theta^* ,\Gamma_t^*,X_{miss}^*, U_{miss}^*;X_t,U_t\right) >\mathcal L\left(\Theta^{opt},\Gamma_t^{opt}, X^{opt}_{miss}, U^{opt}_{miss};X_t,U_t\right)$}%
				{
				\lnl{lines:g*}$\Gamma_t^* = \Gamma_t^{opt}$\;
				\lnl{lines:x*}$X_{miss}^* = X_{miss}^{opt}$\;
				\lnl{lines:u*}$U_{miss}^* = U_{miss}^{opt}$\;
				\lnl{lines:t*}$\Theta^* = \Theta^{opt}$\;
				}
		}
\BlankLine
\end{algorithm}
\DecMargin{1em}
\end{small}
%
As already discussed, the functional (\ref{eq:FEMMVARX_inverse}) is not convex and there exists no analytical global solution of the constrained minimization problem (\ref{eq:FEMMVARX_inverse}-\ref{eq:FEMMVARX_inverse_contr}). Instead, a local optimal solution can be found through alternating optimization (AO) with respect to $\Theta,\Gamma_t, U_{miss}, X_{miss}$ as described in Algorithm~\ref{algo:FEM-annealing}, lines \ref{lines:gold} to \ref{lines:subspace4}. After the random initialization of $\Gamma^{opt}_t$, line \ref{lines:gold}, $X^{opt}_{miss}$ and $U^{opt}_{miss}$, line \ref{lines:XUtold}, and the first estimate of the model parameter $\Theta^{opt}$, line \ref{lines:told}, the actual AO starts. The four involved alternating steps are described below.\\%
In \textbf{Step1}, line~\ref{lines:subspace1}, the optimal switching process $\Gamma_t$ is estimated. Herewith, $\Gamma_t$ is discretized by Finite Elements as proposed in~\citep{horenko2010finite}, the BV constraints are incorporated by a set of linear inequalities. The resulting optimization problem is a linear constrained minimization with respect to $\Gamma$ with linear convex constraints~\citep{horenko2011nonstationarity,metzner2012analysis}. This problem can be solved with standard methods like the simplex method.\\%
In \textbf{Step2}, line~\ref{lines:subspace3}, the missing values in $X_t$ are reconstructed by solving either a constrained or an unconstrained QP. The existence and the accuracy of the solution depends on the regularity of the matrix $Z_{miss}$, defined in equation (\ref{eq:VARX_qp_miss}).\\%
In \textbf{Step3}, line~\ref{lines:subspace4}, the missing values in $U_t$ are estimated exactly analogue to the third step. The existence and the accuracy of the solution depends on the regularity of the matrix $D_{miss}$, defined in equation (\ref{eq:VARX_qp_Umiss}).\\%
In \textbf{Step4}, line~\ref{lines:subspace2}, the local VARX model parameters $\Theta^{opt}=\left(\theta^{opt}_1,\dots,\theta^{opt}_K\right)$ are estimated. Exploiting the fact that the averaged model distance functional (\ref{eq:FEMMVARX_inverse}-\ref{eq:FEMMVARX_inverse_contr}) is uncoupled with respect to $\theta_k$ for different $k=1,\dots,K$, each $\theta_k$ can be estimated separately by solving: %
\begin{align}
	\theta_k^{opt} = \argmin\limits_{\theta_k}\sum\limits_{t = t_{st}}^T\gamma_t^{(k)}g\left(\theta_k, U_{miss}, X_{miss}; X_t, U_t\right), \quad k=1,\dots,K. %
\end{align}
for fixed $X_{miss}$ and $U_{miss}$. Here the objective functional $g\left(.\right)$, defined as the Euclidian distance, is convex and the optimal $\theta_k$ can be estimated analytically. In some cases illustrated bellow this estimation may be ill-conditioned and would require an appropriate regularization (i.e., an L2-regularization~\citep{Hastie2009Elements}). \\
The convergency of the AO is achieved if the value $\mathcal L \left(\Theta,\Gamma_t, U_{miss}, X_{miss}; X_t, U_t\right)$ stops decreasing significantly. The AO procedure is repeated $maxRestart$-times with different initializations, line~\ref{lines:annealing}. The parameters with the smallest value of the objective functional are returned, lines \ref{lines:g*} to \ref{lines:t*}.\\%
In general, the convergency of AO can be guaranteed theoretically for some special cases, e.g., if the objective functional is strictly convex, twice smooth differentiable, and the second derivative is positive definite, then the AO approach is converging to the global optimal solution using any initialization~\citep{bezdek2003convergence}. %
Here we refer to a less restrictive case: Under the assumption that (a) the objective function is twice continuously differentiable with respect to all variables, (b) the optimum exists and is unique and (c) the solution space for each variable is compact, AO is converging to either a local, a saddle-point or a global optima~\citep{bezdek2002some}. To follow this convergency result, we will assume that condition (b) holds. Condition (c) is fulfilled, as a compact solution space can always be provided. What remains is to check condition (a): Condition (a) is satisfied with respect to $\Theta$, $X_{miss}$ and $U_{miss}$, as the corresponding objective functions are defined as Euclidean distances and hence are twice differentiable. Further, condition (a) is also valid for the constraint minimization with respect to $\Gamma_t$. In particular, because the BV constraint, that produces a nondifferentiable problem, can be transformed in a set of linear constraints as was shown in~\citep{metzner2012analysis}, such the minimization with respect to $\Gamma_t$ results in solving a linear problem with linear constraints, which is twice differentiable.\\ %
The accuracy of estimators in Step2, Step3, and Step4 depends on the condition number of the matrices to be inverted. In some cases the condition of the corresponding matrix is bad or the matrix is not even regular. %
The condition of the corresponding matrices, and thus, the accuracy of the solution can be improved by incorporating additional information, i.e., by regularizing the problem. FEMM-VARX approach incorporates the Ridge and Lasso shrinkage techniques. In Step2 Ridge shrinkage is imposed on the parameter $X_{miss}$. That is, the QP problem in (\ref{eq:VARX_qp}) is solved with respect to %
\begin{align}
	\label{eq:ridge_xt_}
	\| X_{miss}\|_2 \leq C_{X_{miss}}, 
\end{align}
where $C_{X_{miss}}$ is a fixed constant. By incorporating the Ridge constraint into the QP via a Lagrange multiplier $\lambda$ we obtain again a QP problem:%
\begin{align}
	\mathcal L \left(\Theta,\Gamma_t,X_{miss}; X_t, U_t\right) = X_{miss}^{\dagger}Z_{miss}X_{miss} + F^{\dagger}X_{miss} + \lambda\|X_{miss}\|_2.
\end{align}
Also in Step3 Ridge shrinkage is imposed on the parameter $U_{miss}$. Analogue to the above case, the Ridge shrinkage constraint is incorporated into the equation (\ref{eq:VARX_qpu}) via the Lagrange multiplier and a QP has to be solved. %
Please note, the regularization of the parameters $X_{miss}$/$U_{miss}$ focuses on Ridge shrinkage, rather then Lasso shrinkage. This choice is supported by two reasons. First, Lasso is not considered for $X_{miss}$/$U_{miss}$ because as their dimension increases, the corresponding optimization problem becomes computationally expensive: In order to obtain a differentiable optimization problem, the L1-norm constraint is rewritten into linear constraints by introducing slack variables and this almost doubles the parameter space. Second, Lasso forces not significant elements towards zero, while Ridge shrinkage provides a smooth shrinkage of all the elements. The latter seems to be more appropriate in the context of $X_{miss}/U_{miss}$ as the significance of external factors and of the regressive part is approached by the VARX parameter $\Theta$.\\%
In Step4 FEMM-VARX framework enables regularization of the model parameter $\Theta$ by incorporating Lasso shrinkage, i.e.:%
\begin{align}
	\label{eq:lasso_theta_}
	\| \theta_k\|_1 \leq C_{\Theta}, 
\end{align}
for $k=1,\dots,K$, where $C_{\Theta}$ is a fixed constant. In this case Lasso shrinkage is preferred to Ridge because it forces elements that are not significant towards zero. And, as the dimension of $\theta_k$ is usually small in comparison to $X_{miss}$/$U_{miss}$, application of Lasso shrinkage does not impose computational drawbacks.%
\subsection{FEMM-VARX: Conceptual comparison with standard techniques} 
\label{subsec:femm_varx}

In the following the FEMM-VARX is compared methodologically to some of the standard methods which are discussed in Section~\ref{sec:intro}.

FEM-framework describes the nonstationary dynamics by a set of $K$ local models and a nonstationary, nonparametric switching process. The case when there is just one model, i.e., $K=1$ and $C = 0$, corresponds to a stationary model. The presence of different models, i.e., $K > 1$, indicates the presence of "systematically missing information". The main conceptual advantage of FEM in its general form over the Bayesian mixture models, e.g., Hidden Markov Models (HMM), is that the involved nonstationarity is resolved in a nonparametric way and allows a robust nonparametric regularization of the underlying regime-switching process $\Gamma_t$. The HMM is a purely parametric approach with a priori assumptions (e.g., Gaussianity, Markovianity, homogeneity). In the case of HMM, the involved hidden process is parametrized by a probabilistic model, e.g., time-homogenous Markov process, and requires an initial hidden probability. The two assumptions involved in FEM are: (i) $\Gamma_t$ is considered as a function in some (very broad) function space, i.e, the BV-space and (ii) $\Gamma_t$ is persistent/regular in the corresponding function space~\citep{metzner2012analysis}. FEM approach accounts for discontinuous functions and provides for $K>1$ a nonlocal extension of the nonparametric smoothing approach, where the nonstationary switching process connects all observations that belong to the similar dynamics (i.e., assigned to the same cluster) to a single ensemble~\citep{metzner2012analysis}. %
FEMM-VARX is a member of the FEM family and inherits so the above described features. FEMM-VARX reconstructs $X_{miss}$ and $U_{miss}$ without a priori probabilistic or deterministic assumption about their underlying dynamics, but through an optimization of the infinite-dimensional distance functional. In contrast, standard computational techniques, e.g., Gaussian imputation, reconstruct the missing data by making a priori assumption about the parametric distribution of the data. Nonstationary regression based imputation exploits spline regression, and imposes additional locality and continuity/differentiability assumptions on the parameter process. FEMM-VARX provides an approximation to the nonstationary state space formulation, as being an extension of the FEM-VARX approach. As explained above, state space models can not handle $X_{miss}$ and $U_{miss}$ simultaneously, but require two different uncoupled state space formulations~\citep{naranjo2013extending}. \\%
An alternative to FEMM-VARX is to reconstruct first the missing values and then to apply FEM-VARX to the full data for optimal model parameter estimation. For instance, MI or methods based on spectral decomposition could be used. As these methods rely on a priori assumptions like stationarity or Gaussian behavior, the reconstruction of the missing data might be biased, for example, if the underlying dynamics exhibit switching regimes. Following, FEM-VARX will provide a biased description of the underlying dynamics. This conceptual comparison is illustrated practically on a set of generic numerical test cases in Section~\ref{sec:numerical_examples}. %
%
%
%
\section{Numerical Examples} 
\label{sec:numerical_examples}
In this section the performance of FEMM-VARX is demonstrated and compared to parametric and nonparametric state-of-the-art approaches on a set of generic test cases. For the comparison with parametric methods,

we refer to a multiple imputation based approach described in Section~\ref{sec:intro}. In particular, 

 we apply the framework AmeliaII which is available in the statistics toolbox R~\citep{honaker2011amelia, manualR}.
  AmeliaII handles MAR or MCAR missing data by assuming
   Gaussian distribution of observed and missing data. AmeliaII assumes a flat prior for the Gaussian distribution and deploys the EM
   
    algorithm for finding the posterior according to equation (\ref{eq:posterior_ign}). The posterior is then used to draw/impute the missing data~\citep{honaker2011amelia}. To handle missing values in a time series, AmeliaII involves temporal variability by creating a series of polynomials or cubic splines and/or incorporates autoregressive behavior. AmeliaII provides the possibility to set a wide range of configurations, here the focus lies on the main ones: (a) the argument \verb|m| indicates the number of imputed datasets to create, (b) the temporal behavior is described by cubic smoothing splines of time, its dimension is controlled by the argument \verb|splinetime|, (c) the arguments \verb|lags| and \verb|leads| enable to involve future and past values of each single variable into the imputation model, (d) the argument \verb|priors| initializes all the missing values. The involved EM algorithm converges after the value of the corresponding log-likelihood is increasing less as the argument \verb|tolerance|. \\%
For a standard nonparametric approach we refer to SSA based methods described in Section~\ref{sec:intro}. A collection of SSA methods, including missing data reconstruction, is available in the R package RSSA~\citep{Korobeynikov2010,Golyandina2014, manualR}. RSSA is based on the decomposition of the temporal embedded data into the Principal Components (PCs), which describe the principal/major characteristics like an oscillating behavior or a trend of the underlying dynamics. Consequently, the performance of the reconstruction is dependent on the embedding and an appropriate choice of the PCs. Following configurations for RSSA are considered here: (a) the argument \verb|L| controls the temporal embedding, (b) the argument \verb|groups| provides the list of PCs to be considered, (c) the argument \verb|clust| is the number of clusters in which the corresponding PCs are assigned, hereby each cluster represents a different type of dynamics, for example, while one cluster represents the trend in the data, an other might be responsible for the periodical behavior. RSSA is an iterative approach, the number of iterations is limited by the argument \verb|maxiter|. \\%
In the next sections the FEMM-VARX approach, AmeliaII and RSSA are compared on a set of generic test cases. First, we define the input data: Starting with full $X_t$ and full $U_t$ series we introduce to each series $[5\%,\dots,95\%]$ of MCAR missing values such that we obtain 11 sets of $U_t$ and 11 sets of $X_t$. Second, we define three cases, each with 10 tests: (a) $X_t$ with $[5\%,\dots,95\%]$ missing values, $U_t$ full, (b) $X_t$ full, $U_t$ with $[5\%,\dots,95\%]$ missing values, (c) $X_t$ and $U_t$ each with $[5\%,\dots,95\%]$ missing values. In total we obtain 30 data sets, on which we will compare the candidates. %
The comparison of the methods is performed as follows: FEMM-VARX is directly applied for simultaneous reconstruction of the missing data and estimation of the model parameters. RSSA is used to reconstruct the missing data first, then the reconstructed full data is provided to FEM-VARX for model parameter estimation. This procedure is denoted in the following by RSSA-FEM-VARX. AmeliaII represents a MI approach, such that the outcome of its application results in $m$ reconstructed data sets. Each reconstructed set is provided to FEM-VARX for model parameter estimation, such that the MSE between the original data and model parameters and the estimated ones can be computed. The final MSE for model parameters and the final MSE for reconstructed data is averaged over the $m$ different MSE, respectively. AmeliaII in combination with FEM-VARX is denoted in the following by Amelia-FEM-VARX. Please note, for all three methods the FEM-VARX parameters $K$ and $C$ are fixed. %
In further work, the FEMM-VARX approach will be extended and the optimal combination of K and C will be chosen according to Information Criteria for model selection in the presence of missing data, like the ones proposed in~\citep{cavanaugh1998akaike,celeux2006deviance,shimodaira2018information}. %
The comparison of these three methods is performed with respect to quality of missing values reconstruction and the mean square error between the original model parameters and their estimators. %

%
%
\subsection{Generic Test-Case} 
\label{sub:test_case}
\begin{figure}[ht]
  \begin{center}	 
	  \includegraphics[scale=0.4]{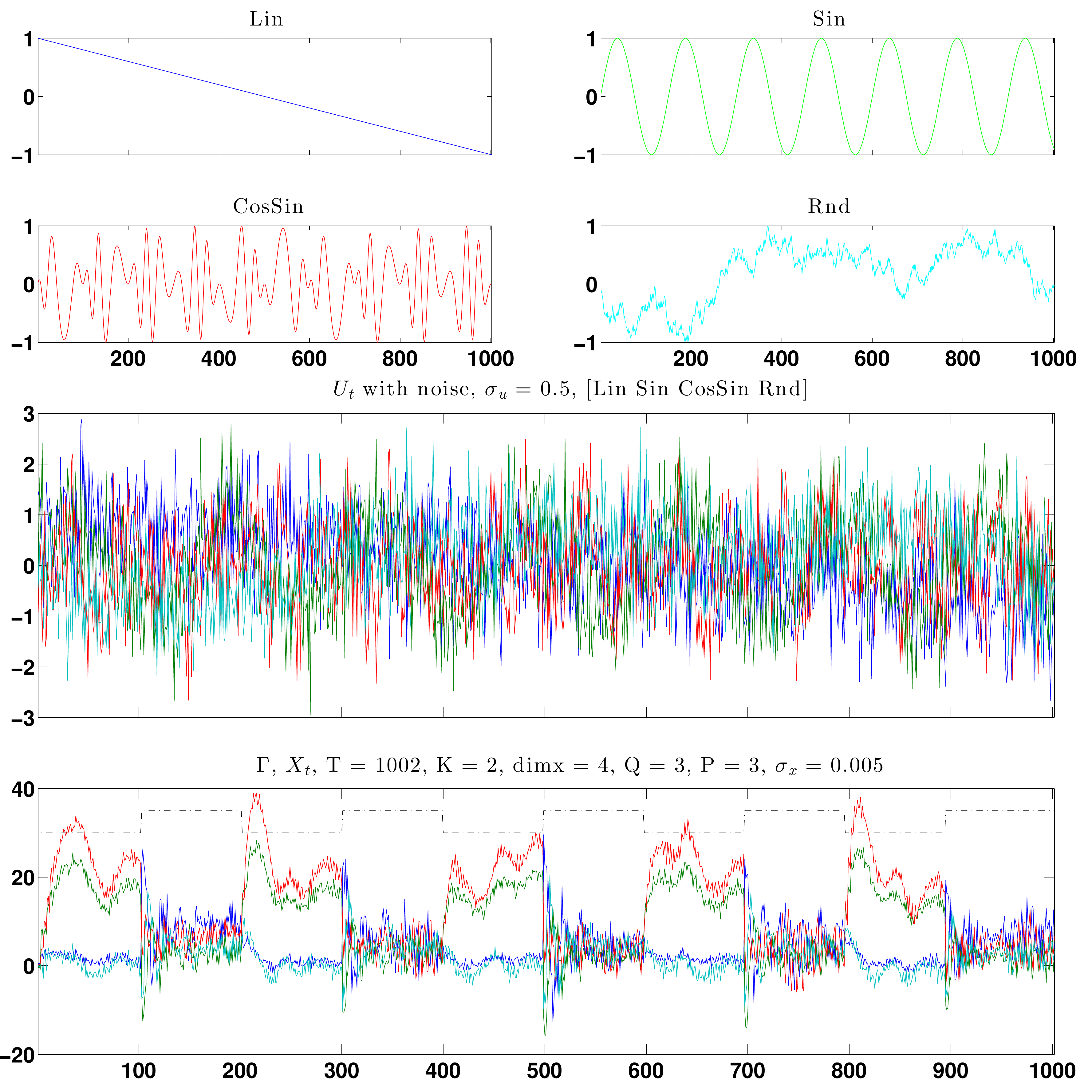}
  \end{center}
  \caption{The first four upper panels show the four components of the external factor $U_t$: the linear trend, the periodical/oscillating- and the random behavior. The second from below shows the noisy $U_t$, i.e., $U_t + \mathcal N(0,\sigma_u)$, which was used to generate $X_t$. The last panel presents the noisy $X_t$, i.e., $X_t + \mathcal N(0,\Sigma)$ and the switching process $\Gamma(t)$ (dash-dot black line). The remaining parameters $T,\,K,\,d,\,Q,\,P$ are described in equation~(\ref{eq:generateXt}). }%
  \label{fig:plot_data_descr}
\end{figure}

\begin{figure}[ht]
  \begin{center}	 
	  \includegraphics[scale=0.33]{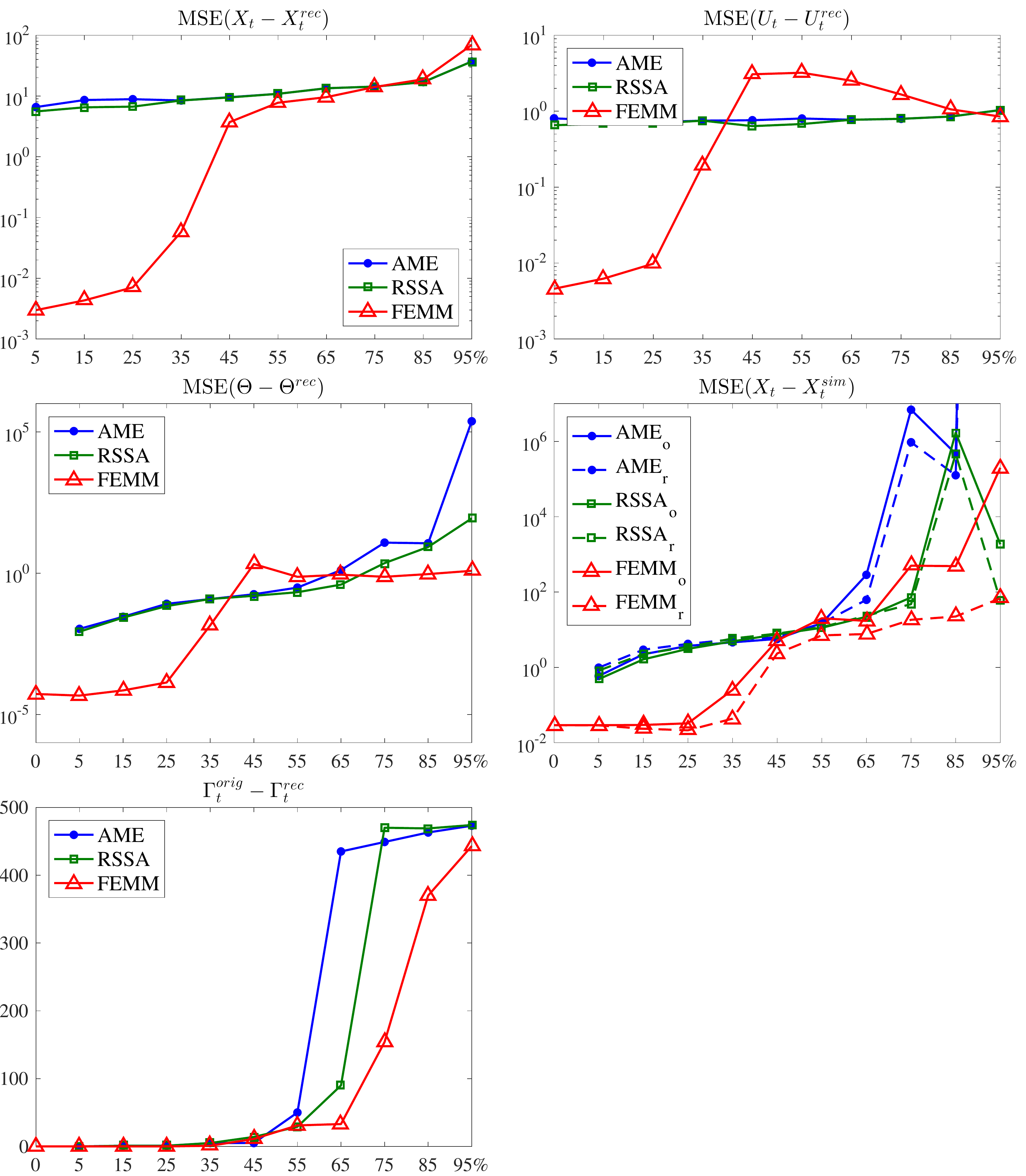}
  \end{center}
  \caption{Results for missing values in $X_t$ and $U_t$. In all panels the lines marked with circles, squares and triangles, correspond to Amelia-FEM-VARX, RSSA-FEM-VARX, and FEMM-VARX, respectively. The MSE panels show the results on a logarithmic scale. In the middle right panel the dashed lines correspond to $X_t^{sim}$ obtained for $\left(\Theta,\Gamma_t,U_t^{rec}\right)$ and solid lines correspond to $X_t^{sim}$ obtained for $\left(\Theta,\Gamma_t, U_t\right)$. The lower panel shows the absolute difference between $\Gamma_t^{orig}$ and $\Gamma_t$. The $0\%$ mark, if present, represents the performance of FEM-VARX on the complete data.}%
  \label{fig:plot_missxtut}
\end{figure}

This section considers a test case, where the underlying dynamics is governed by a discrete switching process and some predefined external factors. This test case emphasises the difficulties arising from the nonstationarity and non-smoothness due to an involved switching process. The artificial series $X_t\in\mathbb R^{4},\,t=4:1002$ is generated according to a mixture VARX model with $K=2,\,Q=3,P=3$: %
\begin{align}
	\label{eq:generateXt}
	&X_t \sim \sum\limits_{k=1}^K\gamma_k(t)\left(\mu^{(k)} + \sum\limits_{q=1}^Q A_q^{(k)}X_{t-q}^{\vphantom{(k)}} +  \sum\limits_{p=0}^P B_p^{(k)}U_{t-p}^{\vphantom{(k)}} + \epsilon^{(k)}\right),  \epsilon^{(k)} \sim\mathcal N(0, \Sigma),\\%
	&\text{with }\mu^{(k)}\in\mathbb R^{4\times 1},\,A_q^{(k)}\in\mathbb R^{4\times 4},\,B_p\in\mathbb R^{4\times 4}\text{ and }\Sigma = 0.005\cdot\mathds{1}, %
\end{align}
where $\mathds{1}\in\mathbb R^{4\times4}$ is the identity matrix. The external factors are given by $U_t = \left(u_1(t),u_2(t),u_3(t),u_4(t)\right)$ with%
\begin{align}
	&u_1(t) = -\frac{2t}{T}+1 + \eta,\label{eq:ut1}\\
	&u_2(t) = \sin\left(\frac{2\pi}{150}t\right)+ \eta,\label{eq:ut2}\\ 
	&u_3(t) = \sin\left(\frac{2\pi}{200} t\right)\cos\left(\frac{2\pi}{40}t + \frac{\pi}{2}\sin\left(\frac{2\pi}{120}t\right)\right)+ \eta,\label{eq:ut3}\\%
	&u_4(t) = \text{random walk}+ \eta,\label{eq:ut4}%
\end{align}
where $\eta\sim\mathcal N(0,\sigma_u)$, with $\sigma_u=0.5$. This choice of the external factors represents the seasonal variations, the linear warming trend and the random behavior often observed in meteorological applications. The values of the involved matrices and the generation of $u_4(t)$ can be found in Appendix~\ref{sec:test_case_fem_varx_matrices}. The model parameters are summarized by $\theta_k=\left(\mu^{(k)},A_1^{(k)},\dots,A_3^{(k)},B_0^{(k)},\dots,B_3^{(k)}\right)$, for $k=1,2$. The corresponding assignment for $\theta_k$ as well as the starting values $X_1,\,X_2,\,X_3$ are outlined in Appendix~\ref{sec:test_case_fem_varx_matrices}. The generated time series $X_t$, the deployed covariates and the switching process $\Gamma(t)$ are shown in Figure~\ref{fig:plot_data_descr}.\\%
In the second step MCAR missing values are inserted in: (a) $X_t$, (b) $U_t$ and (c) $X_t$, $U_t$. There are no missing data in the first and the last time steps of the series and/or the covariates, as RSSA requires complete start and final observations. AmeliaII receives additionally a linear trend which is required by the cubic spline smoothing for describing the temporal behavior. Lastly, applying FEMM-VARX, RSSA-FEM-VARX, and Amelia-FEM-VARX to cases (a), (b), (c), the aim is to reconstruct the missing values and to estimate the parameters $\Gamma_t=\left(\gamma^{(1)}_t,\dots,\gamma_t^{(K)}\right)$ and $\Theta=\left(\theta_1,\dots,\theta_K\right)$. In the following the configurations for the three approaches are set.\\ %
The FEMM-VARX framework is configured with $K=2$, $C=9$, the number of annealing steps is fixed to $500$, the maximal number of the subspace iterations is set to $100$ and the convergency criterion is set to $0.0005$. The regularization with respect to $X_{miss}$ and $U_{miss}$ is involved, for both cases a set of different Lagrange multipliers is considered, denoted by $\lambda_{list}=\lbrace 0, 0.000001, 0.0001, 0.005, 0.1\rbrace$. %
The configurations for AmeliaII are as follows: \verb|m = 4|, \verb|lags = FALSE| and \verb|leads = FALSE|, i.e., no lead or lags were involved into the model. In cases when those parameters are activated, AmeliaII terminates with an error pointing to highly correlated variables in the model and so to a singular covariance matrix~\citep{honaker2011amelia}. We consider a range of spline dimensions with \verb|splinetime_list = 1:1:6| and set \verb|tolerance = 5.0e-08|. The missing values are initialized by applying the R function \verb|na.approx()|, referring to linear interpolation, such that \verb|priors = na.approx(mdata, rule = 2)|. %
The configurations for RSSA are as follows: a range of possible values for the embedding parameter \verb|L| is considered, denoted in the following by \verb|L_list|, with \verb|L_list = [10:20:90, 100:50:300, 400:100:700, 750]|. The range of possible values for \verb|groups| is denoted by \verb|groups_list| with \verb|groups_list = [10:10:40]|. The possible number of considered clusters is \verb|clust_list =[2:1:10]|. The maximal number of involved iterations was set to \verb|maxiter = 1000|. Analog to AmeliaII the missing values were first initialized using linear interpolation. The optimal configurations for each of the approaches, i.e., the ones that lead to the best result, are chosen with respect to the overall averaged mean square error between the original and the reconstructed series.\\%
For the inference of the results some notations are required: The full, original time series is denoted by $X_t$, the series of external factors is denoted by $U_t$. $X_t$ and $U_t$ with reconstructed missing values are denoted in the following by $X_t^{rec}$ and $U_t^{rec}$, respectively. The true model parameters are denoted by $\Theta^{orig}$ and $\Gamma_t^{orig}$. The model parameters estimated by applying FEM-VARX, FEMM-VARX, RSSA-FEM-VARX or Amelia-FEM-VARX are denoted by $\Theta$ and $\Gamma_t$. 
Lastly, the simulated time series obtained by evaluating equation (\ref{eq:generateXt}) for estimated $\Theta$ and $\Gamma_t$, and $U_t$ or $U_t^{rec}$, is denoted by $X_t^{sim}$. \\%
%
%
We start with case (a), i.e., missing data in $X_t$ only. The optimal model configurations are outlined in Table~\ref{tab:a_optconfi_femmvarx} ($\lambda$ is the Lagrange multiplier referring to Ridge regularization on $X_{miss}$), Table~\ref{tab:a_optconfi_amelia} and Table~\ref{tab:a_optconfi_rssa}. %
\begin{table}[ht!]
\caption{Case (a): Optimal configuration for FEMM-VARX.}%
\label{tab:a_optconfi_femmvarx}
\begin{center}
\begin{tabular}{  c | c  c  c  c  c  c  c  c  c  c }
  & 5\% & 15\% & 25\% & 35\% & 45\% & 55\% & 65\% & 75\% & 85\% & 95\% \\%
\cline{1-11}
$\lambda$  & 0 & 0 & 0 & 0 & 0 & 0 & 0 & 0 & 0 & 0 \\%
\end{tabular}
\end{center}
\vskip -0.1in
\end{table}
%
%
\begin{table}[ht!]
\caption{Case (a): Optimal configuration for AmeliaII.}%
\label{tab:a_optconfi_amelia}
\begin{center}
\begin{tabular}{  c | c  c  c  c  c  c  c  c  c  c }
  & 5\% & 15\% & 25\% & 35\% & 45\% & 55\% & 65\% & 75\% & 85\% & 95\% \\%
\cline{1-11}
splinetime  & 4 & 2 & 1 & 2 & 4 & 6 & 4 & 2 & 1 & 3  \\%
\end{tabular}
\end{center}
\vskip -0.1in
\end{table}
%
%
\begin{table}[ht!]
\caption{Case (a): Optimal configuration for RSSA.}%
\label{tab:a_optconfi_rssa}
\begin{center}
\begin{tabular}{  c | c  c  c  c  c  c  c  c  c  c }
  & 5\% & 15\% & 25\% & 35\% & 45\% & 55\% & 65\% & 75\% & 85\% & 95\% \\%
\cline{1-11}
L       & 20 & 20 & 20 & 20 & 20 & 6 & 10 & 2 & 2 & 150  \\%
groups  & 6  & 6  & 6  & 6  & 6  & 6 & 2  & 2 & 2 & 3  \\%
clust   & 2  & 2  & 2  & 2  & 2  & 6 & 2  & 2 & 2 & 2  \\%
\end{tabular}
\end{center}
\vskip -0.1in
\end{table}
%
\begin{figure}[ht]
  \begin{center}	 
	  \includegraphics[scale=0.31]{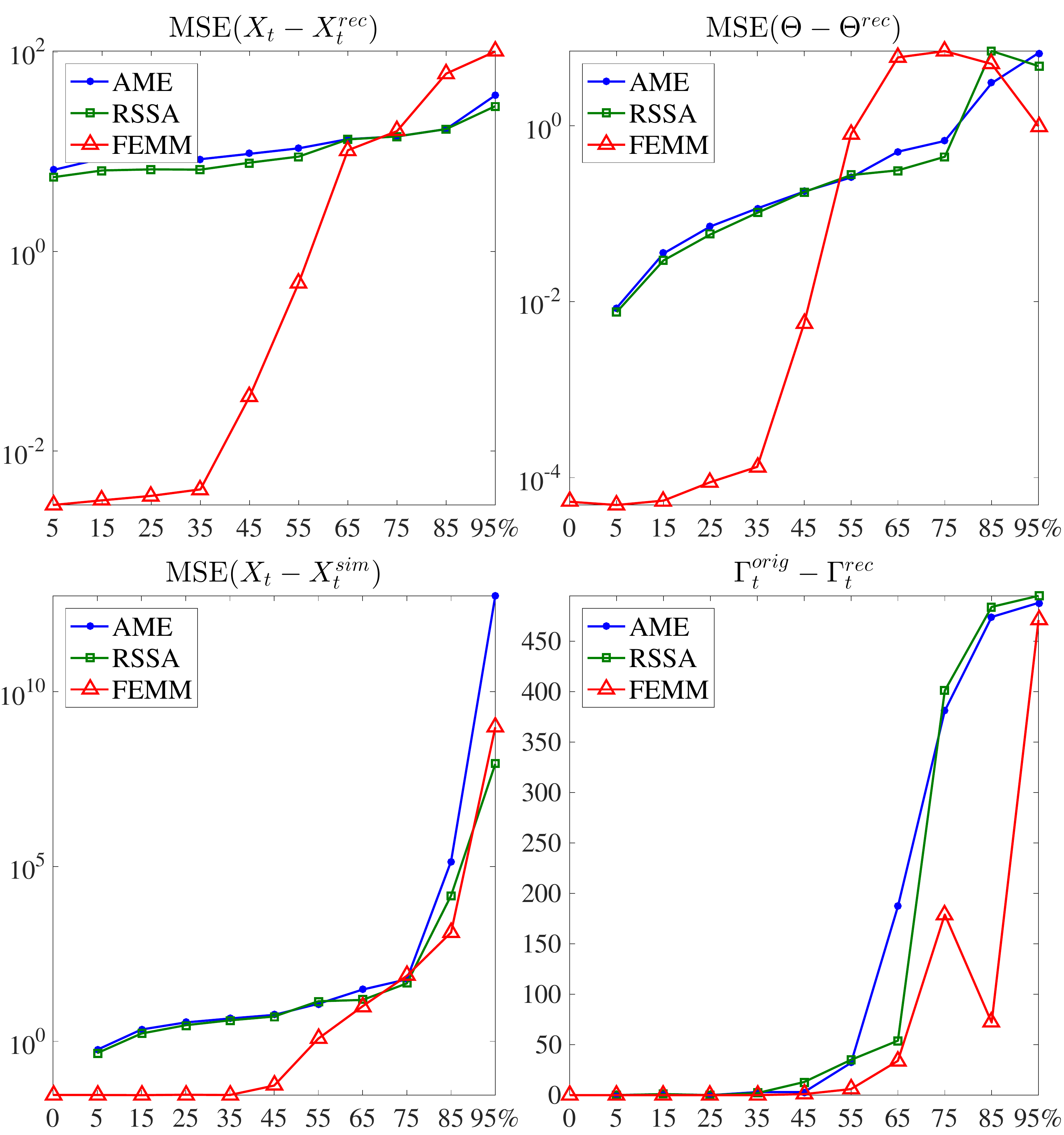}
  \end{center}
  \caption{Results for missing values in $X_t$. In all panels the lines marked with circles, squares and triangles, correspond to Amelia-FEM-VARX, RSSA-FEM-VARX, and FEMM-VARX, respectively. The MSE panels show the results on a logarithmic scale. The lower right panel shows the absolute difference between $\Gamma^{orig}_t$ and $\Gamma_t$. The $0\%$ mark, if present, represents the performance of FEM-VARX on the complete data.} 
  \label{fig:plot_missxt}
\end{figure}
The performance of FEMM-VARX, Amelia-FEM-VARX, and RSSA-FEM-VARX is shown in Figure~\ref{fig:plot_missxt}: The top left panel displays the reconstruction accuracy of $X_{miss}$ measured by the MSE between $X_t$ and $X_t^{rec}$. %
The top right panel shows the estimation accuracy of the model parameters measured by the MSE between $\Theta^{orig}$ and $\Theta$. The lower left panel displays the MSE between the original series $X_t$ and the simulated series $X_t^{sim}$ which is obtained using $\left(\Theta,\Gamma_t, U_t\right)$. The lower right panel shows the number of misfits between $\Gamma_t^{orig}$ and $\Gamma_t$. The zero percent mark, if present in the panels, represents the performance of FEM-VARX on the full data. %
The performance of Amelia-FEM-VARX is slightly better than the performance of RSSA-FEM-VARX. In both cases the reconstruction/estimation accuracy is steadily decreasing with an increasing percentage of missing values. FEMM-VARX outperforms both Amelia-FEM-VARX and RSSA-FEM-VARX until the $45\%$ border of missing values is reached. From $45\%$ of missing values in total, FEMM-VARX performs in some of the considered measures worse. In such scenarios half of $X_t$ is missing and it becomes difficult to reconstruct the data correctly without additional assumption/information about the underlying dynamics, as it is done by the opponents. FEMM-VARX shows an irregular behavior of MSE for some of the measures, in contrast to the steadily increasing MSE values for Amelia-FEM-VARX and RSSA-FEM-VARX. The behavior of FEMM-VARX can be explained as follows: FEMM-VARX estimates the model parameters and reconstructs the missing values in an alternating order, such that the accuracy of $\Theta$, $X_t^{rec}$, and $\Gamma_t$ is interdependent. The correct estimation of the switching process $\Gamma_t$ is "crucial", otherwise the data is assigned to different models and the estimation of $\Theta$ gets distorted (and consequently the simulation of $X_t^{sim}$). With an increasing percentage of missing values in $X_t$, their occurrence/pattern merges to completely missing blocks. This matters if such a block covers a transition between different regimes: Then, the estimation of $\Gamma_t$ becomes biased. 
However, we would like to point out, that this example should be generalized by referring to different "missing value" patterns. For instance, using different random generator initialization we could generate different MCAR patterns and compare the approaches on them. This remains for future work.\\ 
%
%
We continue with the results obtained for the case (b), i.e., missing data in $U_t$ only. The optimal model configurations are outlined in Table~\ref{tab:b_optconfi_femmvarx} ($\eta$ is the Lagrange multiplier referring to Ridge regularization on $U_{miss}$), Table~\ref{tab:b_optconfi_amelia} and Table~\ref{tab:b_optconfi_rssa}. %
\begin{table}[ht!]
\caption{Case (b): Optimal configuration for FEMM-VARX.}%
\label{tab:b_optconfi_femmvarx}
\begin{center}
\begin{tabular}{  c | c  c  c  c  c  c  c  c  c  c }
 & 5\% & 15\% & 25\% & 35\% & 45\% & 55\% & 65\% & 75\% & 85\% & 95\% \\%
\cline{1-11}
$\eta$  & 0.0050  & 0.0050  & 0.0050  & 0.0050  & 0.0050  & 0.0050  & 0.0050  & 0.0050  & 0.0050  & 0.0050   \\%
\end{tabular}
\end{center}
\vskip -0.1in
\end{table}
%
%
\begin{table}[ht!]
\caption{Case (b): Optimal configuration for AmeliaII.}%
\label{tab:b_optconfi_amelia}
\begin{center}
\begin{tabular}{  c | c  c  c  c  c  c  c  c  c  c }
 & 5\% & 15\% & 25\% & 35\% & 45\% & 55\% & 65\% & 75\% & 85\% & 95\% \\%
\cline{1-11}
splinetime  & 6 & 1 & 2 & 1 & 1 & 6 & 3 & 6 & 5 & 3  \\%
\end{tabular}
\end{center}
\vskip -0.1in
\end{table}
%
%
\begin{table}[ht!]
\caption{Case (b): Optimal configuration for RSSA.}%
\label{tab:b_optconfi_rssa}
\begin{center}
\begin{tabular}{  c | c  c  c  c  c  c  c  c  c  c }
  & 5\% & 15\% & 25\% & 35\% & 45\% & 55\% & 65\% & 75\% & 85\% & 95\% \\%
\cline{1-11}
L       & 40 & 20 & 30 & 20 & 30 & 30 & 30 & 50 & 80 & 150\\%
groups  & 3  & 2  & 2  & 2  & 2  & 2  & 2  & 2  & 2  & 2  \\%
clust   & 2  & 2  & 2  & 2  & 2  & 2  & 2  & 2  & 2  & 2  \\%
\end{tabular}
\end{center}
\vskip -0.1in
\end{table}
%
%
\begin{figure}[ht]
  \begin{center}	 
	  \includegraphics[scale=0.33]{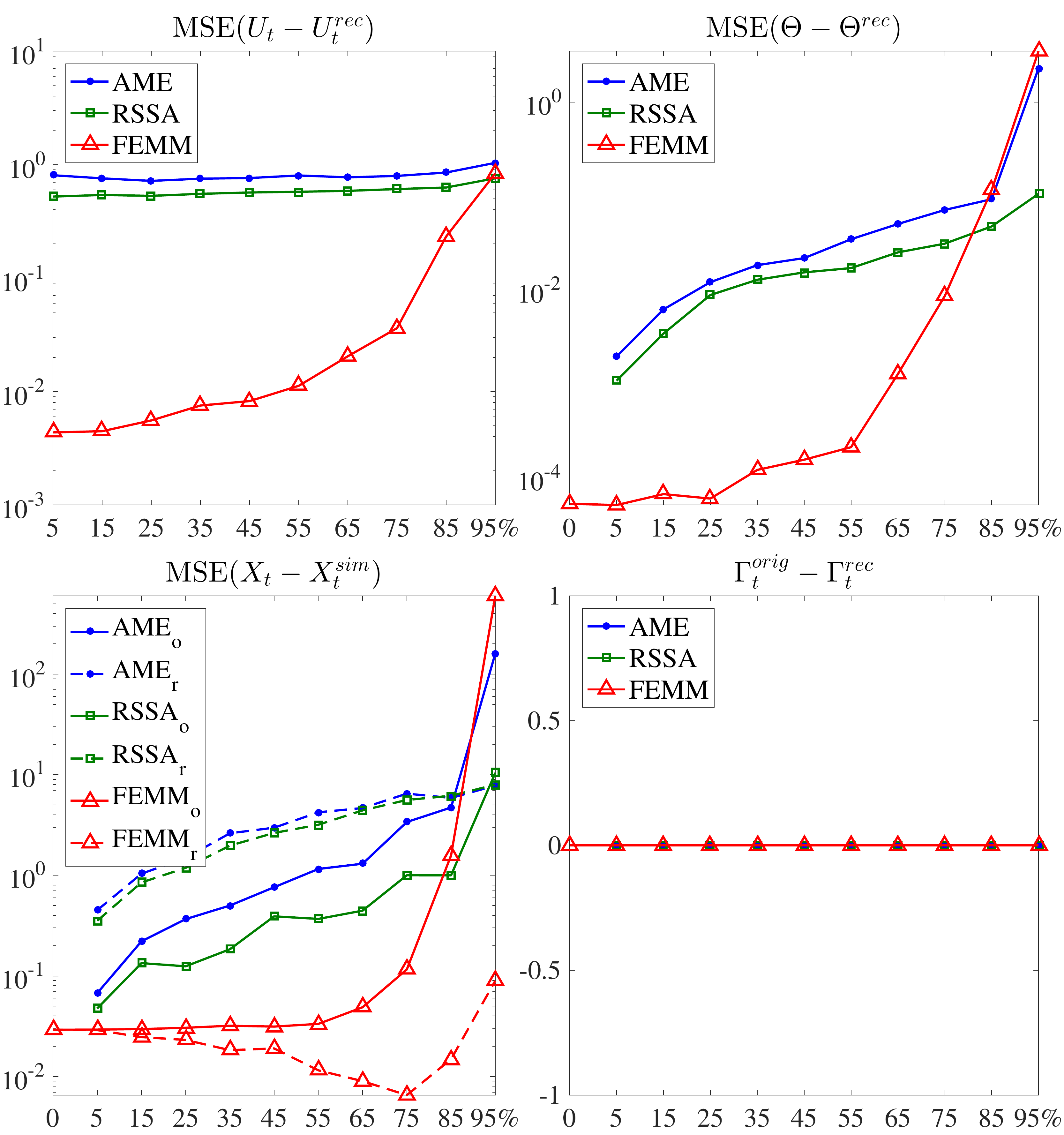}
  \end{center}
  \caption{Results for missing values in $U_t$. In all panels the lines marked with circles, squares and triangles, correspond to Amelia-FEM-VARX, RSSA-FEM-VARX, and FEMM-VARX, respectively. The MSE panels show the results on a logarithmic scale. In the lower left panel the dashed lines correspond to $X_t^{sim}$ obtained for $\left(\Theta,\Gamma_t, U_t^{rec}\right)$ and solid lines correspond to $X_t^{sim}$ obtained for $\left(\Theta,\Gamma_t, U_t\right)$. The lower right panel shows the absolute difference between $\Gamma_t^{orig}$ and $\Gamma_t$. The $0\%$ mark, if present, represents the performance of FEM-VARX on the complete data.}%
  \label{fig:plot_missut}
\end{figure}
The performance of FEMM-VARX, Amelia-FEM-VARX, and RSSA-FEM-VARX is shown in Figure~\ref{fig:plot_missut}. The top left panel displays the reconstruction accuracy of $U_{miss}$ measured by the MSE between $U_t$ and $U_t^{rec}$. The top right panel shows the estimation accuracy of the model parameters measured by the MSE between $\Theta^{orig}$ and $\Theta$. The lower right panel shows the number of misfits between $\Gamma_t^{orig}$ and $\Gamma_t$. The lower left panel displays the MSE between the original series $X_t$ and the simulated series $X_t^{sim}$ obtained for two different scenarios: %
\begin{enumerate}
	\item $X_t^{sim}$ is evaluated according to equation (\ref{eq:generateXt}) for $\left(\Theta,\Gamma_t, U_t^{rec}\right)$ (dashed lines and the subscript $r$, for reconstructed, in the legend),%
	\item $X_t^{sim}$ is evaluated according to equation (\ref{eq:generateXt}) for $\left(\Theta,\Gamma_t, U_t\right)$ (solid lines and the subscript $o$, for original, in the legend).%
\end{enumerate}
These two different scenarios allow to study how the parameters $\Theta,\Gamma_t$ obtained in presence of missing values in $U_t$ perform with true data. For example, this applies when $\Theta$ and $\Gamma_t$ are used for predictions with complete future covariates.\\%
Amelia-FEM-VARX and RSSA-FEM-VARX obtain an accurate $\Theta$ and estimate exactly $\Gamma^{orig}_t$. However, they both return a smaller MSE for the second scenario, i.e, the estimated model parameters and the original $U_t$ approximate the dynamics of $X_t$ better than the estimated model parameters and $U_t^{rec}$. That is, AmeliaII and RSSA provide an $U_t^{rec}$ which describes the underlying dynamics worse than the original $U_t$. Additionally, $U_t^{rec}$ obtained through AmeliaII and RSSA can be considered as a disturbed/noisy $U_t$. In this case, these two scenarios allow to study the sensitivity of the FEM-VARX framework towards disturbances/noise in the input data. Following the results, FEM-VARX seems to be stable and estimates in presence of disturbed data a set of parameters which provide a more accurate simulation of the underlying dynamics with original data. However, for more general conclusion further tests are required.\\%
For FEMM-VARX we see the opposite behavior. This is explained by the following facts: FEMM-VARX refers to $U_{miss}$ as model parameter in addition to $\Theta,\Gamma_t$. Thereby, $\Theta,\Gamma_t$ describe the trend behavior of the data without the noise. And, in the moment when FEMM-VARX is estimating $U_{miss}$ the residual noise in $X_t$ is modeled as well by minimizing the overall MSE with respect to $U_{miss}$. Overall, with an increasing percentage of missing values (i.e, increasing dimension of $U_{miss}$) the FEMM-VARX functional gets more degrees of freedom for describing the residual noise in $X_t$, but at the same time gets faster over-fitted in comparison to Amelia-FEM-VARX and RSSA-FEM-VARX. This explains the decreasing MSE behavior of $X_t^{sim}$ obtained for $\left(\Theta,\Gamma_t,U_t^{rec}\right)$ for FEMM-VARX. \\%
In total, the results for case (b) are better compared to case (a), where missing data in $X_t$ are considered. This is due to the fact that $X_t$ contains more information about the underlying dynamics: it reflects the switching process and the involved external factors $U_t$, please compare equation (\ref{eq:generateXt}). Thus, if $X_t$ is complete FEMM-VARX estimates the underlying switching process and the corresponding model parameter $\Theta$ more accurate.\\%
%
%
Lastly, the results obtained for the case (c), i.e, missing data in $X_t$ and $U_t$, are discussed. The optimal model configurations are outlined in Table~\ref{tab:c_optconfi_femmvarx}, Table~\ref{tab:c_optconfi_amelia} and Table~\ref{tab:c_optconfi_rssa}. %
\begin{table}[ht!]
\caption{Case (c): Optimal configuration for FEMM-VARX.}%
\label{tab:c_optconfi_femmvarx}
\begin{center}
\begin{tabular}{  c | c  c  c  c  c  c  c  c  c  c }
  & 5\% & 15\% & 25\% & 35\% & 45\% & 55\% & 65\% & 75\% & 85\% & 95\% \\%
\cline{1-11}
$\lambda$  & 0 & 0 & 0 & 0 & 0 & 0 & 0 & 0 & 0 & 0  \\%
$\eta$  & 0.0050 & 0.0050 & 0.0050 & 0.0050 & 0.0050 & 0.0050 & 0.0050 & 0.0050 & 0.0050 & 0.0050  %
\end{tabular}
\end{center}
\vskip -0.1in
\end{table}
%
%
\begin{table}[ht!]
\caption{Case (c): Optimal configuration for AmeliaII.}%
\label{tab:c_optconfi_amelia}
\begin{center}
\begin{tabular}{  c | c  c  c  c  c  c  c  c  c  c }
  & 5\% & 15\% & 25\% & 35\% & 45\% & 55\% & 65\% & 75\% & 85\% & 95\% \\%
\cline{1-11}
splinetime  & 2 & 1 & 4 & 6 & 4 & 6 & 1 & 1 & 4 & 6  \\%
\end{tabular}
\end{center}
\vskip -0.1in
\end{table}
%
%
\begin{table}[ht!]
\caption{Case (c): Optimal configuration for RSSA.}%
\label{tab:c_optconfi_rssa}
\begin{center}
\begin{tabular}{  c | c  c  c  c  c  c  c  c  c  c }
 & 5\% & 15\% & 25\% & 35\% & 45\% & 55\% & 65\% & 75\% & 85\% & 95\% \\%
\cline{1-11}
L       & 20 & 20 & 20 & 3 & 10 & 10 & 4 & 2 & 3 & 3 \\%
groups  & 6  & 6  & 6  & 3 & 2  & 2  & 4 & 2 & 3 & 2 \\%
clust   & 2  & 2  & 2  & 2 & 2  & 2  & 2 & 2 & 2 & 2 \\%
\end{tabular}
\end{center}
\vskip -0.1in
\end{table}
The performance and comparison of FEMM-VARX, Amelia-FEM-VARX, and RSSA-FEM-VARX is presented in Figure~\ref{fig:plot_missxtut}. In all panels the lines marked with circles, squares and triangles, correspond to Amelia-FEM-VARX, RSSA-FEM-VARX, and FEMM-VARX, respectively. %
The top left and the top right panels display the reconstruction accuracy for $X_{miss}$ and $U_{miss}$, respectively. The middle left panel shows the MSE between the original model parameters and the estimated ones. The middle right panel displays the MSE between the original series $X_t$ and the simulated series $X_t^{sim}$ which is obtained first using $\left(\Theta,\Gamma_t, U_t^{rec}\right)$ (dashed lines and the subscript $r$ in the legend), and second using $\left(\Theta,\Gamma_t, U_t\right)$ (solid lines and the subscript $o$ in the legend). The corresponding MSE does not show the same behavior as observed in case (b). Here, with increasing percentage of missing data in $X_t$ the estimation of $\Gamma_t$ and thus of $\Theta$ becomes by magnitudes worse. Leading to bigger differences between $X_t^{sim}$ and $X_t$. Also in this case the estimated $U_{miss}$, as additional degrees of freedoms, accounts for the residual noise in $X_t$ even if not as "good" as in case (b). In total, as expected, the performance of all the three approaches is worse in the case missing data is present in $X_t$ and in $U_t$. However, FEMM-VARX outperforms clearly for the first $35\%$ Amelia-FEM-VARX and RSSA-FEM-VARX. Please also note, that $X_t^{sim}$ obtained via Amelia-FEM-VARX runs away for $95\%$ of missing values in total.\\%
In all the three test cases the $X_{miss}$ and/or $U_{miss}$ remains acceptable until the border of $65\%$ of missing data in total. From here, it becomes for every method difficult to reconstruct the dynamics appropriate. The reconstruction of the missing values was carried out till $95\%$ missing value, in order to see how the methodologies handle this kind of missing data problems.%
%
%
%
%
%
\section{Conclusion} 
\label{sec:conclusion}
We proposed a new framework for handling the missing data problem for nonstationary time series which can be described by Vector Auto-Regressive models with eXogenous factors (VARX). The framework is an extension of the Finite Element Methodology for VARX-models with Bounded Variation (BV) of the model parameters (FEM-BV-VARX)~\citep{horenko2010identification,horenko2011nonstationarity}. The resulting approach, denoted by FEMM-VARX, estimates simultaneously the FEM-VARX model parameters and reconstructs the the missing values in the time series $X_t$, denoted by $X_{miss}$, as well as the missing values in the involved exogenous factors $U_t$, denoted by $U_{miss}$. \\%
FEMM-VARX is a semi-parametric approach: Being part of the FEM family the proposed framework describes the underlying nonstationary dynamics by a set of $K$ local VARX models and a nonstationary, nonparametric regime-switching process. FEMM-VARX reconstructs $X_{miss}$ and $U_{miss}$ exploiting alternating optimization (AO) without a need to impose additional parametric assumptions (e.g., Gaussianity). %
The performance of the approach is compared to state-of-art methodologies for handling the missing data problem on a set of test cases with an MCAR missing data pattern. As state-of-the-art methodologies we focused on a parametric multiple imputation based approach, implemented in the R package AmeliaII~\citep{honaker2011amelia, manualR}, and on the nonparametric Singular Spectral Analysis (SSA) based approach for missing data reconstruction, available in the R package RSSA~\citep{Korobeynikov2010, Golyandina2014, manualR}. The considered test cases with regime-switching behavior and involving external factors/covariates were constructed to cover a wide variety of the missing data patterns and scenarios. Three different cases were considered: (a) missing data in the time series, (b) missing data in the involved covariates and (c) missing data both in the time series and in the involved covariates. In all three cases we have investigated situations with different percentages of the missing values: starting with $5\%$ till $95\%$ (in intervals of ten percent). \\%
In case (a), FEMM-VARX outperforms both Amelia-FEM-VARX and RSSA-FEM-VARX until the level of $45\%$ of missing values is reached. From $45\%$ of missing values in total, FEMM-VARX performs in some of the considered measures worse. With an increasing percentage of missing values in $X_t$, the occurrence of missing values merges to blocks. This implies that the switching behavior in $X_t$ can get lost and consequently the estimation of the switching process becomes worse. With more than $55\%$ of missing values in $X_t$ it becomes difficult to reconstruct the data correctly without additional assumption/information about the underlying dynamics as in the case of FEMM-VARX. In contrast, for Amelia-FEM-VARX and RSSA-FEM-VARX the performance is slightly better for cases with more than $55\%$ of missing values caused by the underlying Gaussian assumption (which is true for this particular test case). \\%
In case (b), FEMM-VARX outperforms both Amelia-FEM-VARX and RSSA-FEM-VARX in all measures until $75\%$ of missing values in total. The better performance in comparison to case (a) can be explained as follows. First, for a complete $X_t$ FEMM-VARX estimates the underlying switching process and the corresponding model parameter $\Theta$ more accurately. Second, the new degrees of freedom given by $U_{miss}$ account for the noisy behavior in $X_t$. Following, an increasing dimension of $U_{miss}$ provides additional flexibility to account for the noise in $X_t$, but at the same time it can lead to a faster overfitting.\\%
In case (c), all of the three considered approaches perform worse as compared to the first two cases. FEMM-VARX outperforms both Amelia-FEM-VARX and RSSA-FEM-VARX until the level of $35\%$ of missing values is reached.\\%
The overall performance demonstrated the advantages of FEMM-VARX compared with the state-of-the-art methodologies for the reconstruction of missing data when the underlying dynamics is nonstationary. The stationarity and the Gaussian assumption of the considered competitors lead to averaging over the nonstationarity induced by the switching process rather then reconstructing the underlying dynamics.\\%
However, AmeliaII and RSSA approaches are faster (in terms of the CPU time) and less memory-consuming as compared to the current implementation of the FEMM-VARX framework. The current larger CPU time consumption of the framework results from the two following facts. First, the optimization of the non-convex functional with respect to four variables (in the worst case) requires a wide exploration of the parameter space, i.e., it requires some number of the AO with different, random initializations. However, as these trials are independent from each other, this procedure can be parallelized easily. Second, as the number of missing parameters increases, their reconstruction becomes more computationally expensive since a linear system, or a constrained quadratic problem has to be solved. This can be improved by deploying parallel solvers, e.g., the linear system can be solved by the parallel conjugate gradient method available in the PETSc toolkit. The memory consumption can be improved by assembling the resulting QP problems with respect to the missing data in a more efficient way. Additionally, the framework needs to be tested on further test cases and real applications. These aspects are the subject of ongoing and future work.%
%
%
\section{Acknowledgments} 
\label{sec:acknowledgments}
I. Horenko is partly funded by the Swiss Platform for Advanced Scientific Computing, Swiss National Research Foundation Grant 156398 MS-GWaves, and the German Research Foundation (Mercator Fellowship in the Collaborative Research Center 1114 Scaling Cascades in Complex Systems). O. Kaiser is supported by the Swiss Platform for Advanced Scientific Computing, Swiss National Research Foundation Grant 156398 MS-GWaves.%
%
%
\appendix
%
\section{FEMM-VARX: missing data in $X_t$} 
\label{app:qp_xt_all}
In this section the FEM-VARX model distance function is rewritten with respect to $X_t$ as the referred variable. Let us first consider the FEM-VARX model distance function for a fixed time step $t$: %
\begin{align}
	&g\left(\theta(t);X_t, U_t\right) = || X_t - \mu - \sum\limits_{q=1}^QA_{q}X_{t-q} - \sum\limits_{p=0}^PB_{p}U_{t-p}||_2^2
\end{align}
For simplicity, neglect the sum over different local models $k=1,\dots,K$ and so the index $(k)$. For writing out the Euclidean norm, define new matrices %
\begin{align}
	&\tilde A =\begin{pmatrix} -A_Q & \dots & -A_1 & \mathbb{1} \end{pmatrix}, %
	\tilde B =\begin{pmatrix}  B_0 & B_1 & \dots & B_P	\end{pmatrix},\label{app_eq:ABtilde}\\%
	&\tilde X_t = \begin{pmatrix}
		X_{t-Q}\\
		\vdots\\
		X_{t-1}\\
		X_t
	\end{pmatrix},
	\tilde U_t = \begin{pmatrix}
		U_{t}\\
		U_{t-1}\\
		\vdots\\
		U_{t-P}
	\end{pmatrix}.\label{app_eq:XUtilde}%
\end{align}
where $\mathbb{1}\in\mathbb R^{dimx\times dimx}$ is the identity matrix. Then, we get: 
\begin{align}
	&|| X_t - \mu - \sum\limits_{q=1}^QA_{q}X_{t-q} - \sum\limits_{p=0}^PB_{p}U_{t-p}||_2^2\\%
	&=\left(\tilde A\tilde X_t - \mu -\tilde B \tilde U_{t}\right)^{\dagger}\left(\tilde A\tilde X_t - \mu -\tilde B \tilde U_{t}\right)\\%
	&=\left(\tilde X_t^{\dagger}\tilde A^{\dagger} - \mu^{\dagger} - \tilde U^{\dagger}_{t}\tilde B^{\dagger}\right)\left(\tilde A\tilde X_t - \mu -\tilde B \tilde U_{t}\right)\\%
	&=\tilde X_t^{\dagger} \tilde A^{\dagger}\tilde A \tilde X_t - \mu^{\dagger}\tilde A\tilde X_t - \tilde U_t^{\dagger}\tilde B^{\dagger}\tilde A\tilde X_t \\%
	&\quad- \tilde X_t^{\dagger}\tilde A^{\dagger}\mu + \mu^{\dagger}\mu + \tilde U_t^{\dagger}\tilde B^{\dagger}\mu \nonumber\\%
	&\quad- \tilde X^{\dagger}_t\tilde A^{\dagger}\tilde B\tilde U_t + \mu^{\dagger}\tilde B\tilde U_t + \tilde U_t^{\dagger}\tilde B^{\dagger}\tilde B \tilde U_t.\nonumber%
\end{align}
Now this equation is considered with respect to $\tilde X_t$ as a new variable. Such that we have a quadratic, a linear, and a constant term with respect to $\tilde X_t$, each defined as follows, respectively,%
\begin{align}
	&\bar Z_t =\tilde A^{\dagger}\tilde A\\%
	&\bar F_t = -2\tilde A^{\dagger}\left(\mu+\tilde B\tilde U_t\right)\\%
	&\bar c_t = \mu^{\dagger}\mu + 2 \mu^{\dagger}\tilde B\tilde U_t + \tilde U_t^{\dagger}\tilde B^{\dagger}\tilde B \tilde U_t%
\end{align}
Following we get: 
\begin{align}
	&|| X_t - \mu - \sum\limits_{q=1}^QA_{q}X_{t-q} - \sum\limits_{p=0}^PB_{p}U_{t-p}||_2^2 = \tilde X_t^{\dagger} \bar Z_t \tilde X_t^{\dagger} +  \bar F_t^{\dagger}\tilde X_t  + \bar c_t.%
\end{align}
In the next step, the sum over the local models $k=1,\dots,K$ is induced into the construction of the quadratic, the local, and the constant terms, such that we get:  %
\begin{align}
	&\sum\limits_{k=1}^K\gamma^{(k)}_t|| X_t - \mu^{(k)} - \sum\limits_{q=1}^QA^{(k)}_{q}X_{t-q} - \sum\limits_{p=0}^PB^{(k)}_{p}U_{t-p}||_2^2 \\%
	&= \tilde X_t^{\dagger} Z_t \tilde X_t^{\dagger} +  F_t^{\dagger} \tilde X_t  + c_t,%
\end{align}
where 
\begin{align}
	&Z_t = \sum\limits_{k=1}^K\gamma^{(k)}_t\tilde A^{k\dagger}\tilde A^k=
		\sum\limits_{k=1}^K\gamma^{(k)}_t\begin{bmatrix} -A^{k\dagger}_{Q}\\ \dots\\ -A^{k\dagger}_{1}\\\mathbb{1}\end{bmatrix} %
		\begin{bmatrix} -A^k_{Q},\dots, -A^k_{1},\mathbb{1}\end{bmatrix},\label{app_eq:matCt_vecft}\\%
	&F_t =  \sum\limits_{k=1}^K\gamma^{(k)}_t\left(-2\tilde A^{k\dagger}\left(\mu^k+\tilde B^k\tilde U_t\right)\right) =	-2 \sum\limits_{k=1}^K\gamma^{(k)}_t\begin{bmatrix} -A^{k\dagger}_{Q}\\ \dots\\ -A^{k\dagger}_{1}\\\mathbb{1}\end{bmatrix}\left(\mu^{(k)} + \begin{bmatrix} B^k_{0},\dots, B^k_{P}\end{bmatrix} \begin{bmatrix} U_{t}\\ \vdots\\ U_{t-P}\end{bmatrix}\right),\label{app_eq:matCt_vecft2}\\%
	&c_t = \sum\limits_{k=1}^K\gamma^{(k)}_t \bar c^k_t 
\end{align}
In the final step, we consider the FEM-VARX model distance function for all time steps and rewrite it into matrix vector form with respect to $X = \left(X_{1}^{\dagger},\dots,X_T^{\dagger}\right)^{\dagger}$: %
\begin{align}
	\label{app_eq:QP_X}
	&\sum\limits_{t=t_{st}}^T \tilde X_t^{\dagger} Z_t \tilde X_t +  F_t^{\dagger} \tilde X_t  + c_t = X^{\dagger}Z X + F^{\dagger}X + c.%
\end{align}
The construction of matrix $Z$ and vector $F$ is outlined in Figure~\ref{fig:ZF}, $c$ is the corresponding constant term and can be omitted.%
%
%
%
\subsection{Assembling of the QP for $X_{miss}$} 
\label{app:qp_xt_miss}
In the following, the QP (\ref{app_eq:QP_X}) is reduced to a QP with respect to the missing values, i.e., $X_{miss}$, only. For this purpose we define two binary masks:%
\begin{align*}
	&m \in\mathbb N^{dimxT},~~m_i = 
	\begin{cases}
		1 & \text{if } X_i = NaN\\
		0 & \text{else}
	\end{cases}\\
	&o = \neg m.
\end{align*}
That is, $m$ indicates all the missing and $o$ the observed dimensions in $X$. Further, we define a slicing operator for any matrix $M$ and any vector $v$:%
\begin{align*}
	&M^{[:,m]} \text{ all rows of matrix $M$ and columns $i$ of $M$ whit $m_i = 1$};\\
	&M^{[m,m]} \text{ rows $i$ of matrix $M$ with $m_i = 1$ and columns $i$ of $M$ with $m_i = 1$};\\
	&v^{[m]}   \text{ dimensions $i$ of vector $v$ with $m_i = 1$}.
\end{align*}
Please note, the slicing operator $[.]$ always applies before the transpose operator $\dagger$. With this definitions the vector $X$ can be decomposed into $X = [X^{[m]}\, X^{[o]}]$ where $X^{[m]}$ corresponds to $X_{miss}$. Using this decomposition, the components of the QP in (\ref{app_eq:QP_X}) can be rewritten to %
\begin{align}
	&X^{\dagger}Z X =  X^{[m]\dagger} Z^{[m,m]} X^{[m]} + X^{[o]\dagger}Z^{[o,m]} X^{[m]} + X^{[m]\dagger}Z^{m,o}X^{[o]} + r_Z\label{eq:VARX_qp_Q}\\%
	&F^{\dagger}X  = F^{[m]\dagger}X^{[m]}+ r_F\label{eq:VARX_qp_L}
\end{align}
The terms $r_Z$ and $r_F$ are the remaining summands which do not depend on the unknown $X^{[m]}$. Inserting (\ref{eq:VARX_qp_Q}) and (\ref{eq:VARX_qp_L}) into (\ref{app_eq:QP_X})  we obtain  %
\begin{align}
	\label{app_eq:VARX_qp_mask}
	X^{\dagger}Z X + F^{\dagger}X
	= X^{[m]\dagger} Z^{[m,m]} X^{[m]} + \left( 2X^{[o]\dagger}Z^{[o,m]} + F^{[m]\dagger} \right) X^{[m]} + r, %
\end{align}
with $r = r_Z + r_F$. As $r$ does not depend on the unknown $X^{[m]}$, it corresponds to a constant and can be neglected during the optimization step. In the last step, the notations $Z_{miss} = Z^{[m,m]} $, $F_{miss} = 2X^{[o]\dagger}Z^{[o,m]} + F^{[m]\dagger}$ are used to obtain the reduced QP with respect to $X_{miss}$: %
\begin{align}
	\label{app_eq:VARX_qp}
	\mathcal L \left(\Theta,\Gamma_t,X_{miss}; X_t, U_t\right) = X_{miss}^{\dagger}Z_{miss}X_{miss} + F_{miss}^{\dagger}X_{miss}.
\end{align}
%
\section{FEMM-VARX: missing data in $U_t$} 
\label{app:qp_ut_all}
In this section the FEM-VARX model distance function is rewritten with respect to $U_t$ as the referred variable in line with Appendix~\ref{app:qp_xt_all}. First, define new matrices as follows: %
\begin{align}
	&\hat A =\begin{pmatrix} -\mu & \mathbb{1} & -A_1 & \dots & -A_Q \end{pmatrix}, %
	\hat B =\begin{pmatrix}  B_P & B_{P-1} & \dots & B_0	\end{pmatrix},\label{app_eq:ABhat}\\%
	&\hat X_t = \begin{pmatrix}
		1\\
		X_{t}\\
		X_{t-1}\\
		\vdots\\
		X_{t-Q}
	\end{pmatrix},
	\hat U_t = \begin{pmatrix}
		U_{t-P}\\
		\vdots\\
		U_{t-1}\\
		U_t
	\end{pmatrix},\label{app_eq:XUhat}%
\end{align}
Then, the model distance function for a fixed time step is rewritten according to (compare also Appendix~\ref{app:qp_xt_all}):%
\begin{align}
	&|| X_t - \mu - \sum\limits_{q=1}^QA_{q}X_{t-q} - \sum\limits_{p=0}^PB_{p}U_{t-p}||_2^2%
	=\left(\hat A \hat X  - \hat B \hat U \right)^{\dagger}\left(\hat A \hat X - \hat B \hat U \right)\\%
	&=\hat X_t^{\dagger} \hat A^{\dagger} \hat A \hat X - 2 \hat X_t^{\dagger} \hat A^{\dagger}\hat B\hat U + \hat U^{\dagger}\hat B^{\dagger}\hat B \hat U\label{app_eq:QP_Umiss_t} %
\end{align}
Now this equation (\ref{app_eq:QP_Umiss_t}) is considered with respect to $\tilde U_t$ as a new variable. Such that we have a quadratic, a linear, and a constant term with respect to $\tilde U_t$, each defined as follows, respectively,%
\begin{align}
	&\bar D_t =\hat B^{\dagger}\hat B\\%
	&\bar G_t = - 2 \hat X_t^{\dagger} \hat A^{\dagger}\hat B\\%
	&\bar v_t = \hat X_t^{\dagger} \hat A^{\dagger} \hat A \hat X.%
\end{align}
By inducing the sum over the local models $k=1,\dots,K$ we get:  %
\begin{align}
	&\sum\limits_{k=1}^K\gamma^{(k)}_t|| X_t - \mu^{(k)} - \sum\limits_{q=1}^QA^{(k)}_{q}X_{t-q} - \sum\limits_{p=0}^PB^{(k)}_{p}U_{t-p}||_2^2 \\%
	&= \hat U_t^{\dagger} D_t \hat U_t^{\dagger} +  G_t^{\dagger} \hat U_t  + v_t,%
\end{align}
with 
\begin{align}
	&D_t = \sum\limits_{k=1}^K\gamma^{(k)}_t\tilde B^{k\dagger}\tilde B^k=
		\sum\limits_{k=1}^K\gamma^{(k)}_t\begin{bmatrix} B^{k\dagger}_{P}\\ \dots\\ B^{k\dagger}_{0}\end{bmatrix}\begin{bmatrix} B^k_{P},\dots, B^k_{0}\end{bmatrix},\label{app_eq:Dt}\\%
	&G_t =  \sum\limits_{k=1}^K\gamma^{(k)}_t\left(- 2 \hat X_t^{\dagger} \hat A^{k\dagger}\hat B^k\right)%
	     = 2 \begin{bmatrix} 1,X^{\dagger}_t, X^{\dagger}_{t-1}, \dots, X^{\dagger}_{t-Q}\end{bmatrix} \sum\limits_{k=1}^K\gamma^{(k)}_t%
				 \begin{bmatrix}\mu^{k\dagger}\\-\mathbb 1\\A^{k\dagger}_1 \\ \vdots \\ A^{k\dagger}_{Q}  \end{bmatrix}%
				 \begin{bmatrix} B^k_{P},\dots, B^k_{0}\end{bmatrix},\label{app_eq:Gt}\\%
	&v_t = \sum\limits_{k=1}^K\gamma^{(k)}_t \bar v^k_t = \sum\limits_{k=1}^K\gamma^{(k)}_t\left( \hat X_t^{\dagger} \hat A^{k\dagger} \hat A^k \hat X\right).%
\end{align}
In the final step, we consider the FEM-VARX model distance function for all time steps and rewrite it into matrix vector form with respect to $U = \left(U_{1}^{\dagger},\dots,U_T^{\dagger}\right)^{\dagger}$: %
\begin{align}
	\label{app_eq:QP_U}
	&\sum\limits_{t=t_{st}}^T \hat U_t^{\dagger} D_t \hat U_t +  G_t^{\dagger} \hat U_t  + v_t = U^{\dagger}D U + G^{\dagger}U + v.%
\end{align}
The construction of matrix $D$ and vector $G$ is performed in line with Figure~\ref{fig:ZF} except that the corresponding shift of $D_t$ and $G_t$ is performed with $m$, $v$ is the corresponding constant term and can be omitted. \\%
In the next step the QP (\ref{app_eq:QP_U}) needs to be reduced with respect to $U_{miss}$: We skip the derivation at this point, as it is performed in exact the same manner as in Appendix~\ref{app:qp_xt_miss}, i.e., by applying the corresponding binary masks and the slicing operator. %
%
\section{Test-Case FEMM-VARX matrices} 
\label{sec:test_case_fem_varx_matrices}
In this section the matrices for generating the series $X_t$ according to equation (\ref{eq:generateXt}) are defined. The matrices for the first model, i.e., $k=1$ are given by%
\begin{align}
	&\mu^{(1)}=\begin{pmatrix*}[r]
	    2\\
	    6\\
	    3\\
	    -1
	\end{pmatrix*},\,
	A_1^{(1)} = \begin{pmatrix*}[r]
     0.03 &  -0.07 &   0.01 &   0.02\\
     0.02 &   0.07 &   0.03 &   0.03\\
    -0.01 &   0.03 &   0.04 &  -0.01\\
     0.07 &   0.01 &  -0.02 &   0.07
	\end{pmatrix*},
\end{align}
\begin{align}
	&A_2^{(1)} = \begin{pmatrix*}[r]
		 0.07 &  -0.03 &   0.04 &   0.05\\
		 0.01 &   0.06 &  -0.01 &   0.02\\
		 0.04 &  -0.02 &   0.01 &   0.07\\
		-0.02 &   0.06 &   0.02 &  -0.01
	\end{pmatrix*},\,
	A_3^{(1)} =\begin{pmatrix*}[r]
		0.20 &   0.10 &  -0.10 &   0.20\\
		0.30 &  -0.20 &   0.50 &   0.30\\
		0.40 &   0.50 &   0.40 &   0.60\\
		0.90 &   0.30 &  -0.30 &   0.20
	\end{pmatrix*},
\end{align}
\begin{align}
	&B_0^{(1)} = \begin{pmatrix*}[r]
		 0.40  &   0.10 &   -0.30 &   0.20\\
		 0.10  &   0.70 &   0.40  &   0.50\\
		 0.20  &  -0.10 &   0.30  &   0.40\\
		-0.20  &   0.70 &   0.60  &  -0.10
	\end{pmatrix*},\,
	B_1^{(1)} = \begin{pmatrix*}[r]
		-0.30 &   0.50 &   0.40 &  -0.20\\
		 0.60 &  -0.30 &   0.30 &   0.40\\
		 0.10 &   0.60 &  -0.10 &   0.20\\
		 0.30 &  -0.70 &   0.40 &  -0.20
	\end{pmatrix*},
\end{align}
\begin{align}
	&B_2^{(1)} = \begin{pmatrix*}[r]
		-0.06 &   0.02 &   0.03 &  -0.03\\
		 0.05 &   0.01 &  -0.01 &   0.08\\
		 0.09 &   0.07 &   0.06 &  -0.02\\
		 0.04 &  -0.04 &  -0.05 &   0.01
	\end{pmatrix*},\,
	B_3^{(1)} = \begin{pmatrix*}[r]
		 0.002 &  -0.002 &   0.004 &   0.005\\
		 0.001 &   0.002 &  -0.004 &   0.002\\
		 0.002 &   0.001 &   0.001 &   0.005\\
		-0.003 &   0.001 &  -0.002 &  -0.002
	\end{pmatrix*}.
\end{align}
and the matrices for the second model, i.e., $k=2$, are given by
\begin{align}
	&\mu^{(2)}=\begin{pmatrix*}[r]
	    5\\
	    4\\
	    1\\
	    2\\
	\end{pmatrix*},\,
	A_1^{(2)} = \begin{pmatrix*}[r]
		0.03 &   0.07 &   0.03 &  -0.04\\
	   -0.06 &   0.04 &   0.07 &   0.02\\
	   -0.09 &   0.01 &   0.10 &   0.05\\
	   -0.08 &  -0.10 &  -0.01 &  -0.05
	\end{pmatrix*},
\end{align}
\begin{align}
	&A_2^{(2)} = \begin{pmatrix*}[r]
	    0.03 &   0.08 &   0.01 &   0.03\\
	    0.01 &  -0.09 &   0.07 &  -0.06\\
	    0.03 &   0.06 &  -0.09 &  -0.02\\
	    0.05 &   0.01 &  -0.01 &   0.09
	\end{pmatrix*},\,
	A_3^{(2)} = \begin{pmatrix*}[r]
	    0.10 &   -0.20 &   0.80 &  -0.90\\
	   -0.30 &   -0.40 &  -0.30 &   0.80\\
	    0.70 &    0.70 &  -0.40 &  -0.20\\
	    0.10 &   -0.70 &   0.20 &   0.60
	\end{pmatrix*}
\end{align}
\begin{align}
	&B_0^{(2)} = \begin{pmatrix*}[r]
		0.40 &  -0.20 &   0.90 &  -1.00\\
	    0.10 &   0.50 &   0.60 &  -0.70\\
	    0.10 &  -0.50 &   0.50 &  -0.30\\
	   -0.10 &   0.50 &   0.60 &   0.50
	\end{pmatrix*},\,
	B_1^{(2)} = \begin{pmatrix*}[r]
	    -0.10 &  -0.20 &  -0.30 &   0.10\\
	     0.20 &   0.10 &  -0.30 &  -0.80\\
	     0.90 &   0.80 &   0.40 &  -0.30\\
	     0.80 &  -0.70 &  -0.30 &   0.20
	\end{pmatrix*},
\end{align}
\begin{align}
	&B_2^{(2)} =\begin{pmatrix*}[r]
	    0.06 &   0.09 &   0.07 &  -0.07\\
	   -0.05 &   0.08 &  -0.01 &   0.02\\
	   -0.09 &  -0.04 &  -0.01 &  -0.05\\
	   -0.07 &   0.09 &   0.03 &   0.07
	\end{pmatrix*},\,
	B_3^{(2)} = \begin{pmatrix*}[r]
	    0.003 &  -0.007 &  -0.008 &  -0.008\\
	    0.001 &   0.005 &   0.007 &   0.004\\
	   -0.004 &  -0.004 &  -0.008 &   0.001\\
	    0.006 &   0.006 &   0.004 &   0.008
	\end{pmatrix*}.
\end{align}
The starting values for generating $X_t$ are given below:
\begin{align}
	X_1=\begin{pmatrix*}[r]
	    0.30\\
	   -0.50\\
	    0.20\\
	    0.10
	\end{pmatrix*},\,
	X_2=\begin{pmatrix*}[r]
    	0.70\\
    	0.10\\
    	0.30\\
   	   -0.30
	\end{pmatrix*},\,
	X_3=\begin{pmatrix*}[r]
	    0.10\\
	   -0.90\\
	    0.40\\
	    0.10
	\end{pmatrix*}.
\end{align}
The fourth external factor $u_4(t)$, for $t=1,\dots,1002$, represents a composition of the random walk $\omega(t)$ and the Gaussian noise $\eta\sim\mathcal N(0,1)$: %
\begin{align}
	u_4(t) = \omega(t) + \eta,
\end{align} 
where $\omega(t)$ is defined by: 
\begin{align}
	\omega(t) = \omega(t-1) + a + (b-a)rand,
\end{align}
with $a=-0.5$, $b=0.5$ and $\omega(1) = 0.5$, while $rand$ represents a uniformly distributed random number generator in the interval $(0,1)$. %
%
%
%
\section{Visualization of the reconstruction} 
\label{sec:visualisation_of_the_reconstruction}
\begin{figure}[ht]
  \begin{adjustbox}{addcode={\begin{minipage}{\width}}{\caption{Case (a) for $p=5\%$: Reconstruction of $X_t$. %
	  }%
  	\label{fig:XT5}
	\end{minipage}},rotate=90,center}
      \includegraphics[scale=.4]{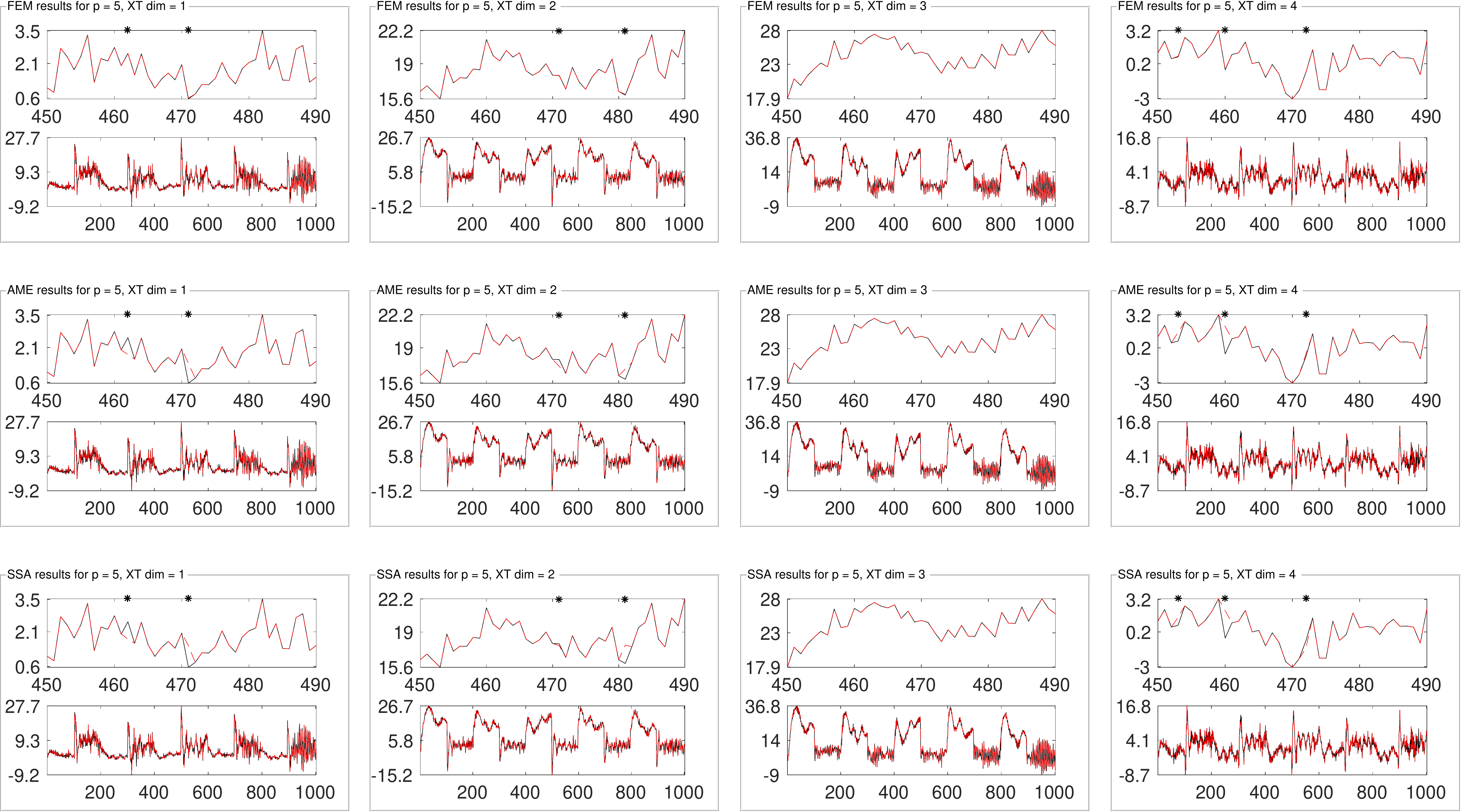}%
  \end{adjustbox}
\end{figure}

\begin{figure}[ht]
  \begin{adjustbox}{addcode={\begin{minipage}{\width}}{\caption{Case (a) for $p=35\%$: Reconstruction of $X_t$. %
	  }%
  	\label{fig:XT35}
	\end{minipage}},rotate=90,center}
      \includegraphics[scale=.4]{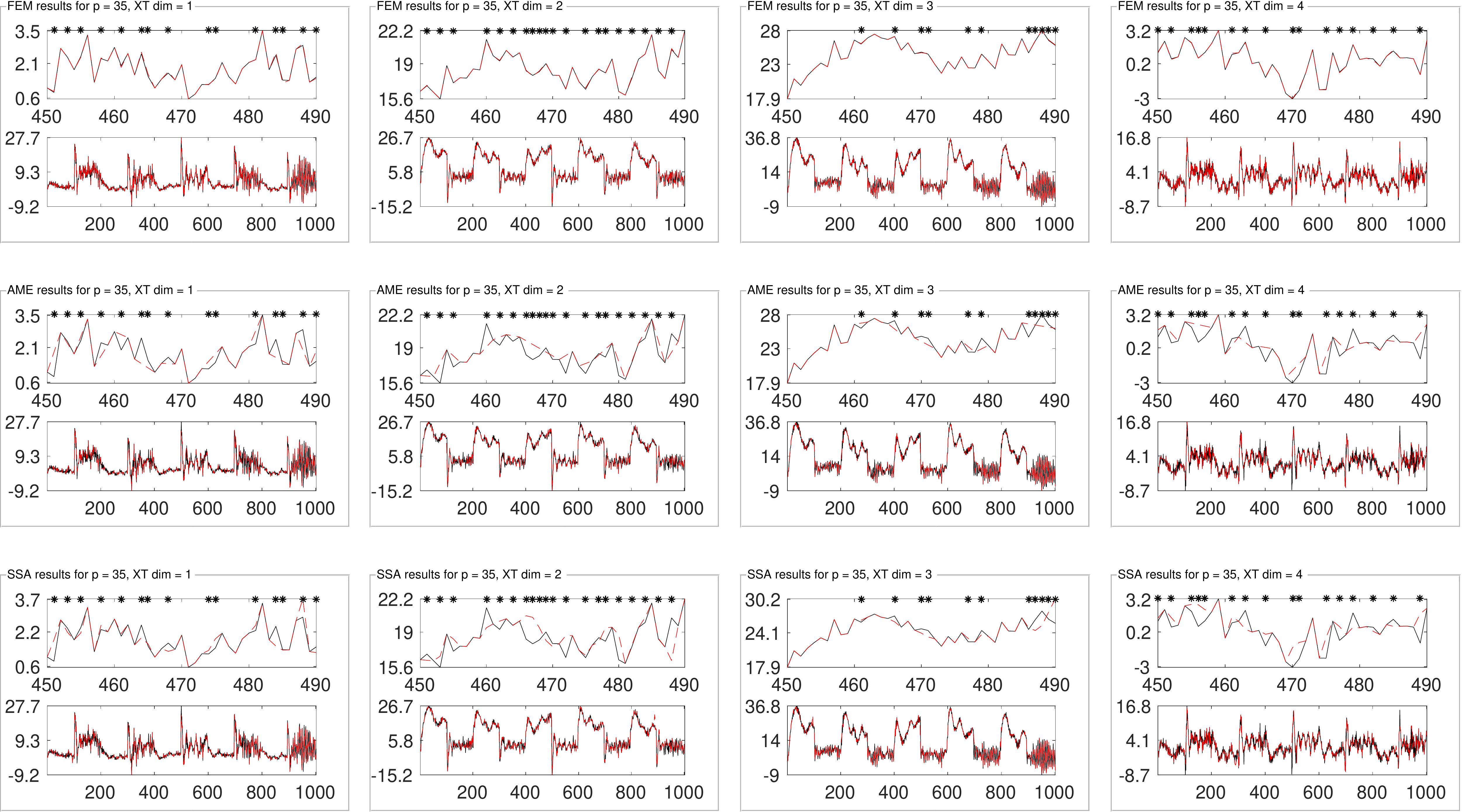}%
  \end{adjustbox}
\end{figure}

\begin{figure}[ht]
  \begin{adjustbox}{addcode={\begin{minipage}{\width}}{\caption{Case (a) for $p=55\%$: Reconstruction of $X_t$. %
	  }%
	    \label{fig:XT55}
  		\end{minipage}},rotate=90,center}
      \includegraphics[scale=.4]{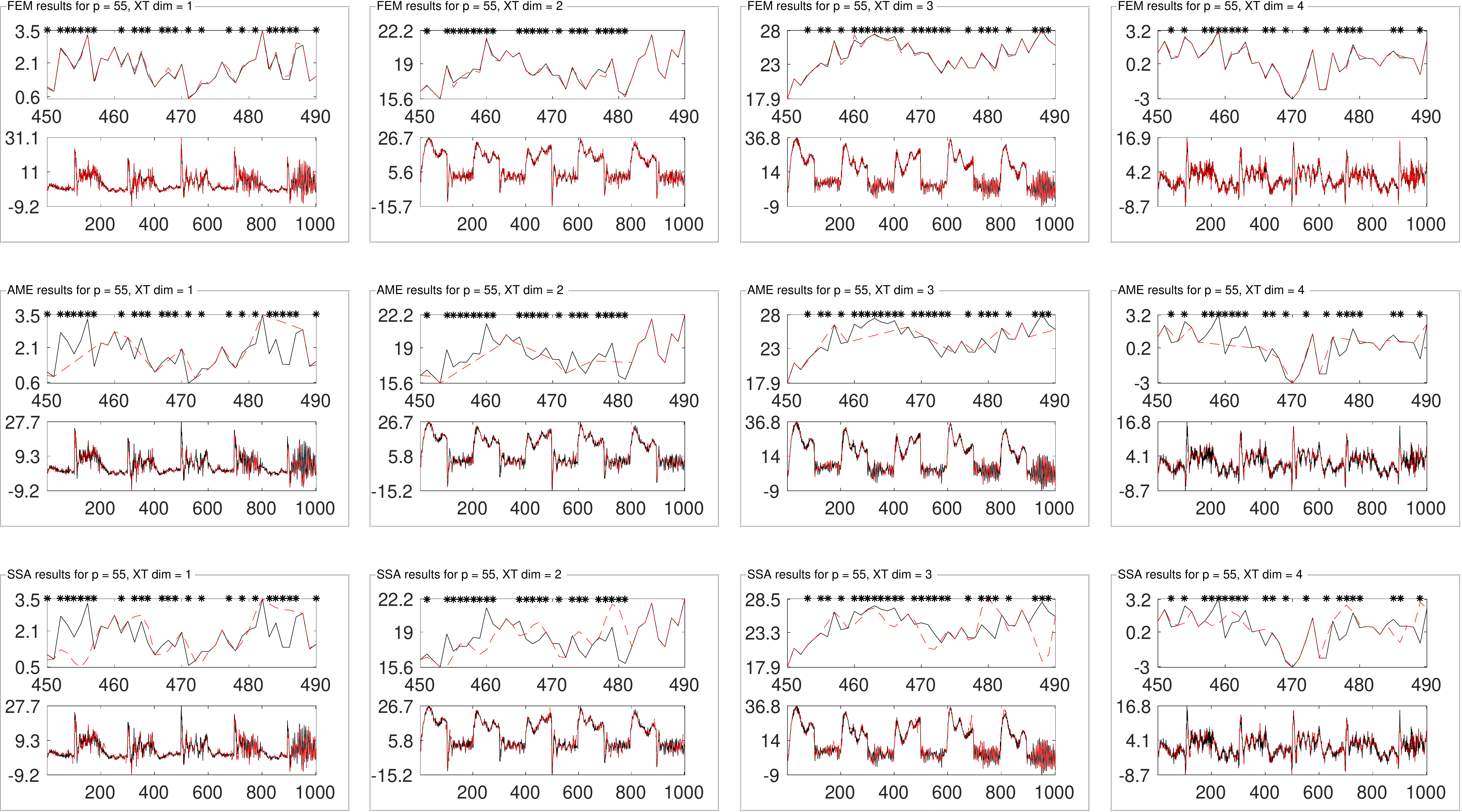}%
  \end{adjustbox}
\end{figure}

\begin{figure}[ht]
  \begin{adjustbox}{addcode={\begin{minipage}{\width}}{\caption{Case (a) for $p=75\%$: Reconstruction of $X_t$. %
	  }%
	  \label{fig:XT75}
  		\end{minipage}},rotate=90,center}
      \includegraphics[scale=.4]{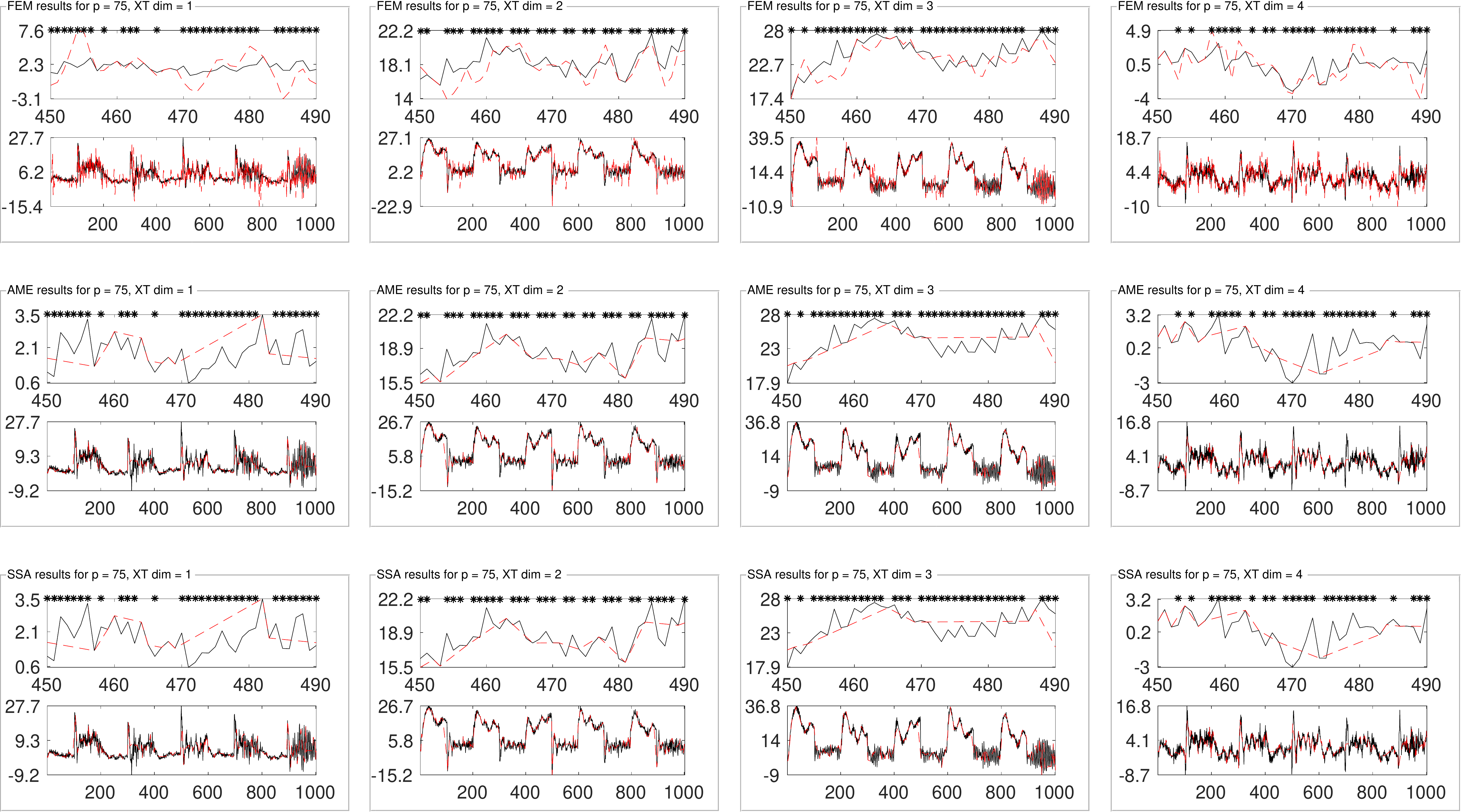}%
  \end{adjustbox}
\end{figure}

\begin{figure}[ht]
  \begin{adjustbox}{addcode={\begin{minipage}{\width}}{\caption{Case (a) for $p=95\%$: Reconstruction of $X_t$. %
	  }%
	    \label{fig:XT95}
  	\end{minipage}},rotate=90,center}
      \includegraphics[scale=.4]{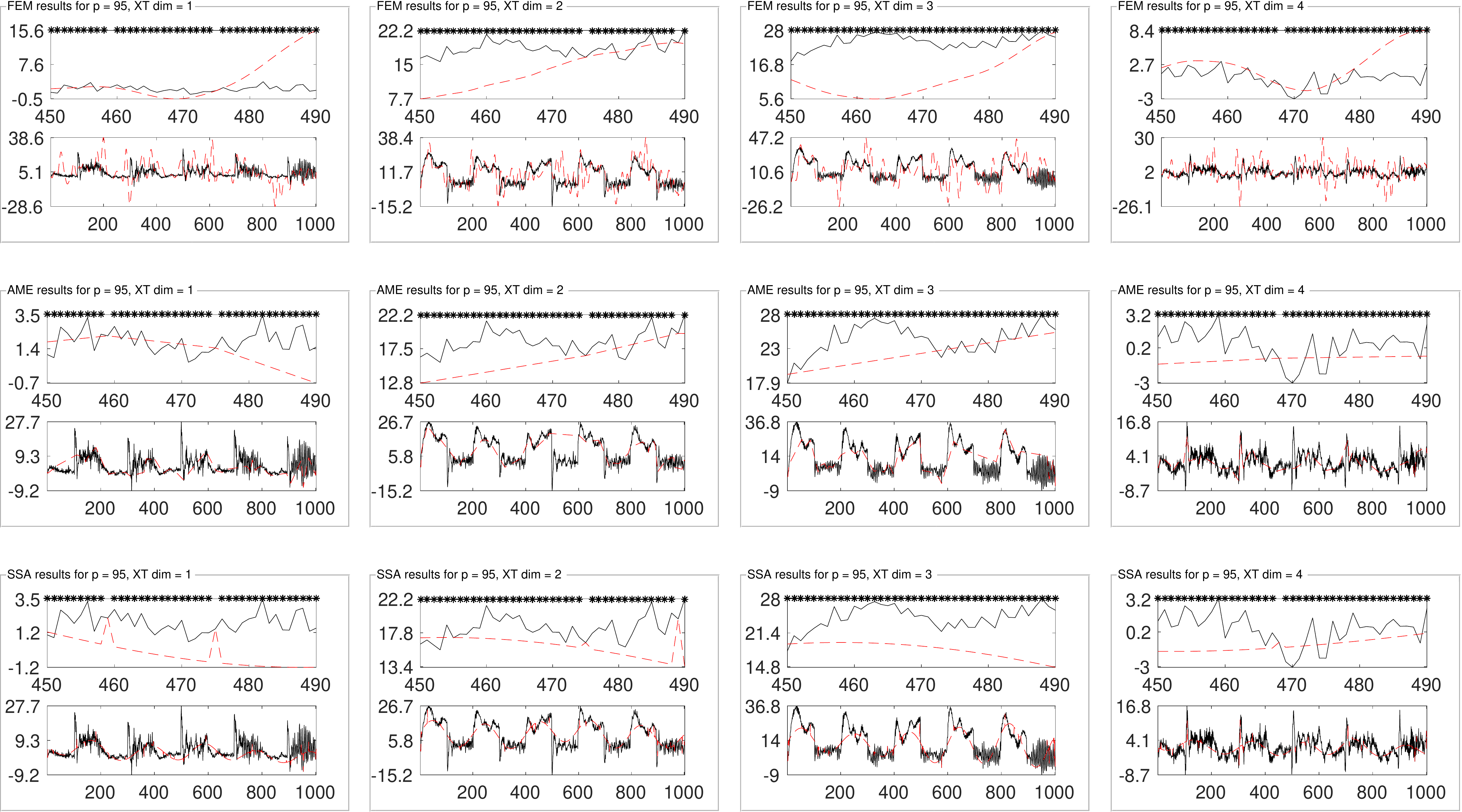}%
  \end{adjustbox}
\end{figure}

\begin{figure}[ht]
  \begin{adjustbox}{addcode={\begin{minipage}{\width}}{\caption{Case (b) for $p=5\%$: Reconstruction of $U_t$. %
	  }%
	    \label{fig:UT5}
  		\end{minipage}},rotate=90,center}
      \includegraphics[scale=.4]{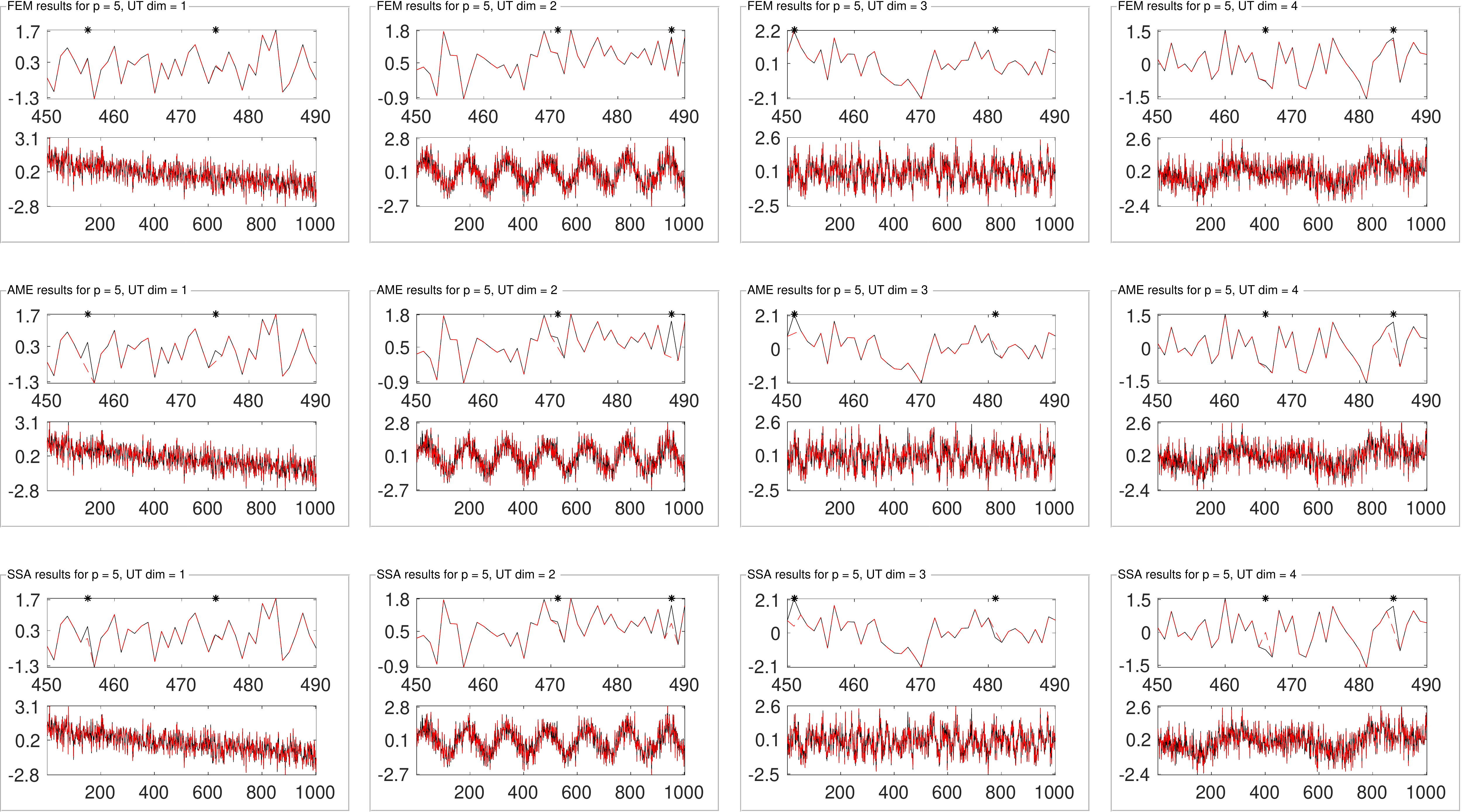}%
  \end{adjustbox}
\end{figure}

\begin{figure}[ht]
  \begin{adjustbox}{addcode={\begin{minipage}{\width}}{\caption{Case (b) for $p=35\%$: Reconstruction of $U_t$. %
	  }%
	    \label{fig:UT35}
  		\end{minipage}},rotate=90,center}
      \includegraphics[scale=.4]{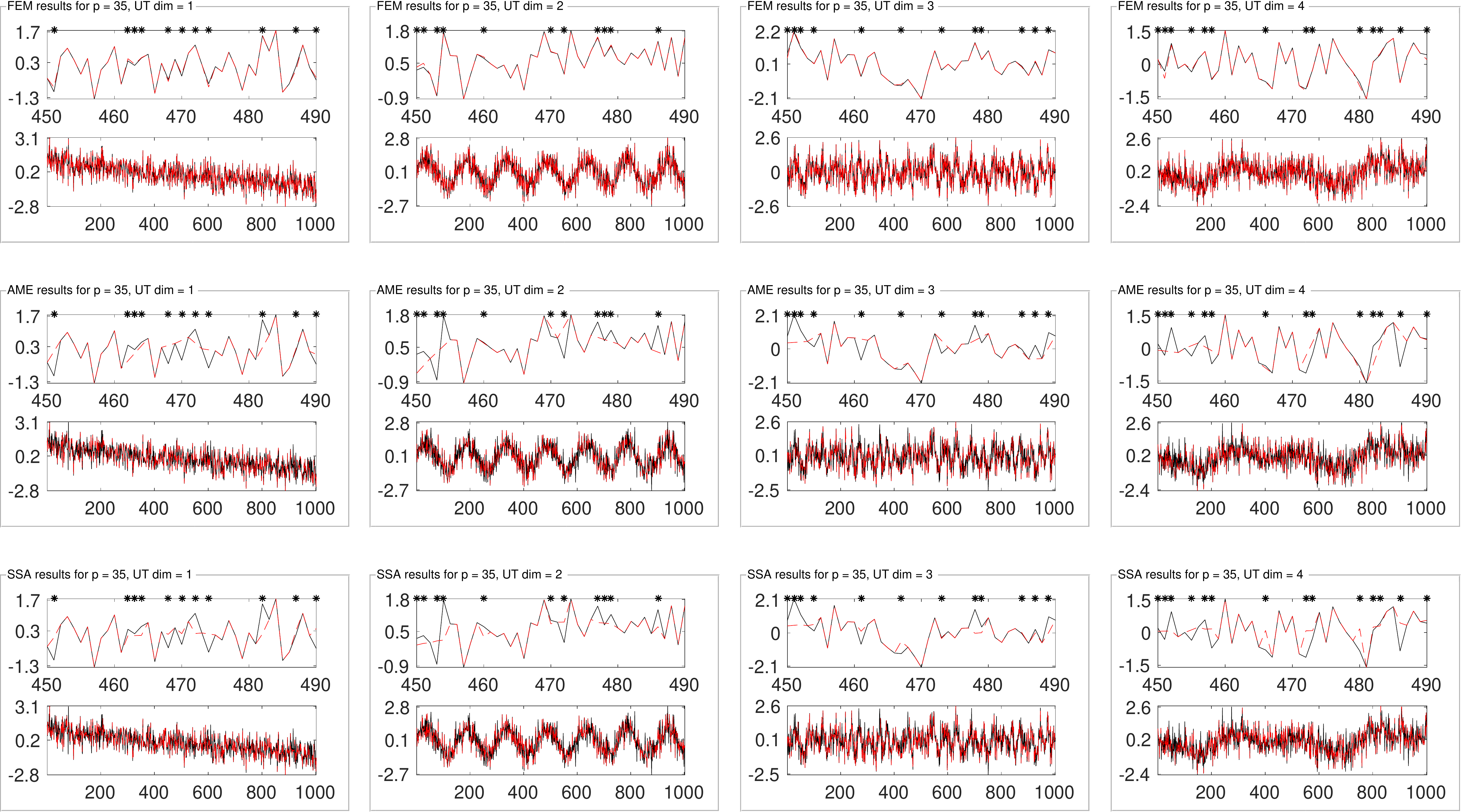}%
  \end{adjustbox}
\end{figure}

\begin{figure}[ht]
  \begin{adjustbox}{addcode={\begin{minipage}{\width}}{\caption{Case (b) for $p=55\%$: Reconstruction of $U_t$. %
	  }%
	    \label{fig:UT55}
  		\end{minipage}},rotate=90,center}
      \includegraphics[scale=.4]{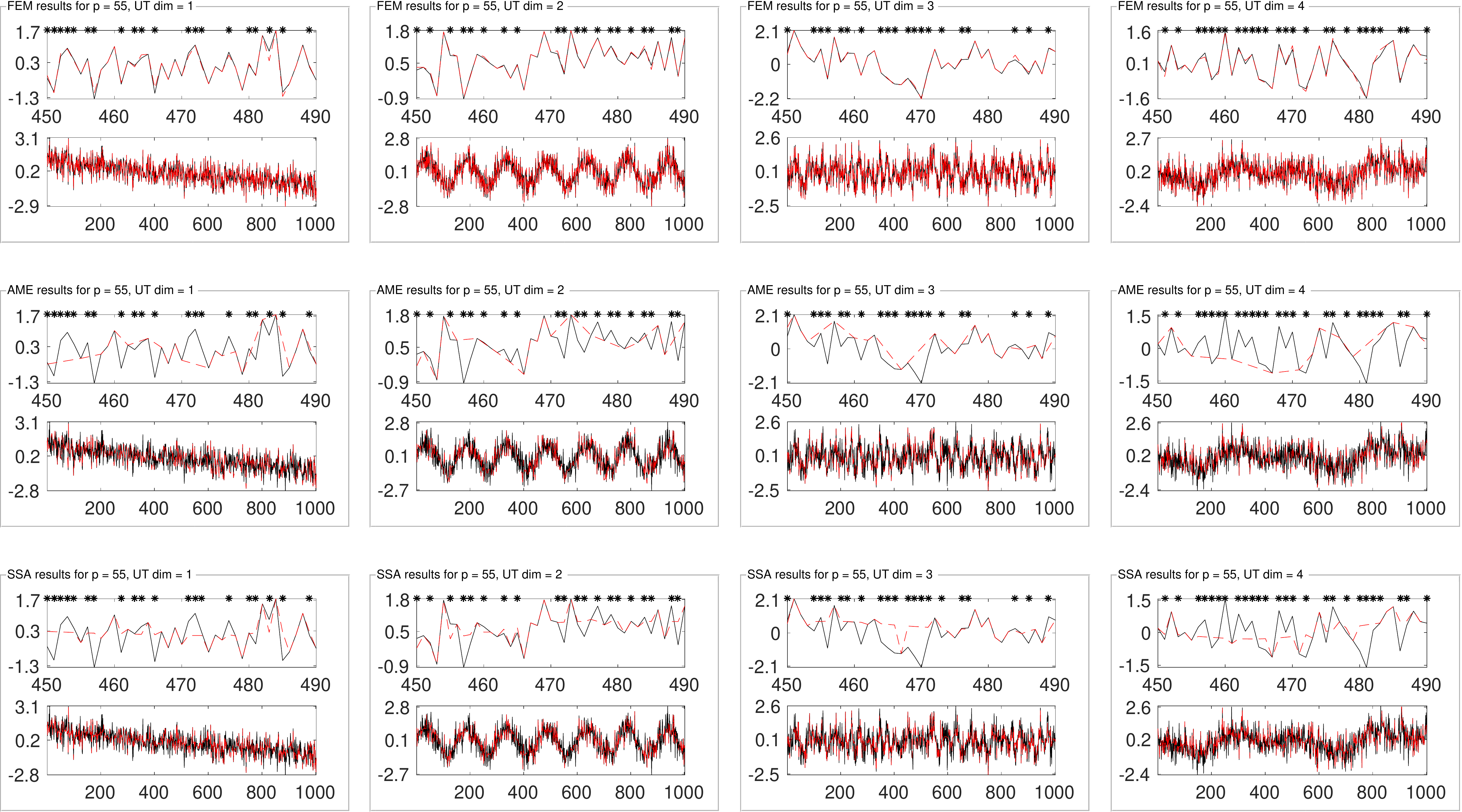}%
  \end{adjustbox}
\end{figure}

\begin{figure}[ht]
  \begin{adjustbox}{addcode={\begin{minipage}{\width}}{\caption{Case (b) for $p=75\%$: Reconstruction of $U_t$. %
	  }%
	    \label{fig:UT75}
  		\end{minipage}},rotate=90,center}
      \includegraphics[scale=.4]{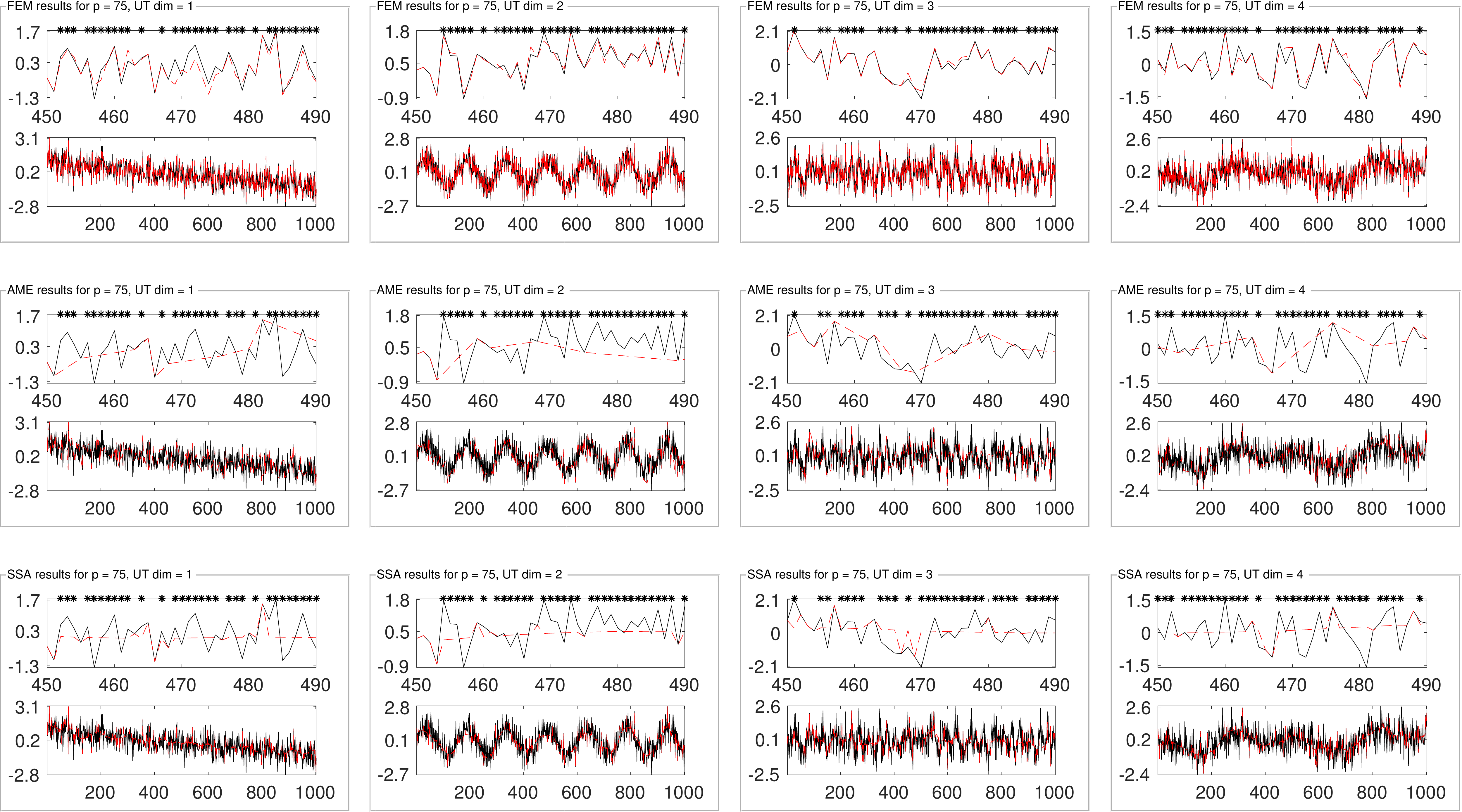}%
  \end{adjustbox}
\end{figure}

\begin{figure}[ht]
  \begin{adjustbox}{addcode={\begin{minipage}{\width}}{\caption{Case (b) for $p=95\%$: Reconstruction of $U_t$. %
	  }%
	    \label{fig:UT95}
  	\end{minipage}},rotate=90,center}
      \includegraphics[scale=.4]{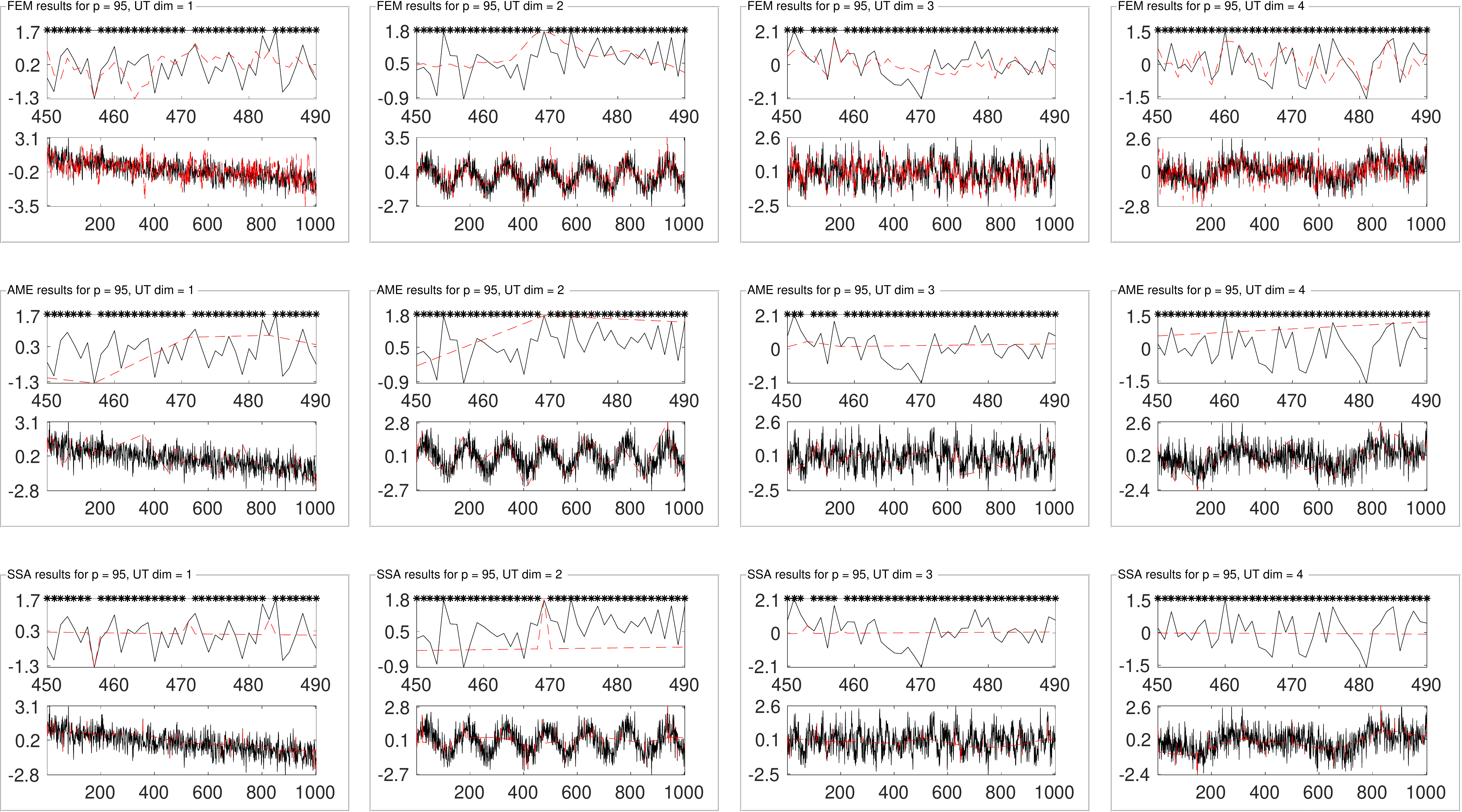}%
  \end{adjustbox}
\end{figure}

\begin{figure}[ht]
  \begin{adjustbox}{addcode={\begin{minipage}{\width}}{\caption{Case (c) for $p=5\%$: Reconstruction of $X_t$. %
	  }%
	    \label{fig:XTUT5_xt}
  		\end{minipage}},rotate=90,center}
      \includegraphics[scale=.4]{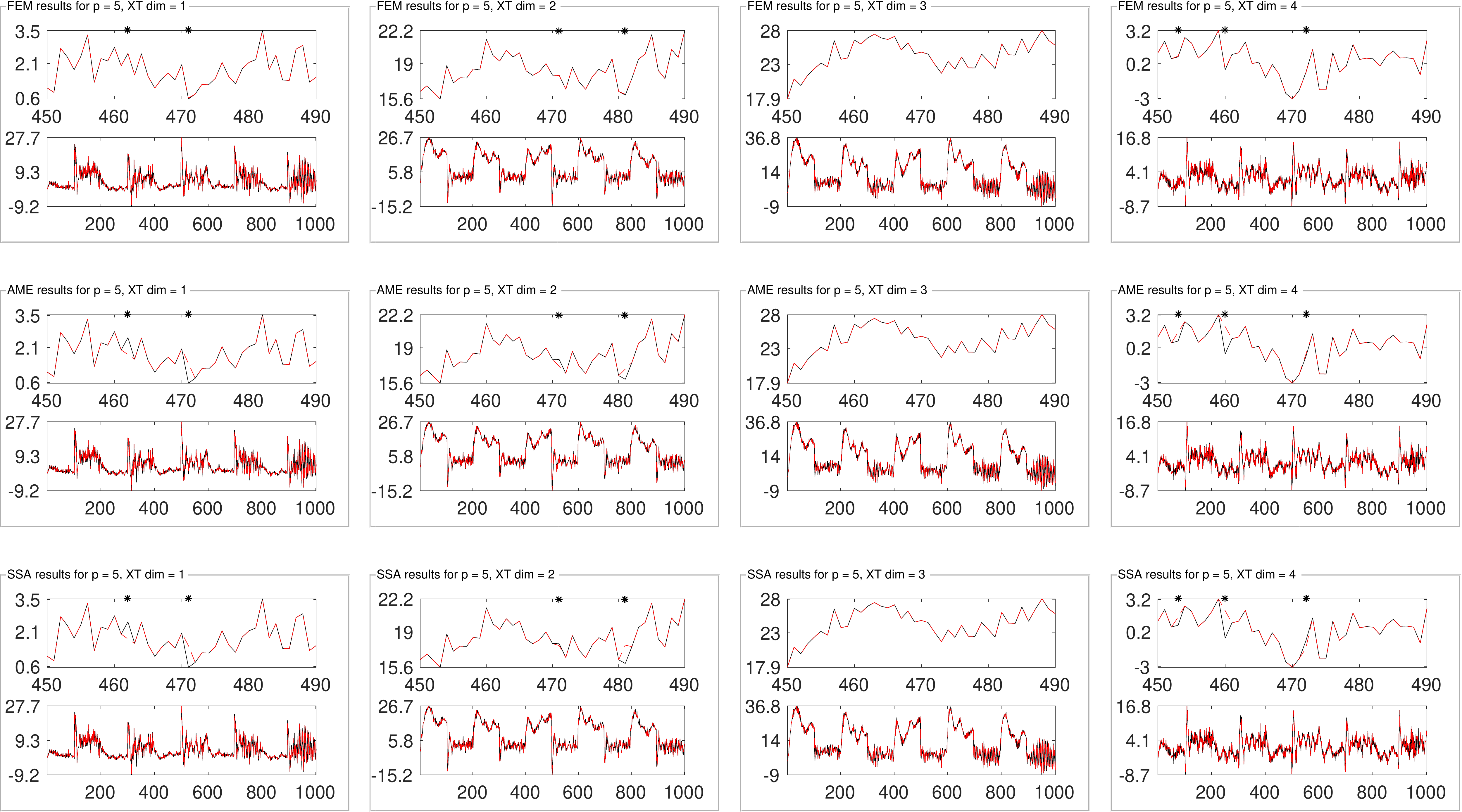}%
  \end{adjustbox}
\end{figure}

\begin{figure}[ht]
  \begin{adjustbox}{addcode={\begin{minipage}{\width}}{\caption{Case (c) for $p=35\%$: Reconstruction of $X_t$.
	  }%
	    \label{fig:XTUT35_xt}
  		\end{minipage}},rotate=90,center}
      \includegraphics[scale=.4]{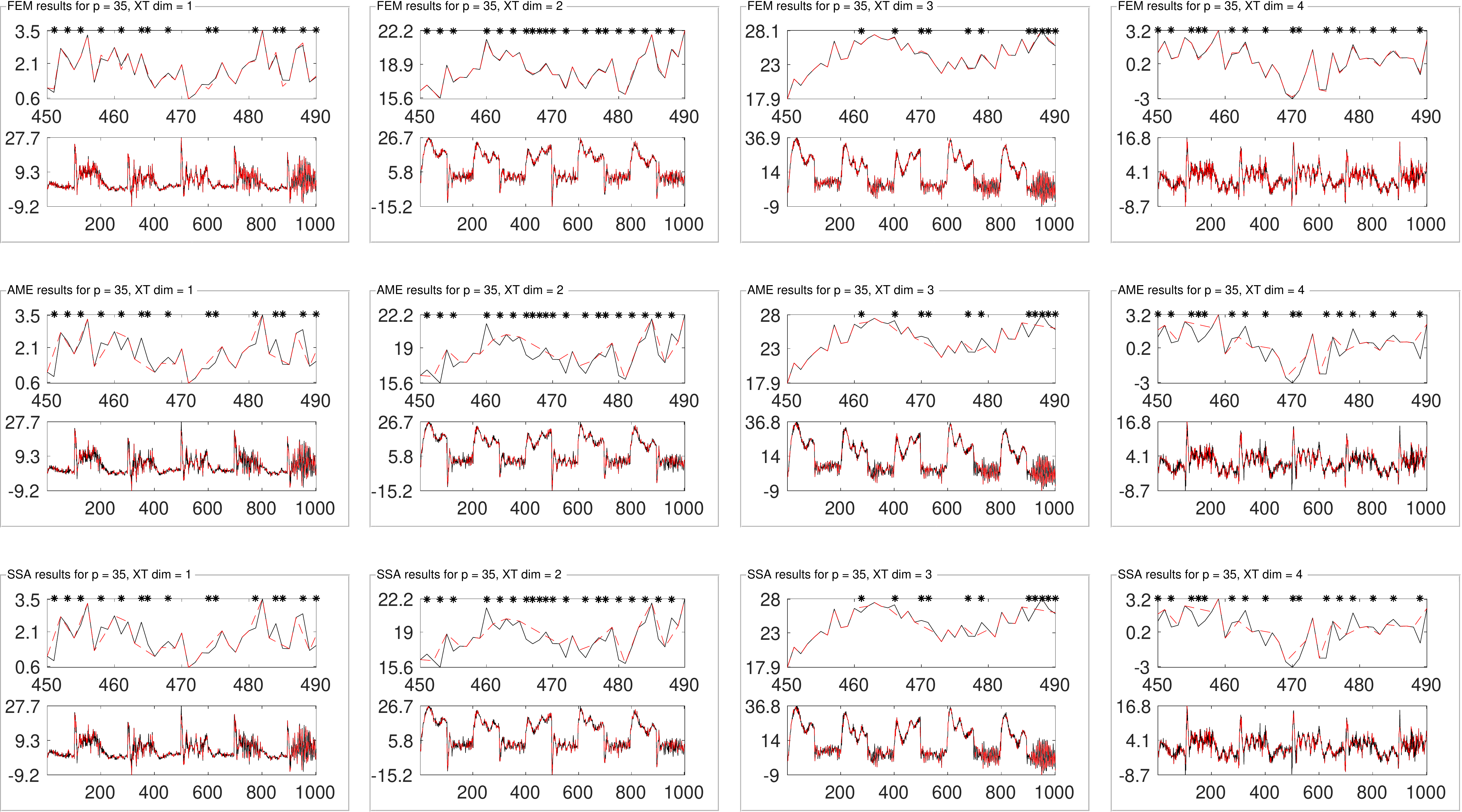}%
  \end{adjustbox}
\end{figure}

\begin{figure}[ht]
  \begin{adjustbox}{addcode={\begin{minipage}{\width}}{\caption{Case (c) for $p=55\%$: Reconstruction of $X_t$. %
	  }%
	    \label{fig:XTUT55_xt}
  		\end{minipage}},rotate=90,center}
      \includegraphics[scale=.4]{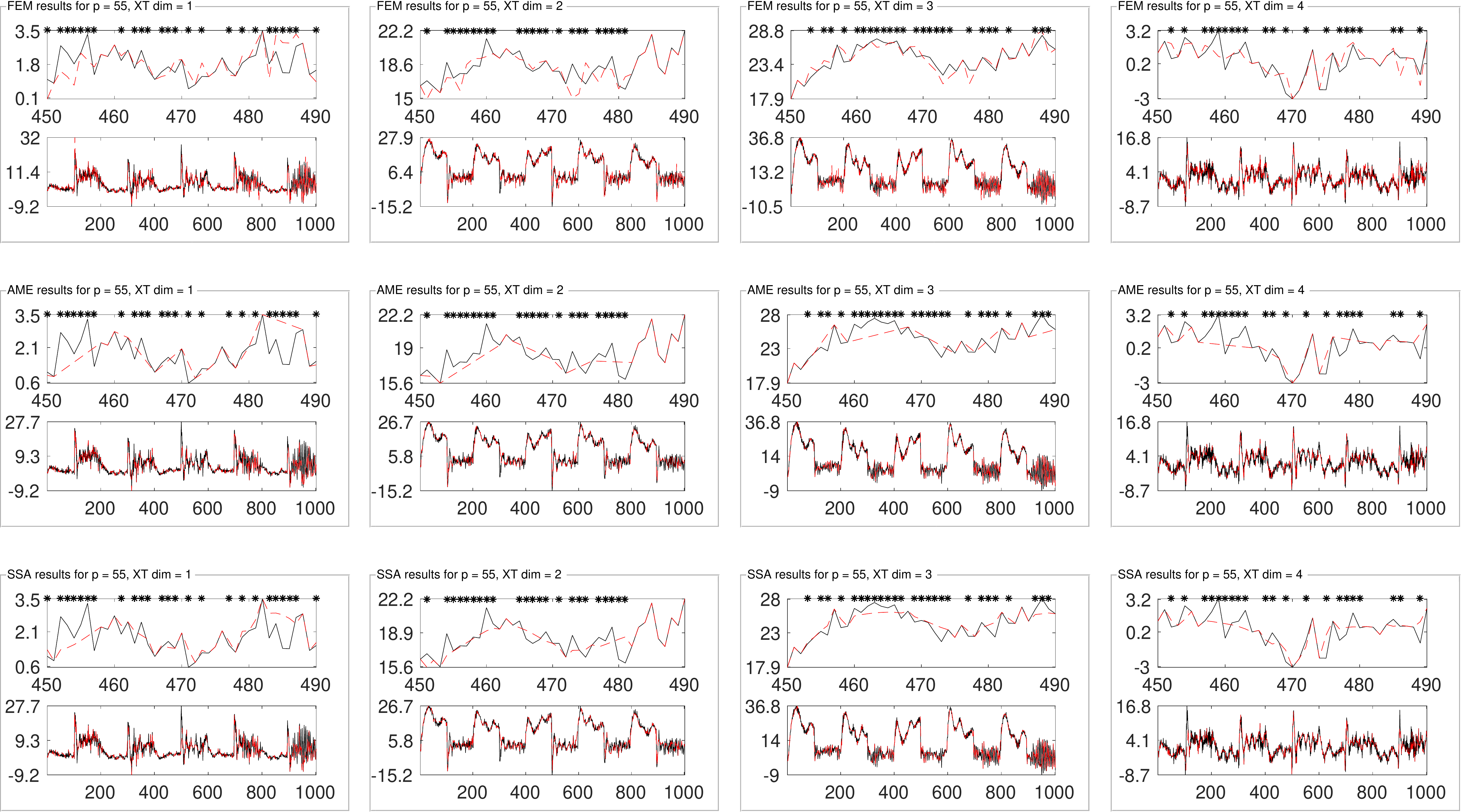}%
  \end{adjustbox}
\end{figure}

\begin{figure}[ht]
  \begin{adjustbox}{addcode={\begin{minipage}{\width}}{\caption{Case (c) for $p=75\%$: Reconstruction of $X_t$. %
	  }%
	    \label{fig:XTUT75_xt}
  		\end{minipage}},rotate=90,center}
      \includegraphics[scale=.4]{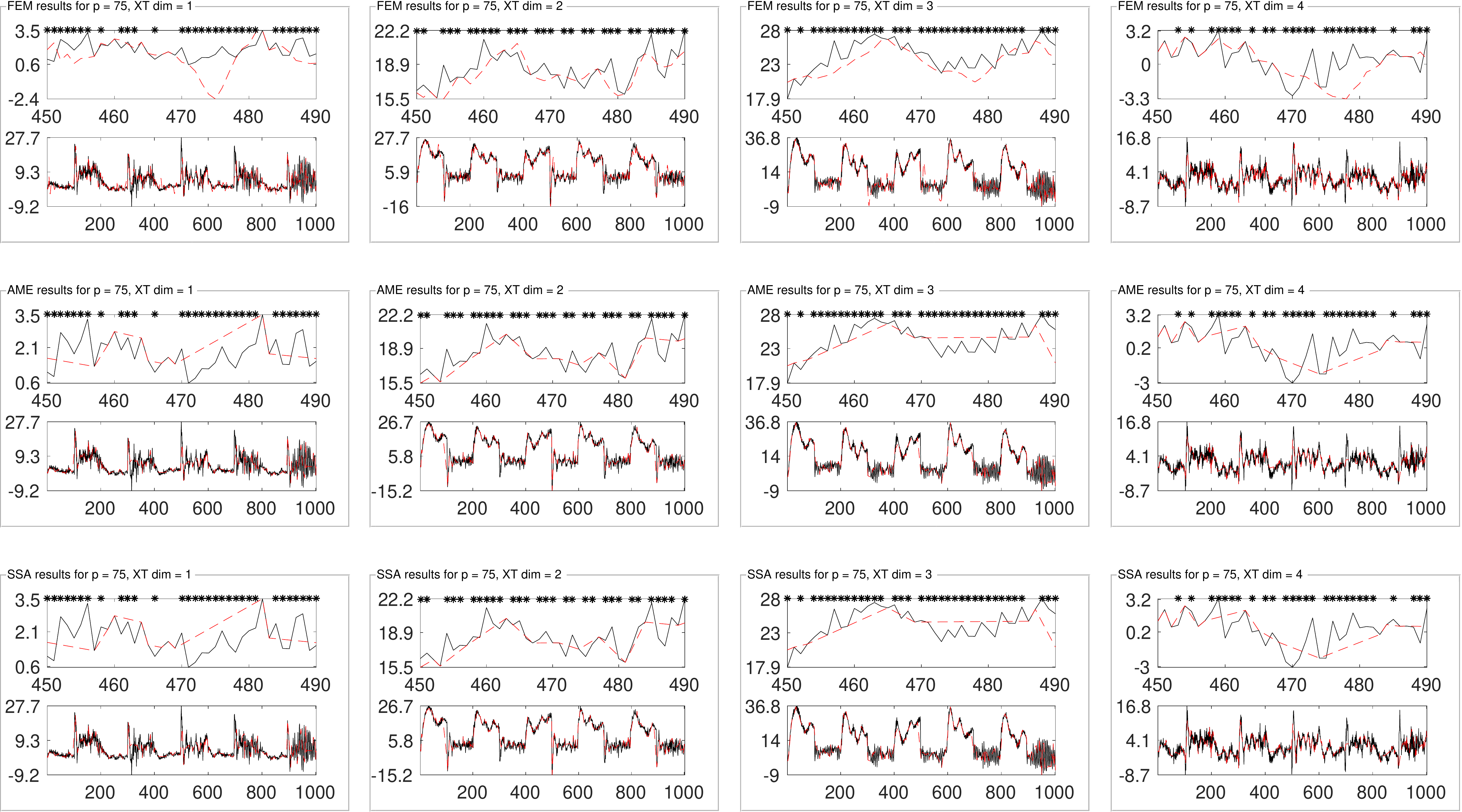}%
  \end{adjustbox}
\end{figure}

\begin{figure}[ht]
  \begin{adjustbox}{addcode={\begin{minipage}{\width}}{\caption{Case (c) for $p=95\%$: Reconstruction of $X_t$. %
	  }%
	    \label{fig:XTUT95_xt}
  	\end{minipage}},rotate=90,center}
      \includegraphics[scale=.4]{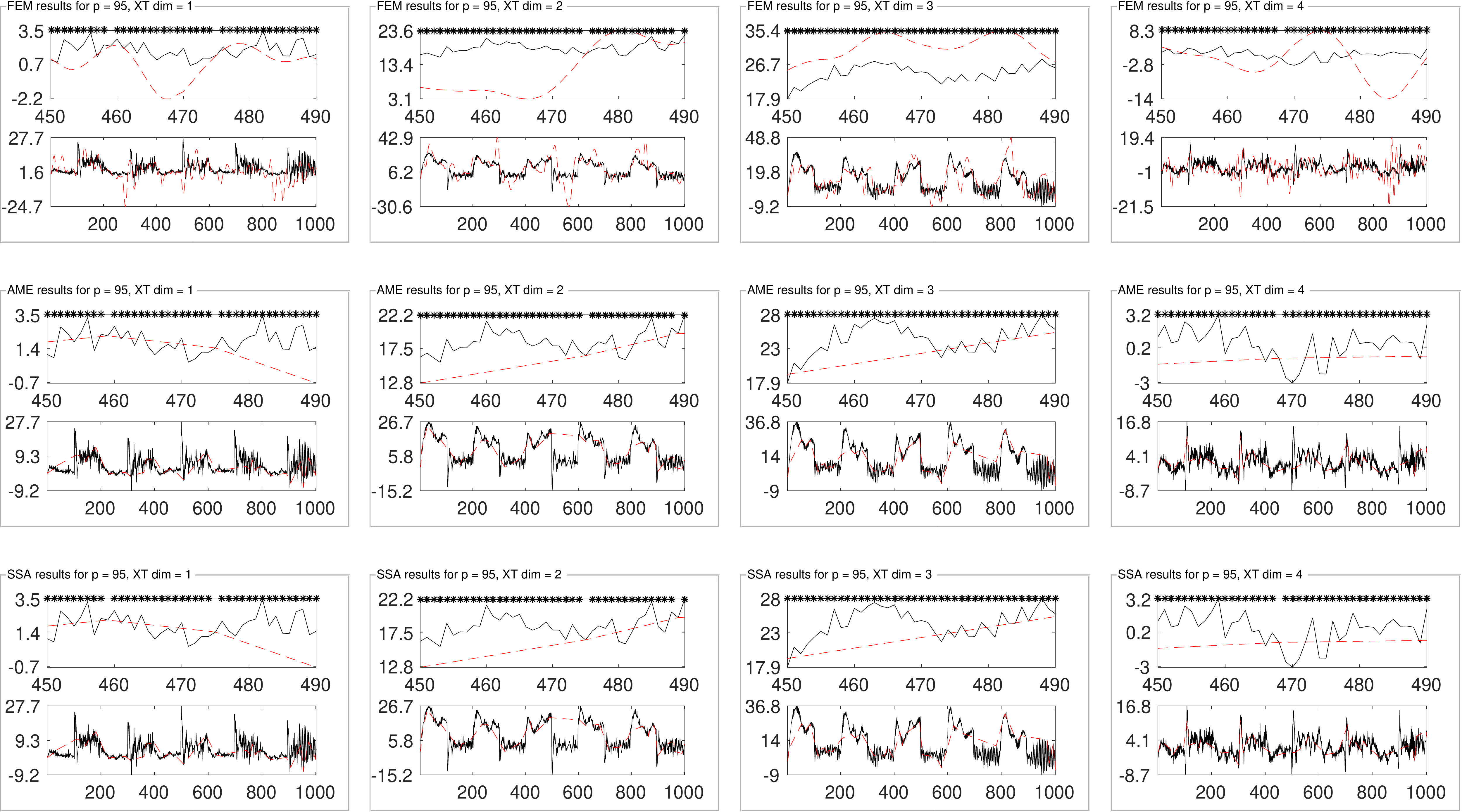}%
  \end{adjustbox}
\end{figure}

\begin{figure}[ht]
  \begin{adjustbox}{addcode={\begin{minipage}{\width}}{\caption{Case (c) for $p=5\%$: Reconstruction of $U_t$.}%
	    \label{fig:XTUT5_ut}
  		\end{minipage}},rotate=90,center}
      \includegraphics[scale=.4]{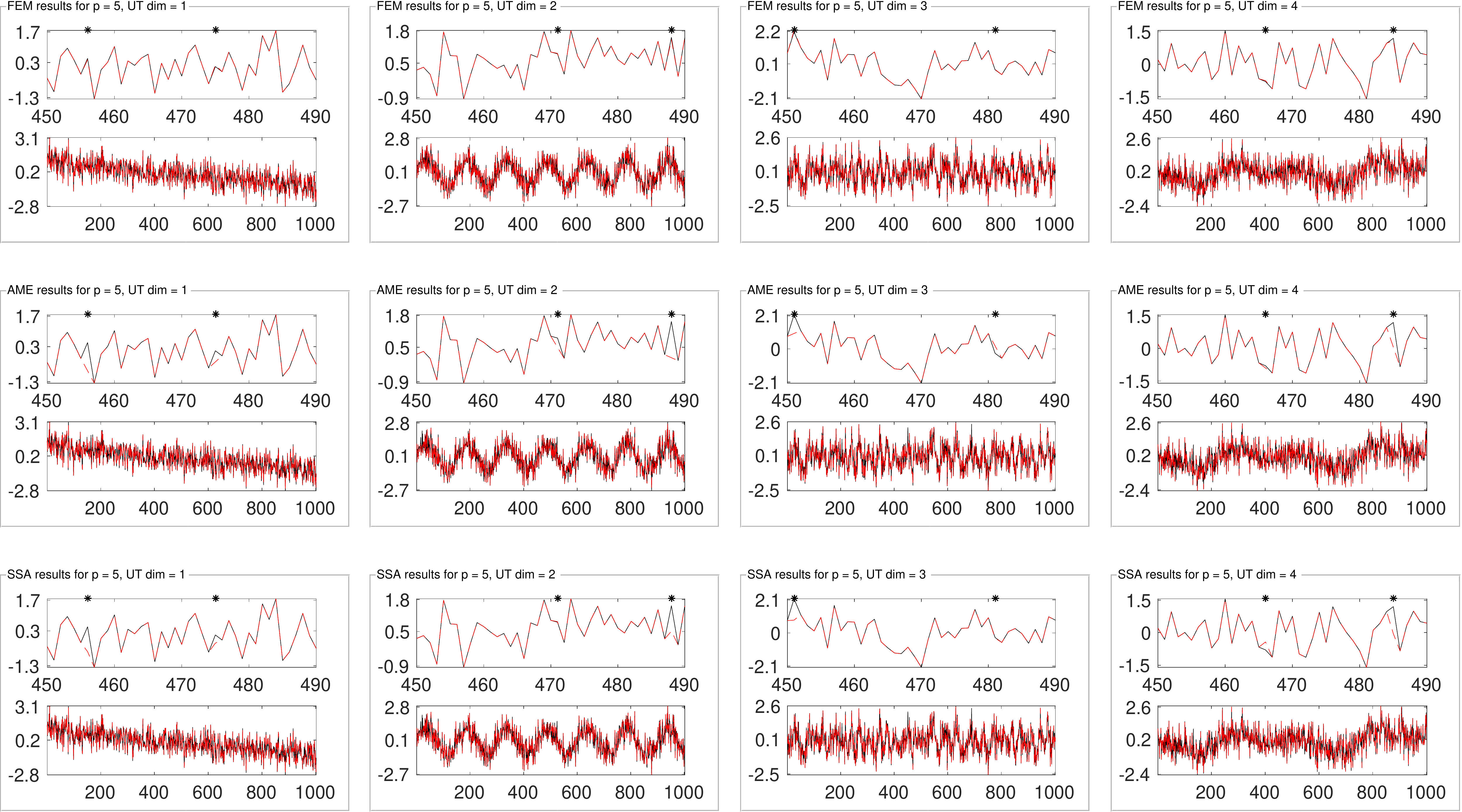}%
  \end{adjustbox}
\end{figure}

\begin{figure}[ht]
  \begin{adjustbox}{addcode={\begin{minipage}{\width}}{\caption{Case (c) for $p=35\%$: Reconstruction of $U_t$.}%
	    \label{fig:XTUT35_ut}
  		\end{minipage}},rotate=90,center}
      \includegraphics[scale=.4]{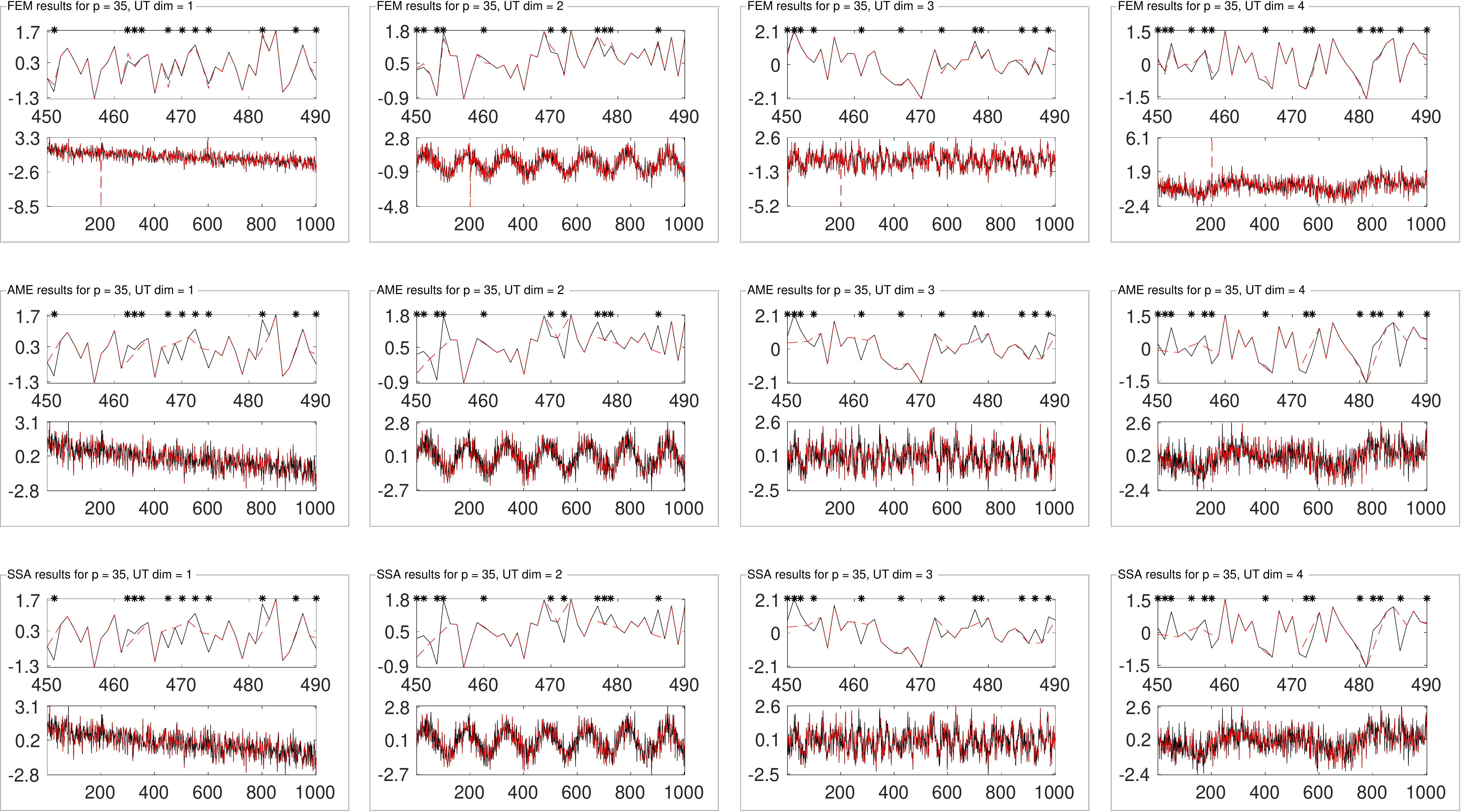}%
  \end{adjustbox}
\end{figure}

\begin{figure}[ht]
  \begin{adjustbox}{addcode={\begin{minipage}{\width}}{\caption{Case (c) for $p=55\%$: Reconstruction of $U_t$. %
	  }%
	    \label{fig:XTUT55_ut}
  		\end{minipage}},rotate=90,center}
      \includegraphics[scale=.4]{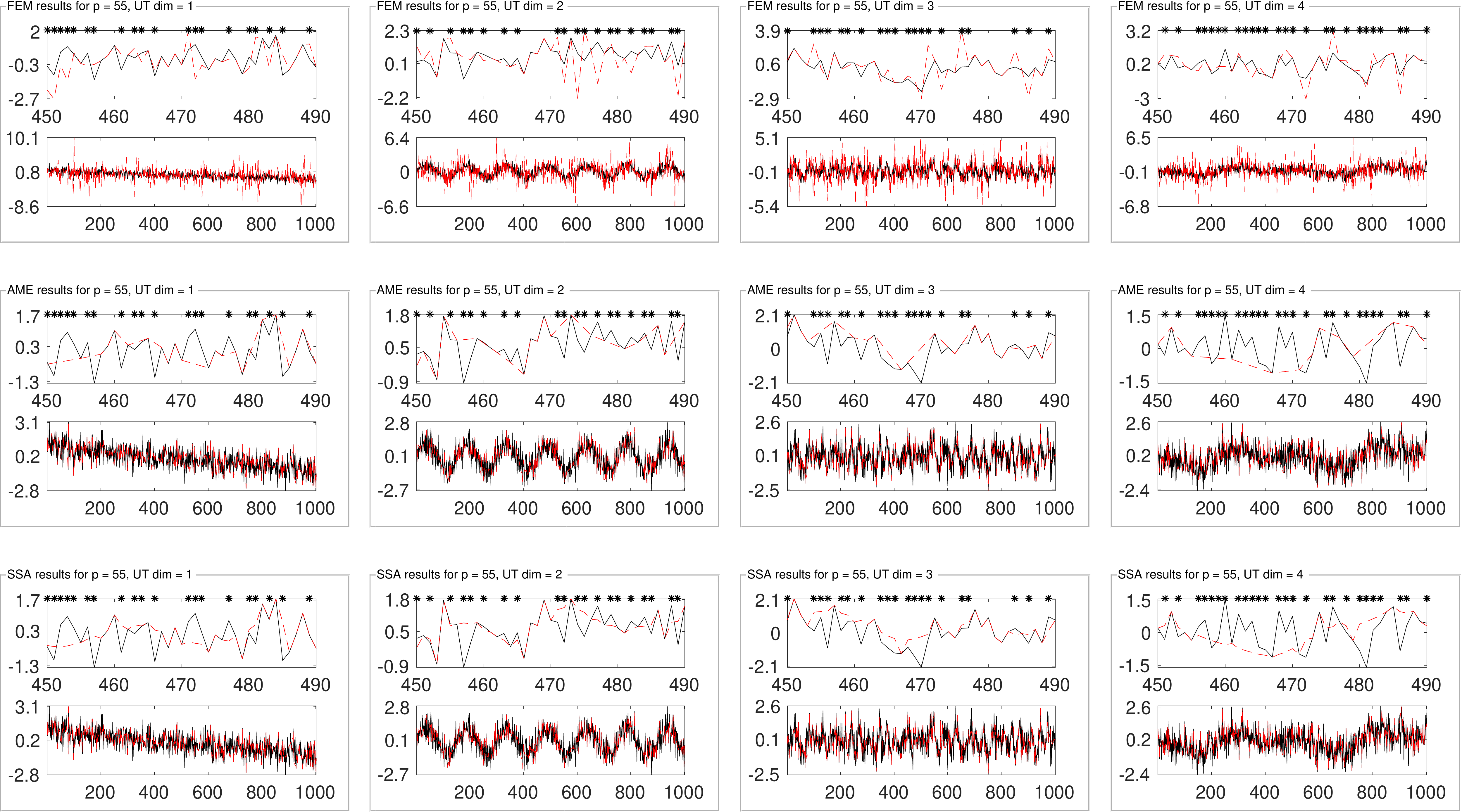}%
  \end{adjustbox}
\end{figure}

\begin{figure}[ht]
  \begin{adjustbox}{addcode={\begin{minipage}{\width}}{\caption{Case (c) for $p=75\%$: Reconstruction of $U_t$. %
	  }%
	    \label{fig:XTUT75_ut}
  		\end{minipage}},rotate=90,center}
      \includegraphics[scale=.4]{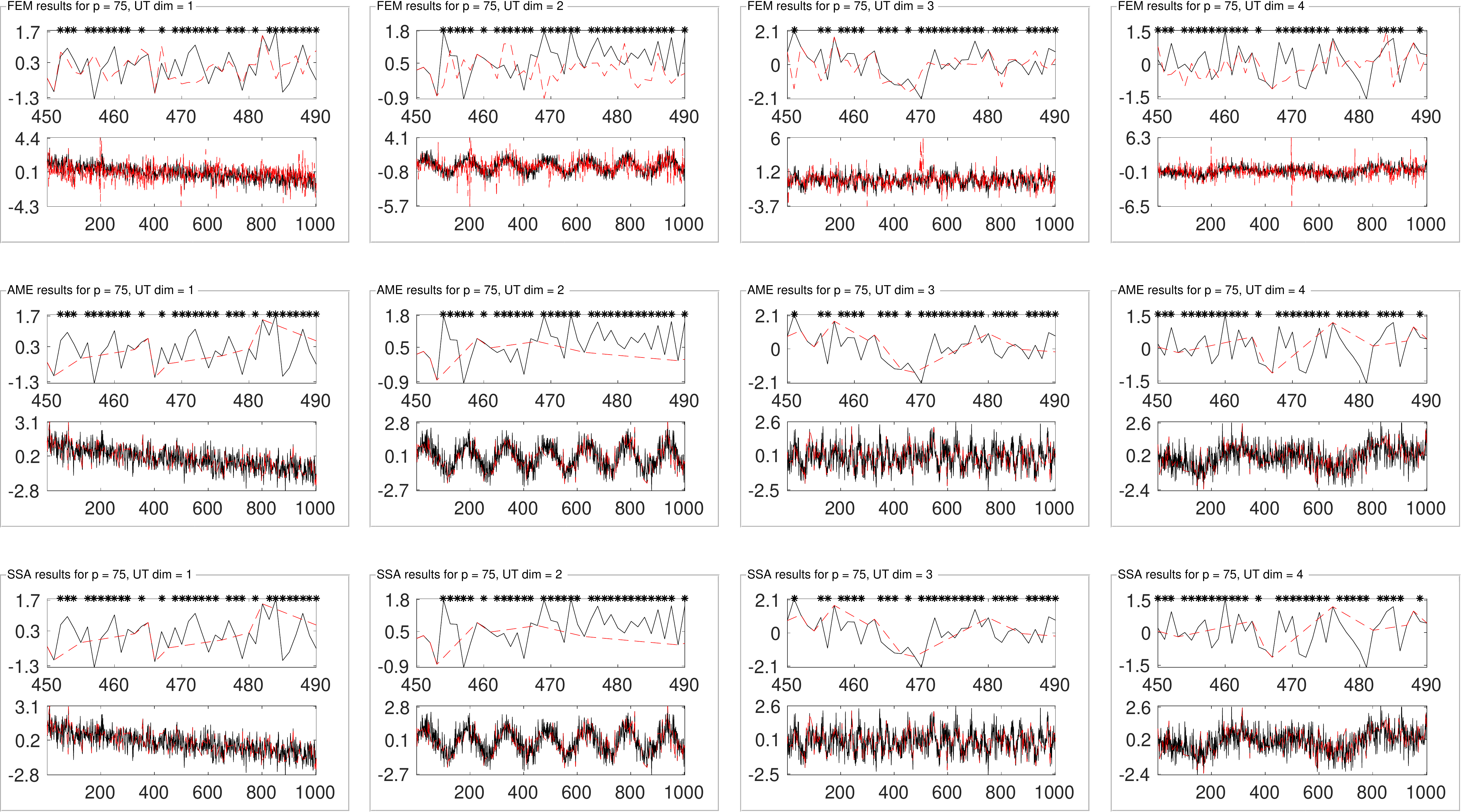}%
  \end{adjustbox}
\end{figure}

\begin{figure}[ht]
  \begin{adjustbox}{addcode={\begin{minipage}{\width}}{\caption{Case (c) for $p=95\%$: Reconstruction of $U_t$. %
	  }%
	    \label{fig:XTUT95_ut}
  	\end{minipage}},rotate=90,center}
      \includegraphics[scale=.4]{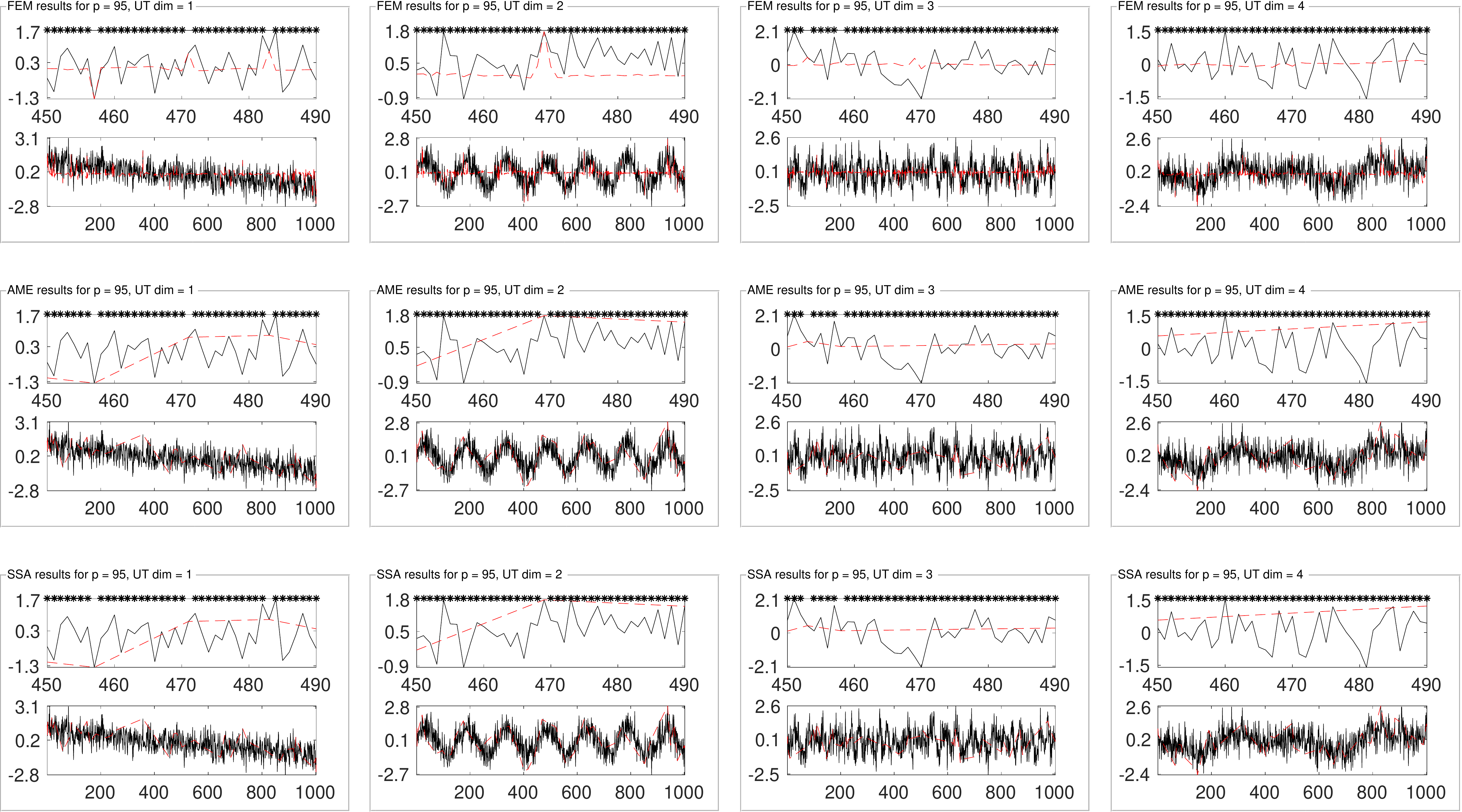}%
  \end{adjustbox}
\end{figure}

In this section the reconstruction of missing values obtained by applying the FEMM-VARX approach, AmeliaII and RSSA is compared explicitly. The reconstruction refers to the in Section~\ref{sec:numerical_examples} defined cases: (a) $X_t$ with $[5\%,\dots,95\%]$ missing values, $U_t$ full, (b) $X_t$ full, $U_t$ with $[5\%,\dots,95\%]$ missing values, (c) $X_t$ and $U_t$ each with $[5\%,\dots,95\%]$ missing values. To keep this section short, the reconstruction of missing values for these three cases is shown only for $[5\%,35\%,55\%,75\%,95\%]$ of missing values. \\%
The reconstruction of the missing values in $X_t$ for $p=[5\%, 35\%, 55\%, 75\%, 95\%]$ is shown in Figures~\ref{fig:XT5},~\ref{fig:XT35},~\ref{fig:XT55},~\ref{fig:XT75},~\ref{fig:XT95}. The reconstruction of the missing values in $U_t$ for $p=[ 5\%, 35\%, 55\%, 75\%, 95\%]$ is shown in Figures~\ref{fig:UT5},~\ref{fig:UT35},~\ref{fig:UT55},~\ref{fig:UT75},~\ref{fig:UT95}. For case (c), the reconstruction of $X_{miss}$ is shown in Figures~\ref{fig:XTUT5_xt},~\ref{fig:XTUT35_xt},~\ref{fig:XTUT55_xt},~\ref{fig:XTUT75_xt},~\ref{fig:XTUT95_xt} and the reconstruction of $U_{miss}$ in Figures~\ref{fig:XTUT5_ut},~\ref{fig:XTUT35_ut},~\ref{fig:XTUT55_ut},~\ref{fig:XTUT75_ut},~\ref{fig:XTUT95_ut}. %
The layout of these figures is described in the following. Each figure shows the reconstruction of the missing values by FEMM-VARX (first row of panels) AmeliaII (second row of panels) and RSSA (third row of panels). Each column corresponds to one dimension of the considered data. Each panel displays the reconstructed data (dashed red line) and the original data (solid black line). Hereby, the lower plot in a panel shows the data for all time steps and the upper plot shows a zoomed view on the data ($40$ time steps, from $t=450$ till $t=490$). The stars on the top of the zoomed view indicate the occurrence of the missing values. Please note, AmeliaII represents a MI approach, for visualization purposes we used one imputation. \\%
In general, FEMM-VARX aims to reconstruct the missing values exactly. In the figures, FEMM-VARX reconstruction shows first visible deviations from the true/exact values for $p = 35\%$. For the opponents, the visible deviations are present already from $p=5\%$. With an increasing $p$ the differences become larger and at a certain $p$ the reconstructed values start drifting apart from the true values, pointing to an increasing ill-posedness of the problem. The incorporated Ridge regularization seems to be not sufficient any more and stronger regularization, i.e., a priori assumptions, might be necessary. For instance, Gaussian assumption about the underlying dynamics of the missing values, as it is done by the opponents. However, FEMM-VARX captures the underlying dynamics quite good almost up to $p=95\%$, depending upon the case. \\%
AmeliaII aims to reconstruct the average underlying dynamics by assuming Gaussian distribution of the data. Reconstruction obtained by AmeliaII tends visually towards linear interpolation between the observed values. The deviations from exact values become worse in cases when multiple missing values occur one by one, especially, when the blocks of missing values cover some peaks or when missing values cover the transitions between the clusters completely.\\ %
RSSA performs qualitatively similar to AmeliaII. Here, the reconstruction of the missing values is based on the principal eigenvectors of the observed data and thus is not necessary linear. Please note, that for case (a) with $p=75\%$ and for case (c) for $p=[35\%,75\%,95\%]$ for both $X_{miss}$ and $U_{miss}$ the reconstruction by RSSA is very similar to the one obtained by AmeliaII. %
%
%
%
%
%
%
%
%


\end{document}